\documentclass[12pt,letterpaper]{JHEP3}
\pdfoutput=1
\usepackage{cite}
\usepackage{graphicx}
\usepackage{url}
\usepackage{subfig}
\usepackage{setspace}
\usepackage{psfrag}
\usepackage{amsmath}

\newskip\zatskip \zatskip=0pt plus0pt minus0pt
\def\matth{\mathsurround=0pt}

\def\atversim#1#2{\lower0.7ex\vbox{\baselineskip\zatskip\lineskip\zatskip
  \lineskiplimit 0pt\ialign{$\matth#1\hfil##\hfil$\crcr#2\crcr\sim\crcr}}}

\def\neu2{\tilde \chi_2^0}
\def\neu1{\tilde \chi_1^0}
\def\jet{\mathrm{jet}}

\def\MET{\ifmmode E_T^{\mathrm{miss}} \else $E_T^{\mathrm{miss}}$\fi}
\def\MEFF{\ifmmode M_{\mathrm{eff}} \else $M_{\mathrm{eff}}$\fi}
\def\MMIN{\ifmmode M_{\mathrm{min}} \else $M_{\mathrm{min}}$\fi}
\def\Re{{\cal R \mskip-4mu \lower.1ex \hbox{\it e}\,}}
\def\Im{{\cal I \mskip-5mu \lower.1ex \hbox{\it m}\,}}
\def\ie{{\it i.e.}}
\def\eg{{\it e.g.}}

\def\etal{{\it et al.}}

\def\sub#1{_{\lower.25ex\hbox{$\scriptstyle#1$}}}
\def\tev{\,{\rm TeV}}
\def\gev{\,{\rm GeV}}

\def\to{\rightarrow}

\def\subw{_{\rm w}}
\def\mh{\ifmmode m\sbl H \else $m\sbl H$\fi}
\def\mch{\ifmmode m_{H^\pm} \else $m_{H^\pm}$\fi}
\def\mt{\ifmmode m_t\else $m_t$\fi}
\def\mc{\ifmmode m_c\else $m_c$\fi}
\def\mz{\ifmmode M_Z\else $M_Z$\fi}
\def\mw{\ifmmode M_W\else $M_W$\fi}
\def\mws{\ifmmode M_W^2 \else $M_W^2$\fi}
\def\mhs{\ifmmode m_H^2 \else $m_H^2$\fi}   
\def\mzs{\ifmmode M_Z^2 \else $M_Z^2$\fi}
\def\mts{\ifmmode m_t^2 \else $m_t^2$\fi}
\def\mcs{\ifmmode m_c^2 \else $m_c^2$\fi}
\def\mchs{\ifmmode m_{H^\pm}^2 \else $m_{H^\pm}^2$\fi}
\def\ztwo{\ifmmode Z_2\else $Z_2$\fi}
\def\zone{\ifmmode Z_1\else $Z_1$\fi}
\def\mtwo{\ifmmode M_2\else $M_2$\fi}
\def\mone{\ifmmode M_1\else $M_1$\fi}
\def\tb{\ifmmode \tan\beta \else $\tan\beta$\fi}
\def\xw{\ifmmode x\subw\else $x\subw$\fi}
\def\ch{\ifmmode H^\pm \else $H^\pm$\fi}
\def\lum{\ifmmode {\cal L}\else ${\cal L}$\fi}
\def\inpb{\ifmmode {\rm pb}^{-1}\else ${\rm pb}^{-1}$\fi}
\def\infb{\ifmmode {\rm fb}^{-1}\else ${\rm fb}^{-1}$\fi}
\def\epem{\ifmmode e^+e^-\else $e^+e^-$\fi}
\def\ppb{\ifmmode \bar pp\else $\bar pp$\fi}
\def\pbp{\ifmmode ~^(\bar p^)p\else $~^(\bar p^)p$\fi}
\def\bsg{\ifmmode B\to X_s\gamma\else $B\to X_s\gamma$\fi}
\def\bsll{\ifmmode B\to X_s\ell^+\ell^-\else $B\to X_s\ell^+\ell^-$\fi}
\def\bstt{\ifmmode B\to X_s\tau^+\tau^-\else $B\to X_s\tau^+\tau^-$\fi}

\def\matth{\mathsurround=0pt}

\author{John A. Conley\\
        Physikalisches Institut, Universit\"at Bonn, Nu\ss allee 12, 53113 Bonn, Germany}
\author{James S. Gainer\\
High Energy Physics Division, Argonne National Laboratory,
Argonne, IL 60439 USA and \\
Department of Physics and Astronomy, Northwestern University, Evanston, IL 60208 USA}
\author{JoAnne L. Hewett, My Phuong Le, Thomas G. Rizzo\\
        SLAC National Accelerator Laboratory, 2575 Sand Hill Rd., Menlo Park, CA  94025, USA} 
\title{Supersymmetry without prejudice at the LHC}
\preprint{SLAC-PUB-14094\\ANL-HEP-PR-10-15\\nuhep-th/10-06\\Bonn-TH-2010-03}

\abstract{The discovery and exploration of Supersymmetry in a model-independent fashion will be a daunting task due to 
the large number of soft-breaking parameters in the MSSM. In this paper, we explore the 
capability of the ATLAS detector at the LHC ($\sqrt s=14$ TeV, 1 fb$^{-1}$) to find SUSY within 
the 19-dimensional pMSSM subspace of the MSSM using their standard transverse missing energy and long-lived particle 
searches that were essentially designed for mSUGRA. To this end, we employ a set of $\sim 71$k 
previously generated model points in the 19-dimensional parameter space that satisfy all of the existing 
experimental and theoretical constraints. Employing ATLAS-generated SM backgrounds and following 
their approach in each of 11 missing energy analyses as closely as possible, we explore all of these 
$71$k model points for a possible SUSY signal. To test our analysis procedure, we first verify 
that we faithfully reproduce the published ATLAS results for the signal distributions for their 
benchmark mSUGRA model points.  We then show that, requiring all sparticle masses to lie below 1(3) 
TeV, almost all(two-thirds) of the pMSSM model points are discovered with a significance $S>5$ in at 
least one of these 11 analyses assuming a 50\% systematic error on the SM background. If this 
systematic error can be reduced to only 20\% then this parameter space coverage is increased. 
These results are indicative that the ATLAS SUSY search strategy is robust
under a broad class of Supersymmetric models.
We then explore in detail the properties of the kinematically accessible model 
points which remain unobservable by these search analyses in order to ascertain problematic cases which 
may arise in general SUSY searches.} 

\begin{document}


\section{Introduction}

The LHC has recently begun operations, providing our first direct glimpse of the Terascale
in a laboratory setting, and new physics discoveries are widely expected.
Supersymmetry (SUSY) is one of the most attractive candidates out of a
litany of potential theories beyond the Standard Model (SM) as it contains a natural
dark matter candidate, addresses the weak hierachy problem, and provides a
framework for unification of the forces \cite{Drees:2004jm,Baer:2006rs}.  
However, evidence for Supersymmetry
has yet to be observed \cite{Amsler:2008zzb}; hence it cannot exist in its most fundamental form 
and must be a broken symmetry.  Various mechanisms for the breaking of
Supersymmetry have been proposed 
\cite{Cremmer:1982vy,Giudice:1998bp,Dine:1995ag,Randall:1998uk,Giudice:1998xp,Chacko:1999mi,Kaplan:1999ac}, 
each predicting a characteristic
sparticle spectrum leading to distinctive signatures in colliders and other
experiments.  Of these, gravity mediated Supersymmetry breaking (mSUGRA) is the
most often studied; it contains 5 parameters at the unification scale and thus 
greatly simplifies the exploration of the vast Supersymmetric parameter space.  In particular,
most searches for Supersymmetry at the Tevatron \cite{:2007ww} and the planned search strategies at 
the LHC \cite{Aad:2009wy} have been designed solely in the context of mSUGRA.

The question then arises of how well mSUGRA describes the true breadth of the Minimal
Supersymmetric Standard Model (MSSM) and its possible collider signatures.  It is well-known
\cite{Berger:2008cq,Alwall:2008ve,Alwall:2008va}
that the D0 constraints on squark and gluino production do not hold within a broader class
of SUSY models and that much lighter sparticles ($\sim 150$ GeV) can easily evade these
searches.  This poses a potentially worrisome prospect for the LHC search strategies and their
effectiveness needs to be checked on an extended class of SUSY models.  This provides the
motivation for our work.

In particular, we base our analysis on the recent study published by the ATLAS detector
collaboration \cite{Aad:2009wy}.  Here, the collaboration performed an
extensive examination of a
set of 7 SUSY benchmark points, all of which are based on mSUGRA,
and constructed most of their SUSY search analysis suite from these investigations.  
In this work, we will simulate the ATLAS search analyses, pass an extensive set of
broad-based SUSY models through each search channel, and determine their observability. 
We believe that the results will be
indicative of the robustness of the planned ATLAS SUSY search analysis suite.
In order to perform this test of the ATLAS searches, we strictly adhere to the analyses
as designed by ATLAS, word for word, cut by cut.  While numerous, and perhaps
improved, SUSY collider search techniques have been discussed in the literature \cite{Pape:2006ar,Weiglein:2004hn}, 
it is not our purpose here to discuss or employ them. Instead, we focus here on the 
planned searches that will be performed on actual data.  

We make use of a recent comprehensive bottom-up exploration of the MSSM performed by Berger 
\etal\cite{Berger:2008cq}. In this work, no reference was made to theoretical
assumptions at the high scale or to the mechanism of Supersymmetry breaking.  The theoretical assumptions
were minimal and included only CP conservation, Minimal Flavor Violation, that
the lightest supersymmetric particle (LSP) be
identified with the lightest 
neutralino and be a thermal relic, and that the first and second sfermion
generations be degenerate in mass with negligible Yukawa couplings.  
Enforcing this minimal list of assumptions results in the pMSSM (phenomenological MSSM)
with 19 real weak-scale parameters.  
A scan of $10^7$ points in this 19-dimensional
parameter space was performed over 
ranges chosen to ensure large sparticle production cross sections at the LHC.
Each model (or point in the 19-dimensional space) was 
subjected to a global set of
constraints from spectrum requirements, electroweak precision data, heavy flavor physics, 
cosmological considerations, and LEP and Tevatron collider searches.  
Approximately 70,000 models in the pMSSM scan survived all of the restrictions and were 
found to be phenomenologically viable.
(Interestingly, subjecting the seven ATLAS mSUGRA
benchmark points to these same constraints results in only one of the points
being consistent with the global data set.)
A wide variety of properties and characteristics
were found in this model sample, with features that imply a very large range of possible
predictions for collider signatures.

Specifically, we set up an analysis for each of the 11 search
channels studied in the ATLAS CSC book\cite{Aad:2009wy} and ensure that we reproduce
the CSC results for each of the ATLAS benchmark points in each channel.  
We will then run the $\sim$70k pMSSM points of Berger et. al. through each analysis
channel and perform a statistical test to ascertain the observability of
each model.  We will find that several pMSSM models cannot be detected by the
ATLAS SUSY analysis suite and we will further examine these special cases and ascertain
which characteristics in the sparticle spectrum render then unobservable.  In many cases we
find that the systematic errors associated with the SM backgrounds are the main cause of the
lack of a statistically viable discovery signal and note that a reduction in these errors
would greatly improve the likelihood of discovering SUSY.   We
will also look at a qualitatively different collider signature, that of stable
supersymmetric particles.  Our model set contains a large number of models
with stable sparticles of various identities, and we will evaluate the
prospects of observing these stable sparticles at the LHC.

The next section describes the generation of our model set,
Section \ref{analysisproc} discusses our procedure and analysis set-up, Section \ref{results1} contains
our main results, Section \ref{stabsec} discusses stable particles in our
model set, and our conclusions can be found in Section \ref{conclusions}.


\section{Review of Model Generation}
\label{modelrevgen}

In this section we provide a brief overview of our previously performed model generation procedure;
full details and all original references are given in Refs.{\cite {Berger:2008cq,Cotta:2009zu,Cotta:2009be}}. 

\subsection{Parameter scans}

In performing our exploration of the 19 soft-breaking parameters of the pMSSM, we first 
determine the ranges that we scan over for these parameters as well as 
how their specific values are selected within these ranges. Recall, as discussed above, that these parameters are defined at the TeV scale.  
In our analysis, we employed two independent scans of the pMSSM parameter space with the ranges being fixed such 
that large production cross sections for SUSY particles are likely at the 14 TeV LHC. This means that we will have two independent 
sets of models to examine for LHC SUSY signatures employing the ATLAS analyses. In the model set generated by the first scan, denoted here as 
the FLAT prior set, $10^7$ n-tuples of the (n=)19 parameters were randomly generated, assuming 
flat priors, where the parameter values were chosen {\it uniformly} throughout the ranges:
\begin{eqnarray}
100 \gev \leq m_{\tilde f} \leq 1\tev \,, \nonumber\\
50\gev \leq |M_{1,2},\mu|\leq 1 \tev\,, \nonumber \\ 
100 \gev \leq M_3\leq 1 \tev\,, \nonumber \\ 
|A_{b,t,\tau}| \leq 1 \tev\,, \\
1 \leq \tan \beta \leq 50\,, \nonumber \\ 
43.5\gev \leq m_A \leq 1 \tev\,. \nonumber  
\end{eqnarray}
Here the absolute value signs are present to allow the soft-breaking parameters to have arbitrary sign. To generate the models 
for the second scan, denoted here as the LOG prior set, $2\times 10^6$ 
n-tuples of the (n=)19 parameters were generated, assuming log priors for ({\it only}) the mass parameters with the modified ranges:
\begin{eqnarray}
100 \gev \leq m_{\tilde f} \leq 3\tev \,, \nonumber\\
10\gev \leq |M_{1,2},\mu|\leq 3 \tev\,, \nonumber \\ 
100 \gev \leq M_3\leq 3 \tev\,, \nonumber \\ 
10\gev\leq |A_{b,t,\tau}| \leq 3 \tev\,, \\
1 \leq \tan \beta \leq 60\,, \nonumber \\ 
43.5\gev \leq m_A \leq 3 \tev\,. \nonumber  
\end{eqnarray}
It is important to note that the parameter $\tan \beta$, being a dimensionless quantity, is still being scanned in a flat prior manner, unlike the other parameters, 
when we generate this model set. The expanded parameter range in this case allows for some access to both very light as well as some heavy 
sparticle states that may only be observed at the SLHC. 
The primary goal of this second scan was to compare these results to those of the flat prior study in order to determine the degree that the resulting 
model properties depend on the scan assumptions and whether any possible bias was introduced. 
We found that both scans yield qualitatively similar results, but that the detailed predictions in the two 
cases can be quantitatively different in several aspects. The physical spectra for the sparticles themselves were generated in all cases using the code 
SuSpect2.34{\cite {Djouadi:2002ze}}. 

\subsection{Constraints}

We now turn to a discussion of the theoretical and experimental constraints that we imposed on the set of models generated from these two scans.

\subsubsection*{Theoretical constraints}

The theoretical restrictions we included are standard and were applied 
while generating the sparticle spectrum with the SuSpect code: ($i$) the spectrum must be tachyon free, ($ii$) the spectrum cannot lead to color or charge 
breaking minima,  
($iii$) electroweak symmetry breaking must be consistent, and ($iv$) the Higgs potential is bounded from below. Furthermore, we employed the assumption 
that ($v$) the WIMP LSP is a conventional thermal relic and is identified as the lightest
neutralino. We also imposed the requirement of ($vi$) Minimal Flavor Violation
(MFV) \cite{D'Ambrosio:2002ex} at the 
TeV scale to reduce the impact of Supersymmetry on flavor physics. In this case, the SUSY contributions to flavor physics are mostly controlled by 
the Yukawa couplings and the CKM matrix.  

\subsubsection*{Constraints from precision measurements}

We then imposed experimental constraints from precision electroweak observables, flavor physics, astrophysical measurements, and collider searches
for SUSY particles.  The code
micrOMEGAs2.21 \cite{Belanger:2007zz,Belanger:2008sj,Belanger:2001fz,Belanger:2004yn,Belanger:2006is} takes the MSSM spectrum output from SuSpect and 
implements the restrictions arising from a number of precision and flavor measurements:  we required that the
precision electroweak constraints obtained via possible shifts in the 
$\rho$ parameter, $\Delta \rho$, as well as the rare decays $b\to s\gamma$ and $B_s \to \mu^+\mu^-$ be consistent with their measured values. 
Given the current theoretical and experimental 
uncertainties for the value of the $g-2$ of the muon, we implemented the 
loose requirement that $(-10 \leq \Delta(g-2)_\mu \leq 40)\cdot 10^{-10}$ in our analysis. In addition to these constraints which are essentially built into the 
micrOMEGAs2.21 code, we demanded consistency with the measured value of the branching fraction for $B\to \tau \nu$ and required that the ratio of first/second to the
third generation squark soft breaking masses (of a given charge and helicity) differ from unity by no more than a factor of $\sim 5$ to satisfy the bounds from 
meson-anti-meson mixing. 

\subsubsection*{Dark matter constraints}

We employed two constraints that arise from the Dark Matter (DM) sector:
first, we required that the calculated DM relic 
density not exceed the limit obtained from  
the 5 year WMAP measurement, thus allowing for the possibility that the DM
sector consists of multiple components besides the lightest neutralino.{\footnote {Note that although we did {\it not} require 
the WMAP bound to be saturated this 
condition is satisfied in a reasonable subset of our resulting models.}} Second, we imposed the search constraints from the DM direct detection experiments, 
allowing for a factor of 4 uncertainty in the calculation of the cross section 
from possible variations in the input parameters and matrix elements. These calculations were also performed with 
the micrOMEGAs2.21 code. 

\subsubsection*{Tevatron constraints}

Collider searches, of course, play an important role in placing constraints on the pMSSM parameter space. Since the Tevatron searches for SUSY  
are closer in spirit to the LHC analysis we present below, we discuss them in more detail than the corresponding investigations from LEP. 
We first consider the restrictions imposed on the squark and gluino sectors 
arising from the null result of the multijet plus missing energy search
performed by D0 \cite{:2007ww} that is based on mSUGRA. In our study, we generalized 
their analysis to render it model independent.  For each of our pMSSM models, we computed the NLO SUSY cross sections for squark and gluino production using 
PROSPINO2.0 \cite{Beenakker:1996ed,Beenakker:1996ch,Beenakker:1997ut,Beenakker:1999xh,Spira:2002rd,Plehn:2004rp}. The decays for these sparticles were  
computed via SDECAY/HDECAY (\ie, SUSYHIT1.1){\cite {Djouadi:2006bz} to obtain the relevant decay chains 
and branching fractions and these results were 
then passed to PYTHIA6.4{\cite {Sjostrand:2006za}} for hadronization and fragmentation. We then used PGS4{\cite {PGS}} to simulate the D0 detector and impose the 
kinematic cuts for the analysis; PGS4 was tuned to reproduce the results and efficiencies for the three benchmark mSUGRA points employed by D0 in their published multijet 
study. For an integrated luminosity of 2.1 $\infb$, we found that the $95\%$ CL upper limit on the number of signal events from combining all of the production 
channels was 8.34, where we employed the statistical method of Feldman and Cousins{\cite {Feldman:1997qc}}. 
Models with event rates larger than this were then removed from further consideration.  
Interestingly, light squarks and gluinos (masses of order 150-200 GeV) with small mass splittings with the LSP survive this analysis\cite{Alwall:2008ve,Alwall:2008va}. 

Analogously, we employed constraints from the CDF search for trileptons plus missing energy{\cite {Aaltonen:2008pv}}, which we also generalized to the pMSSM 
using essentially the same method as in the jets plus plus missing energy analysis described above.  Here, we employed a CDF tune for PGS4 which we obtained by 
reproducing the CDF benchmark point results.  We used the leading order cross section together with a universal K-factor of 1.3 to mimic the full NLO 
cross section. Specifically in this case, we only made use of the `3 tight lepton' analysis from CDF as it is the easiest to implement with PGS4. 
The $95\%$ CL upper bound on a possible SUSY signal in this channel was then found to be  4.65 events assuming a luminosity of 2.02 $\infb$ as used in the 
CDF analysis. Again, pMSSM  parameter sets leading to larger event rates were dropped from the remainder of our analysis. 

In order to satisfy the large number of stop and bottom searches at the
Tevatron\cite{Brigliadori:2008vf,Abazov:2006fe,Buescher:2005he,Abazov:2008rc,Aaltonen:2007sw}, 
we simply required that the masses of the lightest stop and sbottom 
be larger than that of the top quark, $\simeq 175$ GeV. However, an examination of the various sparticle spectra {\it {a posteriori}} reveals that this cut makes 
very little impact on our final model set.

Both CDF{\cite {Abe:1992vr}} and D0{\cite {Abazov:2008qu}} have placed limits on the direct production of heavy stable charged particles. In our analysis we employed the 
stronger D0 constraint which can be taken to have the form $m_{\chi^+}\geq 206 |U_{1w}|^2 +171 |U_{1h}|^2$ GeV at $95\%$ CL in the case of chargino production.  Here, the 
matrix $U$ determines the Wino/Higgsino content of the lightest chargino and was used to interpolate between the separate purely Wino or Higgsino results quoted by D0. 
This resulted in a {\it very} powerful constraint on the pMSSM since chargino-LSP mass degeneracies are common in our model sample, 
particularly when the LSP is nearly a pure Wino or Higgsino or a combination of these two cases. 

\subsubsection*{LEP constraints}

We imposed a large number of constraints arising from the direct searches for both SUSY partners and the extended MSSM Higgs sector from LEP data. 
As for the Tevatron, most of the LEP analyses have been carried out in the mSUGRA framework and thus need careful reconsideration when they are extended to 
cover the more general pMSSM scenario considered here. 
For brevity, we will only mention the details of a few of these here, with a complete discussion of all these constraints being given in our previous 
work{\cite {Berger:2008cq}}. Two of these restrictions arise from $Z$-pole data: 
($i$) we required that the Higgs boson as well as all new charged particles have masses in excess of $M_Z/2$ and also that all new (detector) stable 
charged particles have masses in excess of 100 GeV{\cite {Benelli:2003mq}. Furthermore, ($ii$) we required that $Z$ decays into stable and long-lived neutralinos not 
contribute more than 2 MeV{\cite {LEPEWWG}} to the invisible
width of the $Z$ boson.{\footnote {We note that for the range of sfermion mass  soft breaking parameters we consider, 
$Z$ decay to pairs of sneutrinos is not kinematically allowed so that this final state cannot contribute in any way to the invisible width.}} 

ALEPH{\cite {Barate:1999cn}} has placed a lower limit of 92 GeV on the light squark masses, assuming that the gluino is more massive than the squarks, via their decay 
to a jet+LSP (\ie, jet + missing energy) provided that the mass difference between the squark and the LSP ($\Delta m$) is $\geq 10$ GeV to avoid very soft jets.
We employed this constraint directly, including the $\Delta m$ cut. For light sbottoms, the same sort of decay pattern results in a lower bound of 95 GeV on their mass.
Lower bounds have been placed{\cite {LEPSUSY}} on the masses of right-handed sleptons decaying to leptons plus missing energy of $m\gtrsim 100(95,90)$ GeV for the 
selectron(smuon,stau).  This is, however, only applicable if the slepton masses are at least a few percent larger than that of the LSP, otherwise the final state 
leptons will again be too soft. Our analysis allows for the appearance of this small mass gap. These constraints are also applicable to left-handed sleptons provided 
the corresponding Wino $t-$channel exchange contribution is not very important, an assumption made in our analysis. 
An analogous situation applies to chargino production. If the LSP-chargino mass splitting is $\Delta m >2$ GeV, a direct lower 
limit of 103 GeV on the chargino mass is obtained from LEPII data.  However, if this splitting is $\Delta m <2$ GeV, the 
bound degrades to 95 GeV, {\it provided} that also $\Delta m >50$ MeV, otherwise the chargino would appear as a stable particle in the 
detector and would then be excluded by the stable particle searches discussed above. In the case where the lightest chargino is dominantly Wino, this limit 
is found to be applicable only when the electron sneutrino is more massive than 160 GeV. 

For constraints on the Higgs sector, we imposed the five sets of bounds on the MSSM Higgs sector masses and couplings provided by the LEP Higgs Working 
Group{\cite {LEPHIGGS}}. To do this, we employed the SUSY-HIT routine, recalling that the uncertainty on
the calculated mass of the lightest Higgs boson is approximately 
3 GeV{\cite {Heinemeyer:2004gx}} as determined by SuSpect.

\subsubsection*{Surviving models}

After all of these constraints were imposed, we found that $\sim 68.5\cdot 10^3$ models out of our original sample of $10^7$ pMSSM points in the flat prior set
satisfied all of the restrictions.  In the log prior sample of $2\cdot 10^6$ pMSSM points, only $\sim 2.8\cdot 10^3$ models survived the same constraints.  
As mentioned above, the properties and characteristics of the surviving sets of models from the two scans are qualitatively similar.
We will now examine the production of these $\sim 71.3\cdot 10^3$ viable pMSSM models at the LHC considering the two model sets independently.  
We remind the reader that we refer to each of these points in the pMSSM parameter space as a model.

A wide variety of properties and characteristics were found in this 70k model sample.
In some instances, surprisingly light sparticles (\eg, $\sim 180\gev$ squarks and gluinos)
are still allowed by the data.  The most favored identity of the next-to-lightest
supersymmetric particle (nLSP) was the lightest chargino, followed by the second
lightest neutralino.  However, ten other sparticles (including the right-handed selectron,
the gluino and the up squark) can also play the role of the nLSP
with roughly equal probabilities.  The mass splitting between the LSP and nLSP, a
crucial parameter for collider signals, was found to have a large range spanning
seven orders of magnitude from approximately 100 keV to 100's of GeV.  Over 1100
distinct classification patterns\cite{Feldman:2008hs} were found for the content and ordering
of the four lightest sparticles in the spectrum.  These features imply a very large
range of possible predictions for collider signatures within the
pMSSM.


\section{Analysis Procedure for Inclusive SUSY Production at the LHC}
\label{analysisproc}

As discussed in the Introduction, the single, most important goal of this study is to explore how well the inclusive SUSY searches formulated by the ATLAS
collaboration, designed with mSUGRA in mind, perform when they are applied to the larger and much more general
pMSSM parameter space. To that end, we attempt to follow these analyses as presented by 
ATLAS itself in detail in Ref.{\cite{Aad:2009wy}} as closely as possible. The justification for the choices of specific analysis cuts, the 
size and nature of SM backgrounds and the associated systematics as well as the use of the statistical tests for discovery employed here are the {\it same} 
as those employed by ATLAS and are thus all given and discussed in detail in this reference. These are necessary choices if we are to make a direct comparison to 
the ATLAS mSUGRA study.  

In particular, we examine the eleven
search channels as detailed by ATLAS in this reference:   at least 4(2) jets + \MET~[4(2)j0l],\, at least 4(3,2) jets +
exactly one lepton + \MET~[4(3,2)j1l],\, opposite-sign 
dileptons + at least 4 jets + \MET~[OSDL],\, same-sign 
dileptons + at least 4 jets + \MET~[SSDL],\, three leptons + at least one jet + \MET~[3lj],\,
three leptons + \MET~inclusive [3lm],\, at least one $\tau$ + 4j +
\MET~[$\tau$],\, 
and at least 4 jets with at least two b-tags + \MET~[$b$].  Here, the term listed in brackets for each channel is the
`nickname' that we will use throughout the paper for that analysis.  We considered 85 SUSY production processes that contribute to these 11 signatures.

In order to perform our analysis, we must first 
determine the size and properties of the SM backgrounds to the various analysis signatures listed above.  To this end,
we obtained more details of the results and distributions for the SM backgrounds that were generated by ATLAS itself and was presented in Ref.~{\cite{Aad:2009wy}} 
from the ATLAS SUSY Group~{\cite {ATLAS_SUSY}}. This essentially allowed us to directly employ the ATLAS computed backgrounds in our analysis and 
we did not need to generate any of the SM background ourselves. Provided with these backgrounds we were thus able to perform a better direct comparison of 
our results with the ATLAS mSUGRA studies and this permitted us to concentrate on generating the
expected signal rates for each of these eleven searches for all of the parameter space points in our $\sim$ 71k pMSSM model sample.

\subsection{Generation of the Signal Events}

Several steps were employed in the generation of the signal events for the ATLAS search strategies for our set
of pMSSM models.  First, the SUSY spectra and corresponding 
sparticle and Higgs boson decay tables were generated using a modified version of SUSY-HIT.  As phase space issues can be very important 
in our model set, due to the large number of sparticle near-degeneracies, our modifications included
the incorporation of the light quark (u,d,s,c) and lepton (e and $\mu$) masses in the calculation of branching 
fractions and lifetimes for the various sparticles. For two body decays, we implemented the expressions for the decay with the masses included, while for
three-body decays, we only modified the phase-space cutoff to take into account the mass effects. 
We note that in the case of the light quarks, the hadronization products of the quarks have significantly higher masses than the corresponding bare masses
of the quarks.  We therefore included the mass of the lightest meson of the appropriate type in the relevant phase space cut-offs.
Since it is not uncommon for the mass splitting between $\tilde b_1$ and the LSP to be below the $B$ meson mass, $\simeq 5.3$ GeV, we 
also included the 1-loop processes $\tilde b_1 \to (d,s)+$LSP in the decay tables. Also, since there are many models that have charginos which are 
close in mass to the LSP, we included the full expressions for the chargino decays in the $e\nu$, $\mu \nu$, and 
1-3 pion plus LSP final states \cite{Chen:1996ap,Chen:1999yf}. These were employed for mass splittings below 1~\gev.  

We also
included CKM-suppressed decays of sbottoms, which, as discussed again later, allowed for the decay of bottom squarks with close mass splittings with the LSP.
Another set of modifications was necessary to correctly include four-body final states in the decays of stop squarks with small mass splittings.
SUSY-HIT includes formulae for the decay $\tilde{t}\to\tilde\chi_1^0~b~f~f'$, where $f, f'$ are assumed to be massless fermions.  
We modified the code to compute the decay width to a specific pair of fermions, including phase-space cutoffs, using the appropriate
fermion masses.

In addition, in some cases, the QCD corrections to particular partial widths, most commonly for stop and sbottom decays to Higgs/gauginos and heavy quarks,
were turned off as they led to negative 
branching fractions.  This occurred due to a poor choice of scale and/or a lack of resummation of large QCD correction terms.
Yet another set of corrections to the decay tables was necessary
in order to resolve PYTHIA errors that occurred; see the discussion below for more details. 

Next, the NLO cross 
sections for the $\sim 85$ SUSY production processes we considered were computed using a modified version of 
PROSPINOv.2.1~\cite{Beenakker:1996ch,Beenakker:1997ut,Beenakker:1999xh,Spira:2002rd,Plehn:2004rp} 
that avoided potentially negative K-factors due to sign issues associated with
the neutralino masses.  This modification is now implemented in the current
version of the code.
Processes involving $\tau$-sneutrinos or charged Higgs production are not supported
by the current version of Prospino, so their K-factors are not included.  We note that these processes
tend to have very small cross sections at the LHC, so this has a negligible effect on our results. 
We employed the CTEQ6.6M parton distribution functions \cite{Nadolsky:2008zw}
when performing these calculations, as well as in our event generation. 

PYTHIAv.6.418 was employed for event generation, fragmentation/showering, and hadronization.
In order to apply the K-factors calculated with PROSPINO, we generated individual event samples for each of the 85 SUSY production processes
and scaled each by its K-factor.  
In some subset of the models, problems with PYTHIA arose, \eg, it could not handle the final state hadronic fragmentation
in the decays of colored sparticles with small mass splittings.  To address this, we implemented an additional modification to the
decay tables.  For any sparticle with an unboosted decay length longer than $\sim 20$ m, so that it does not decay within the detector, 
we set the decay width to zero so that PYTHIA treats the sparticle
as absolutely stable and does not attempt the decay.    In addition, we attempted to force a larger decay width in the case of
any colored sparticle with a width less than 1~\gev\ to alleviate issues with hadronizing long-lived colored states, but this exacerbated the
problem and led to more frequent serious PYTHIA errors and so this approach was
dropped.  

We are left with roughly 1\% of our pMSSM model sample where PYTHIA
errors occur that are serious enough to lead to a PYSTOP, {\it i.e.}, a halt in the event generation.  In these models, the production 
cross sections can thus be seriously underestimated. Therefore, 
in the remainder of this work, these ``PYSTOP models'' are generally excluded from our results, except where noted otherwise.
Note that since this is only a very tiny fraction of the models we consider, dropping this small set has essentially no impact on the 
results we quote below. This was explicitly verified for all of the ATLAS analyses we consider below for both flat and log prior model samples.

Events were then passed through an ATLAS-tuned version of PGS4 \cite{PGS} for fast detector simulation, employing the kinematic cuts
for the eleven inclusive search analyses described in detail by ATLAS in Ref.~{\cite{Aad:2009wy}} and given below. 
Here, we matched as closely as possible the set of definitions that ATLAS employed~{\cite{Aad:2009wy}} for their final state `objects'  
such as jets, leptons, $\tau$'s, $b$'s and \MET.  In particular, we replaced
the default PGS object isolation routine with an analysis-level routine
which mimics as much as possible the published ATLAS object identification and
isolation procedure.

\subsection{Analysis Cuts}

In the interest of completeness, we here provide a list of the full set of kinematic cuts for each analysis channel that we employ as given by 
ATLAS\cite{Aad:2009wy}:

\begin{itemize}
\item 4-jet + $\MET$:
  \begin{enumerate}
  \item At least 4 jets with $p_T>50~\gev$, at least one of which has $p_T>100~\gev$.
  \item $E_T^{\mathrm{miss}}>100~\gev$ and $\MET>0.2\MEFF$.
  \item Transverse sphericity $S_T>0.2$.
  \item $\Delta\phi(\jet_{1,2,3}-\MET)>0.2$.
  \item Reject events with an $e$ or a $\mu$.
  \item $\MEFF>800~\gev$.
  \end{enumerate}
\item 2-jet + $\MET$:
  \begin{enumerate}
  \item At least 2 jets with $p_T>100~\gev$, at least one of which has $p_T>150~\gev$.
  \item $E_T^{\mathrm{miss}}>100~\gev$ and $\MET>0.3\MEFF$.
  \item $\Delta\phi(\jet_{1,2}-\MET)>0.2$.
  \item Reject events with an $e$ or a $\mu$.
  \item $\MEFF>800~\gev$.
  \end{enumerate}
\item 1 lepton + 4 jets + $\MET$:
  \begin{enumerate}
  \item Exactly one isolated lepton with $p_T>20~\gev$.
  \item No additional leptons with $p_T>10~\gev$.
  \item At least 4 jets with $p_T>50~\gev$, at least one of which has $p_T>100~\gev$.
  \item $E_T^{\mathrm{miss}}>100~\gev$ and $\MET>0.2\MEFF$.
  \item Transverse sphericity $S_T>0.2$.
  \item Transverse mass $M_T>100~\gev$.
  \item $\MEFF>800~\gev$.
  \end{enumerate}
\item OSDL + 4 jets + $\MET$:
  \begin{enumerate}
  \item Exactly two opposite-sign leptons with $p_T>10~\gev$.
  \item At least 4 jets with $p_T>50~\gev$, at least one of which has $p_T>100~\gev$.
  \item $\MET>100~\gev$ and $\MET>0.2\MEFF$.
  \item Transverse Sphericity, $S_T>0.2$.
  \end{enumerate}
\item Trilepton + jet + $\MET$:
  \begin{enumerate}
  \item At least three leptons with $p_T>10~\gev$.
  \item At least 1 jet with $p_T>200~\gev$.
  \end{enumerate}
\item Trilepton + $\MET$:
  \begin{enumerate}
  \item At least three leptons with $p_T>10~\gev$.
  \item At least one OSSF dilepton pair with $M>20~\gev$.
  \item Lepton track isolation: $p_{T,\mathrm{trk}}^{0.2}<1~\gev$ for electrons and $<2~\gev$ for muons, where
    $p_{T,\mathrm{trk}}^{0.2}$ is the maximum $p_T$ of any additional track within a $R=0.2$ cone around the lepton.
  \item $\MET>30~\gev$.
  \item $M<M_Z-10~\gev$ for any OSSF dilepton pair.
  \end{enumerate}
\item $\tau$ + jets + $\MET$:
  \begin{enumerate}
  \item At least 4 jets with $p_T>50~\gev$, at least one of which has
    $p_T>100~\gev$, and at least one $\tau$.
  \item $E_T^{\mathrm{miss}}>100~\gev$.
  \item $\Delta\phi(\jet_{1,2,3}-\MET)>0.2$.
  \item No isolated electrons or muons.
  \item At least one $\tau$ must have $p_T>40~\gev$ and $|\eta|<2.5$.
  \item  $\MET>0.2\MEFF$.
  \item $M_T>100~\gev$, where $M_T$ is the transverse mass of the hardest
    $\tau$ and $\MET$.
  \end{enumerate}
\item $b$ jets + $\MET$:
  \begin{enumerate}
  \item At least 4 jets with $p_T>50~\gev$. 
  \item At least one of which has $p_T>100~\gev$.
  \item $E_T^{\mathrm{miss}}>100~\gev$.
  \item $\MET>0.2\MEFF$.
  \item Transverse sphericity $S_T>0.2$.
  \item At least 2 jets tagged as $b$ jets.
  \item $\MEFF>1000~\gev$.
  \end{enumerate}
\end{itemize} 

In addition to the 4j1l analysis we also considered 3(2)j1l analyses where the cut on the leading jet is raised to $p_T>150~\gev$, the second(and third) 
jet must have $p_T>100~\gev$, and the $\MET$ cut is harder: $\MET>\max(100~\gev,0.25(0.3)\MEFF)$.

Furthermore, in addition to the OSDL analysis there is also an SSDL analysis with identical kinematic cuts, except that, of course, the two leptons 
must have the same charge and they have a somewhat harder cut: $p_T>20~\gev$.  Also in this case, the cut on transverse sphericity is dropped.

\subsection{Statistical Procedure}

In the analysis below we follow the statistical treatment of signal and backgrounds as employed by ATLAS\cite{Aad:2009wy} as closely as possible in determining 
the significance of the signal over background for each pMSSM model in the eleven different search channels. For completeness, the details of the 
ATLAS approach that we follow will be given here. To this end, we allowed for a 
$50\%$ systematic uncertainty in the calculation of both the SM QCD and electroweak 
backgrounds in order to match the ATLAS analyses.  However, we also considered a reduction to the 
case of $20\%$ systematic errors associated with these SM backgrounds. Such a reduction, as was discussed by ATLAS, may be possible in the future 
using both the data itself as well as improved theoretical calculations of SM processes. Interestingly, we note that ATLAS found that these SM backgrounds for 
SUSY are {\it completely dominated} by contributions from electroweak sources as opposed to those arising from pure QCD. 
As we will discuss below, the former choice of background uncertainty led to better agreement with the
ATLAS results for their mSUGRA benchmark models, but the latter case will be seen
to substantially increase the coverage of the pMSSM model parameter space and is something that may be obtainable in the future. 

Directly following the ATLAS study, we compute the signal significance as described below.
We first total all background and signal events above the
$\MEFF$ cut that is specific to each analysis.  We then compute the
probability $p$ that the background fluctuates by chance to the total number of
measured events or above, assuming that the systematic error on the background
is Gaussian and the statistical error is Poissonian.  This means
\begin{equation}
  \label{prob}
  p=A\int_0^\infty db\, G(b;N_b;\delta N_b)\sum_{i=N_{\rm data}}^\infty
  \frac{e^{-b}b^i}{i!}\;,
\end{equation}
where $N_b$ is the number of background events and $\delta N_b$ is the
associated systematic error on this number, while $N_{\rm data}=N_b+N_{\rm
  signal}$ is the total number of events above the $\MEFF$ cut.  $G$ is a Gaussian
distribution and $A$ is a normalization factor ensuring that the probability
that the background fluctuates to any nonnegative integer is one; therefore
$A=p(N_{\rm data}=0)^{-1}$.  The significance $Z_n$, is then given by
\begin{equation}
  \label{zn}
  Z_n=\sqrt{2}\mathrm{erf}^{-1}(1-2p)\;.
\end{equation}

\subsection{Comparison with ATLAS Benchmark Models}

We must first verify that our analysis for each signature can be trusted. To this end, we determine whether we can reproduce the 
results\cite{Aad:2009wy} obtained by ATLAS for their mSUGRA benchmark points (labeled here as SU1,2,3,4,6,8.1 and 9). 
For each point, the ATLAS collaboration generated a large number of signal events and scaled to a luminosity of 1 $\infb$. 
We followed a similar approach in making our comparisons, generating 10 fb$^{-1}$ of events for each ATLAS benchmark model and then scaling down
to 1 fb$^{-1}$.  Due to computing time restrictions, we put a cap of 10k generated events on any one of the 85 SUSY production processes for each 
benchmark model. In addition, at least 100 events were generated in every channel in order to properly evaluate potentially small cross sections; these events were 
then appropriately rescaled. 

Here it is important to note that our SUSY signal generation, as 
described above, necessarily differs in detail from that performed by ATLAS. In contrast to our inclusive SUSY analysis, ATLAS determined their 
mSUGRA spectra and performed their sparticle decay table calculations using
ISASUGRA versions 7.64-7.71.  They used PROSPINOv2.0.6 \cite{Beenakker:1996ed,Beenakker:1996ch,Beenakker:1997ut,Beenakker:1999xh,Spira:2002rd,Plehn:2004rp}
and the CTEQ6M parton distribution functions \cite{Stump:2003yu}
to obtain the NLO results for strong interaction processes, \ie, squark and gluino pair production 
as well as squark-gluino associated production. NLO corrections were not included for the other channels. 
Event generation, fragmentation/showering, and hadronization were performed using HERWIG \cite{Corcella:2000bw,Corcella:2002jc,Moretti:2002eu} 
and the results were then passed through the full ATLAS GEANT detector simulation.  

The results of our comparison benchmark study, as can be seen in Figures \ref{fig12}-\ref{fig789}, suggest that we are indeed able to faithfully reproduce those  
obtained by ATLAS in the case of their mSUGRA benchmark models for all of the various inclusive analyses.  The one possible exception occurs in the tails of 
the $\MEFF$ distributions, where 
statistics are poor and large fluctuations are to be expected.  This is
an important check to perform, and pass, before we embark on computing these signature channels
for our large model set.

\FIGURE{
  \label{fig12}
  \includegraphics[width=0.45\columnwidth]{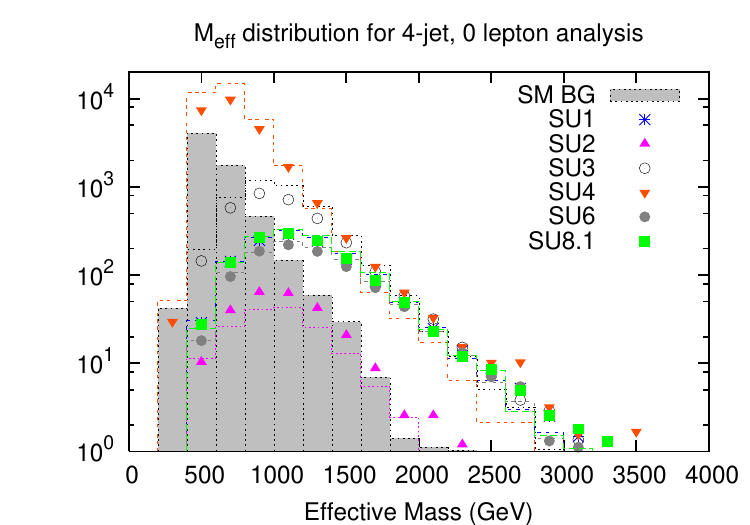}
  \includegraphics[width=0.45\columnwidth]{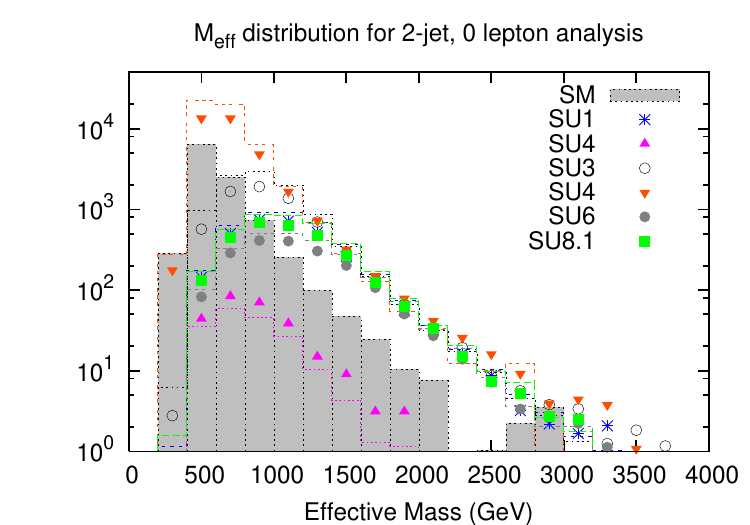}
  \caption{The \MEFF\ distribution for the 4(2) jet, 0 lepton analysis on the
    left(right).  The data points represent our analysis,
    while the color coded lines are the results from the ATLAS study\cite{Aad:2009wy}.}
}
\FIGURE{
    \label{fig45}
    \includegraphics[width=0.45\columnwidth]{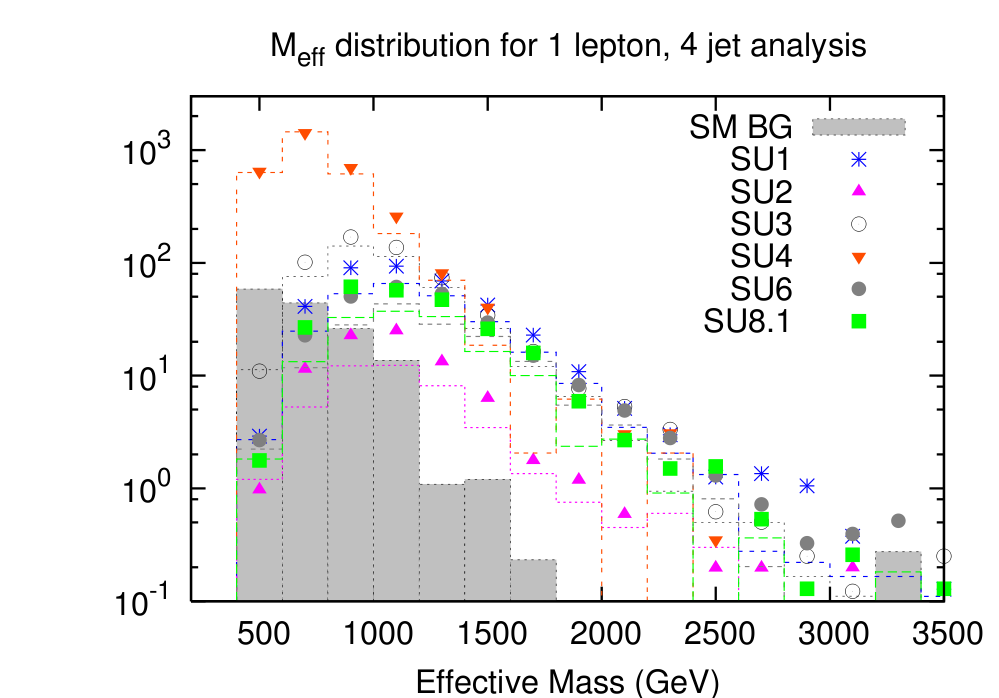} \\
    \includegraphics[width=0.45\columnwidth]{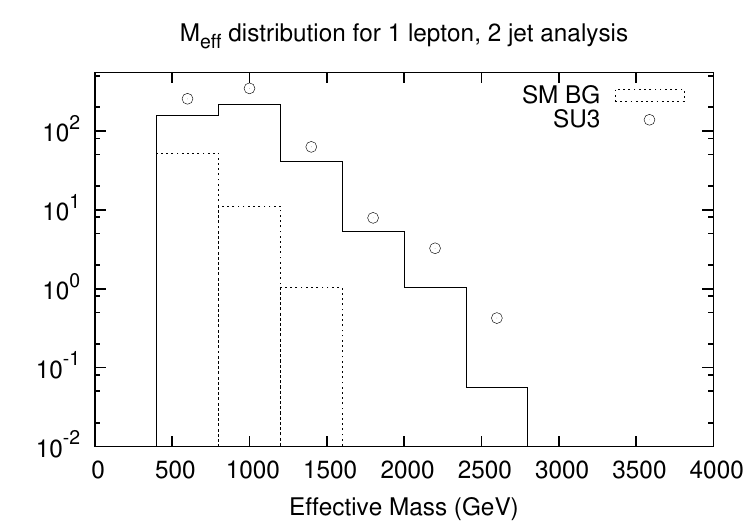}
    \includegraphics[width=0.45\columnwidth]{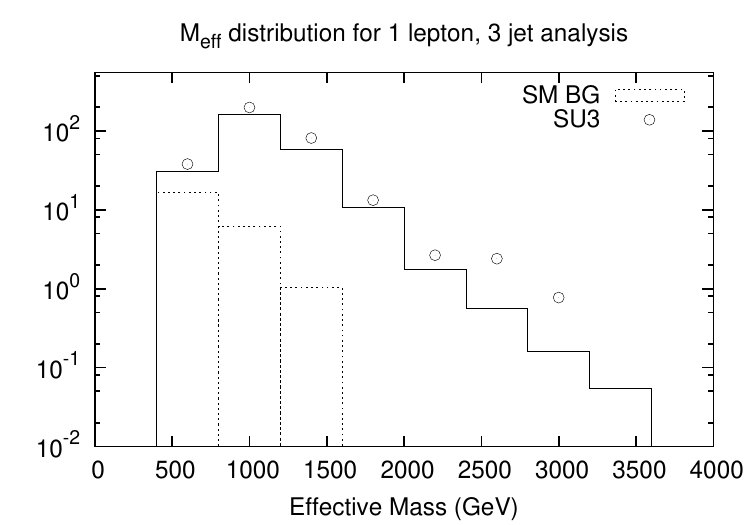}
    \caption{The \MEFF\ distribution for the 1 lepton, 4(2,3) jet analysis on
      the top(bottom left, bottom right).  The data points represent our analysis,
      while the lines are the results from the ATLAS study\cite{Aad:2009wy}.}
}
\FIGURE{
    \label{fig789}
    \includegraphics[width=0.45\columnwidth]{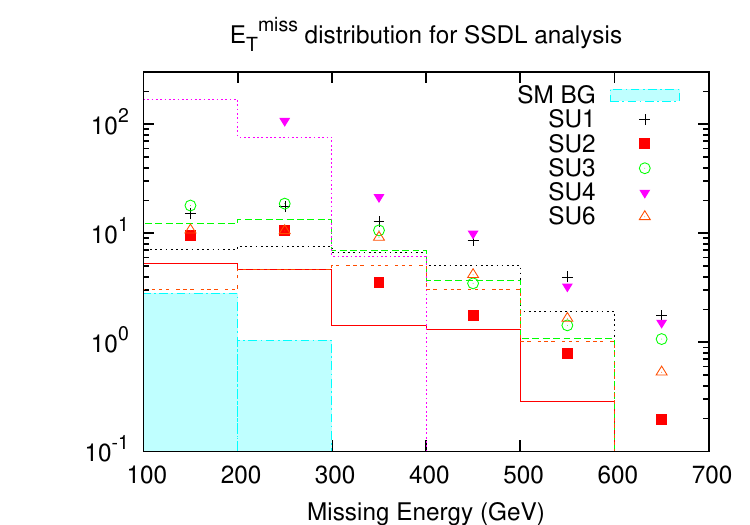}
    \includegraphics[width=0.45\columnwidth]{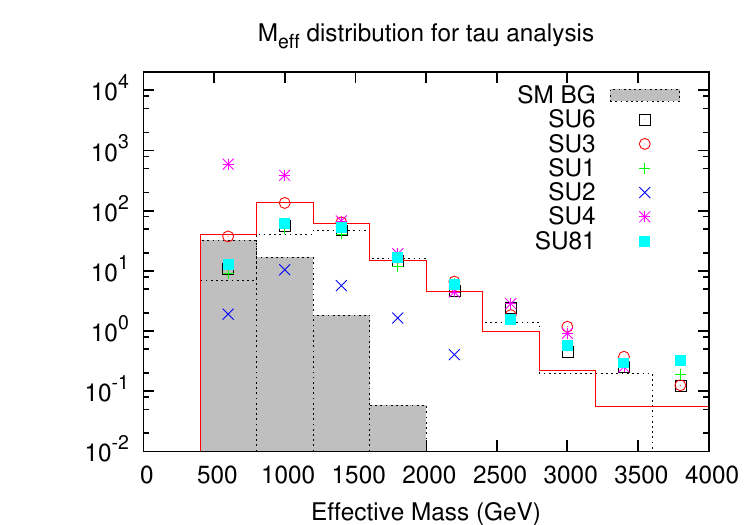}
    \includegraphics[width=0.45\columnwidth]{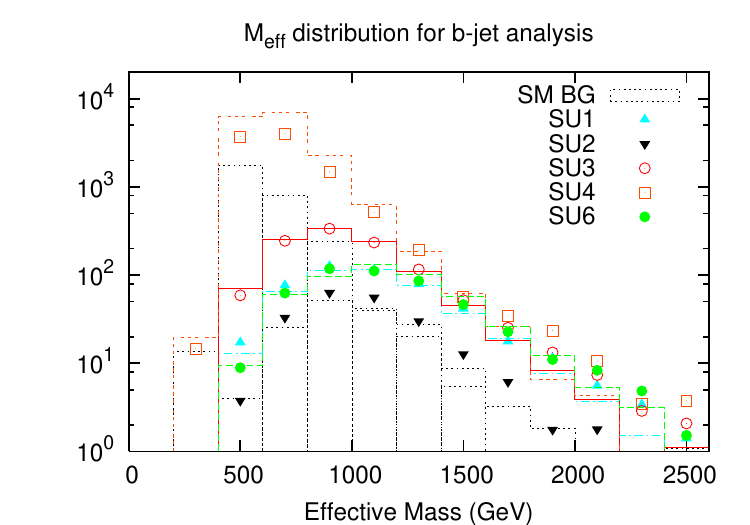}
    \caption{The \MEFF\ (\MET for SSDL) distribution for the same-sign dilepton($\tau$,$b$-jet) analysis
      on the top(bottom left, bottom right).  The data points represent our analysis,
      while the lines are the results from the ATLAS study\cite{Aad:2009wy}.}
}

We should also note that the agreement between our $\tau$ analysis results and that of ATLAS is somewhat suspect because of an issue with the PGS4
$\tau$ fake rate and efficiency.  We find that this fake rate is much higher, and the efficiency much lower, than the values quoted for the ATLAS $\tau$
reconstruction algorithm.  The agreement between our results and those of ATLAS for this analysis are therefore due to some compensation between
these two factors.  In what follows, we will generally show results without the $\tau$ analysis, as we believe its validity is in question.

Having verified the reliability of our event generation, detector simulation,
and data analysis procedure, the last ingredient to check is the statistical
procedure described above. The primary issue is how large a systematic error one should assign to the SM
background due to theoretical uncertainties associated with, \eg,  higher order perturbative calculations.
In \cite{Aad:2009wy}, ATLAS assigns a 50\% error to QCD backgrounds and 20\% to
electroweak backgrounds, and combines these two errors in quadrature.  
What is important, however, is 
that the systematic error represent the true uncertainty in these background
predictions at the time of LHC data analysis.  It is likely that the theoretical
uncertainties on the computation of relevant electroweak backgrounds will continue to be quite
high, especially for processes with additional jets, and could easily be of order 50\%\cite{Lance}.
We therefore adopt a 50\% systematic error on both electroweak and QCD
backgrounds as a conservative default assumption for most of our results that follow.
In many cases, however, we will also show for comparison the effect of reducing the
background systematic error to 20\% for both background samples.  We will return to this point of discussion in more detail below.


\section{Results: ATLAS Inclusive MET Analyses}
\label{results1}

Now that we have convinced the reader that we can do a reasonable job at reproducing the analyses performed by ATLAS    
for their mSUGRA benchmark points, we turn to a discussion of the corresponding analyses for our 71k pMSSM model set. 

\subsection{Global Results}

We first consider some global results. The first, and most important, question we address is what fraction of the two pMSSM model sets would be discovered 
by each of the various inclusive ATLAS analyses assuming an integrated luminosity of $1~\infb$. This will give us a good feel for how good a job the various  
ATLAS MET analyses, designed for mSUGRA, will do at discovering SUSY in the more general pMSSM parameter space. 
The answer to this question for both the flat and log prior model samples can be found in 
Table~\ref{mtfl}. Note that throughout this paper, when we say a model ``passes'' a given analysis, we mean
    the significance $S$ ($Z_n$ in the ATLAS notation) satisfies $S\geq 5$.
Similarly, we say the model is missed or `fails' if $S<5$.  

This Table shows us that the ATLAS MET analyses do a very reasonable job at probing the 
more general pMSSM parameter space and that some searches perform better at this than do others. Specifically, here in this Table we observe a number of 
interesting results: ($i$) The ATLAS search capabilities for the flat and log prior samples are different. 
Clearly, we see that a greater fraction of our pMSSM model points are observable in each of the analyses 
in the case of flat priors than in the case of log priors; there are two obvious reasons for this result. First, in the case of the log 
prior sample, the sparticle spectra generally extend out to far larger masses, $\sim 3$ TeV, rendering them less kinematically 
accessible at the LHC. Second, the models generated by the log prior scan tend to have mass spectra which are somewhat compressed, \ie, more 
sparticles lie in a given mass interval, making it in principle somewhat more difficult 
to produce trigger particles with sufficient $E_T$ to pass the various analysis cuts. We will discuss this issue further below. 

\TABLE{
  \begin{tabular}{ | c || c | c | }
  \hline
  Analysis &  Flat priors & Log priors \\ \hline \hline
     4j0l  &  88.331  &  48.166 \\ \hline
     2j0l  &  87.616  &  47.391 \\ \hline
     4j1l  &  41.731  &  18.371 \\ \hline
     3j1l  &  64.058  &  36.601 \\ \hline
     2j1l  &  62.942  &  33.498 \\ \hline
     OSDL  &  6.0958  &  3.8434 \\ \hline
     SSDL  &  14.774  &  8.8505 \\ \hline
     3lj   &  13.549  &  8.6389 \\ \hline
     3lm   &  2.7406  &  2.8561 \\ \hline
     $\tau$  &  83.510  &  44.006 \\ \hline
     $b$   &  73.983  &  42.948 \\ \hline
\end{tabular}

  \caption{The percentage of the pMSSM model set that passes each analysis, for the flat
    and log prior model sets. This assumes a systematic error of $50\%$ on the SM background.}
  \label{mtfl}
}

($ii$) The (2,4)j0l analyses are, overall, found to be the most powerful of the set of MET analyses in the sense that they lead to 
a discovery for the greatest fraction of our model points for either prior set. We note that the 4j0l analysis is 
found to perform only slightly better than the 2j0l one for both sets of priors when the background systematic error is taken to be 
$50\%$. This is not overly surprising as ATLAS also found the 4j0l analysis to be the most powerful 
in the case of mSUGRA \cite{Aad:2009wy} at $\sqrt s=14$ TeV. In comparison to our results, ATLAS found that for mSUGRA models 
the reach of the 2j0l analysis was much more degraded with respect to the 4j0l case than that found here.{\footnote {We remind the reader that these two 
analyses are not completely mutually exclusive since they are actually requiring {\it at least} 4j and 2j, respectively. ATLAS typically found 
that $\sim 35\%$ of their 2j0l sample also appeared in the corresponding 4j0l
sample \cite{Aad:2009wy}.}} 

($iii$) The (2,3,4)j1l channels do not 
play as important a role in the present study as they did for ATLAS in their analysis of the mSUGRA parameter space. ATLAS determined that 
these three searches were all found to give a somewhat comparable discovery reach in their coverage of the $m_0-m_{1/2}$ plane. 
Here we see that the (2,3)j1l analyses are the relatively more powerful ones in this set of single lepton searches, but are still somewhat 
degraded in relative importance in comparison to the coverage provided by the (2,4)j0l channels. Of course, these two classes of signatures 
provide complementary coverage of most of the model set since the (4,2)j0l search requires
the {\it absence} of leptons. 

($iv$) The $\tau$ 
analysis appears to provide 
almost as large a reach as do the (2,4)j0l channels; here we must recall the warning from the previous section that PGS has simultaneously a low 
$\tau$ efficiency and a high fake rate. It is thus likely that the model coverage offered by this channel is somewhat overestimated. 
However, we note that for large 
$\tan \beta$, ATLAS found the $\tau$ analysis to be a reasonably powerful
channel in the case of mSUGRA. 

($v$) Neither the SSDL nor the 
OSDL searches do particularly well at detecting many models; this is primarily due to the relatively small number of dilepton final 
states in our model sample. We will return to this issue further below. The 3lj and 3lm analyses are also seen to provide poor model coverage 
(as might then be expected due to the low number of final state leptons).  The less inclusive 3lj channel appears to do somewhat better than  
the more inclusive 3lm case, most likely due to reduced SM background. 

What are the reasons that the ATLAS SUSY search analysis channels fail to observe the full pMSSM model sample?  One reason could be
the rather low luminosity, 1~fb$^{-1}$, assumed in this study.  However, recall that we need 
to overcome not only the possible low statistics available in the signal channel but also the large systematic error associated 
with the uncertainties in the SM backgrounds. If these are large, as we'll see below is the case for the 4j0l and 2j0l analyses, then 
increasing the integrated luminosity will actually be of minimal use and in such circumstances 
it is more important to get a better handle on the size of the backgrounds from either direct measurements or refined theoretical calculations. 
To address the issue of how useful increasing the integrated luminosity would be for the cases at hand,  
we display in Table~\ref{mtflh} our results (analogous to those in Table~\ref{mtfl} above) for an integrated luminosity of 
$10~\infb$ while maintaining a 50\% systematic error on the SM backgrounds. Clearly, for all analyses, and for both flat and log priors, 
the fraction of models that could be discovered increases. However, in most cases this increase is seen to be quite modest (in particular,
for the (2,4)j0l channels) compared to what one might expect, 
although some channels show a more significant improvement than others. Although increased luminosity is 
always helpful to some extent, many pMSSM models are clearly missed for physics reasons and not just due to insufficient statistics; 
certainly some of this is due to the large uncertainties in the SM backgrounds. 

\TABLE{
  \begin{tabular}{ | c || c | c | }
  \hline
  Analysis &  Flat priors & Log priors \\ \hline \hline
     4j0l  &  88.578  &  48.080 \\ \hline
     2j0l  &  87.774  &  47.378 \\ \hline
     4j1l  &  44.885  &  20.421 \\ \hline
     3j1l  &  70.907  &  45.975 \\ \hline
     2j1l  &  68.419  &  40.473 \\ \hline
     OSDL  &  6.6796  &  4.2467 \\ \hline
     SSDL  &  25.518  &  15.879 \\ \hline
     3lj   &  17.361  &  11.078 \\ \hline
     3lm   &  2.9135  &  2.9542 \\ \hline
     $\tau$ &  86.505  &  45.606 \\ \hline
     $b$   &  76.939  &  44.572 \\ \hline
\end{tabular}

  \caption{Same as Table~\ref{mtfl} but for an integrated luminosity of 10 $\infb$.}
  \label{mtflh}
}

\subsection{Impact of Background Uncertainties}

We now further quantify the effect of systematic
uncertainties on the observability of a SUSY signal. Figure~\ref{gain} shows how the significance of
several of the ATLAS \MET~searches will scale (in the Gaussian limit) when the integrated luminosity is increased from 
1 $\infb$ to 10~$\infb$ as a function of the systematic error on the associated SM backgrounds. 
Here we see that for channels with large backgrounds, significant improvement in the signal significance is prevented by
sizable systematic errors when the luminosity is increased by a factor of 10, \ie, analyses which have large SM backgrounds lead to searhes 
which are already essentially {\it systematics dominated} at luminosities of order 1 fb$^{-1}$. In particular, we note that substantial gains
in significance for the (4,2)j0l channel are not possible unless the associated systematic errors are substantially reduced.
This is one of the main reasons why a 
significantly larger fraction of our models are not captured by the most powerful (4,2)j0l analyses when the luminosity is increased. 
We also see that analyses with lower SM backgrounds, however, are more statistics limited and will find their search reaches improved as the intergrated 
luminosity increases.

\FIGURE{
  \includegraphics[width=0.7\columnwidth]{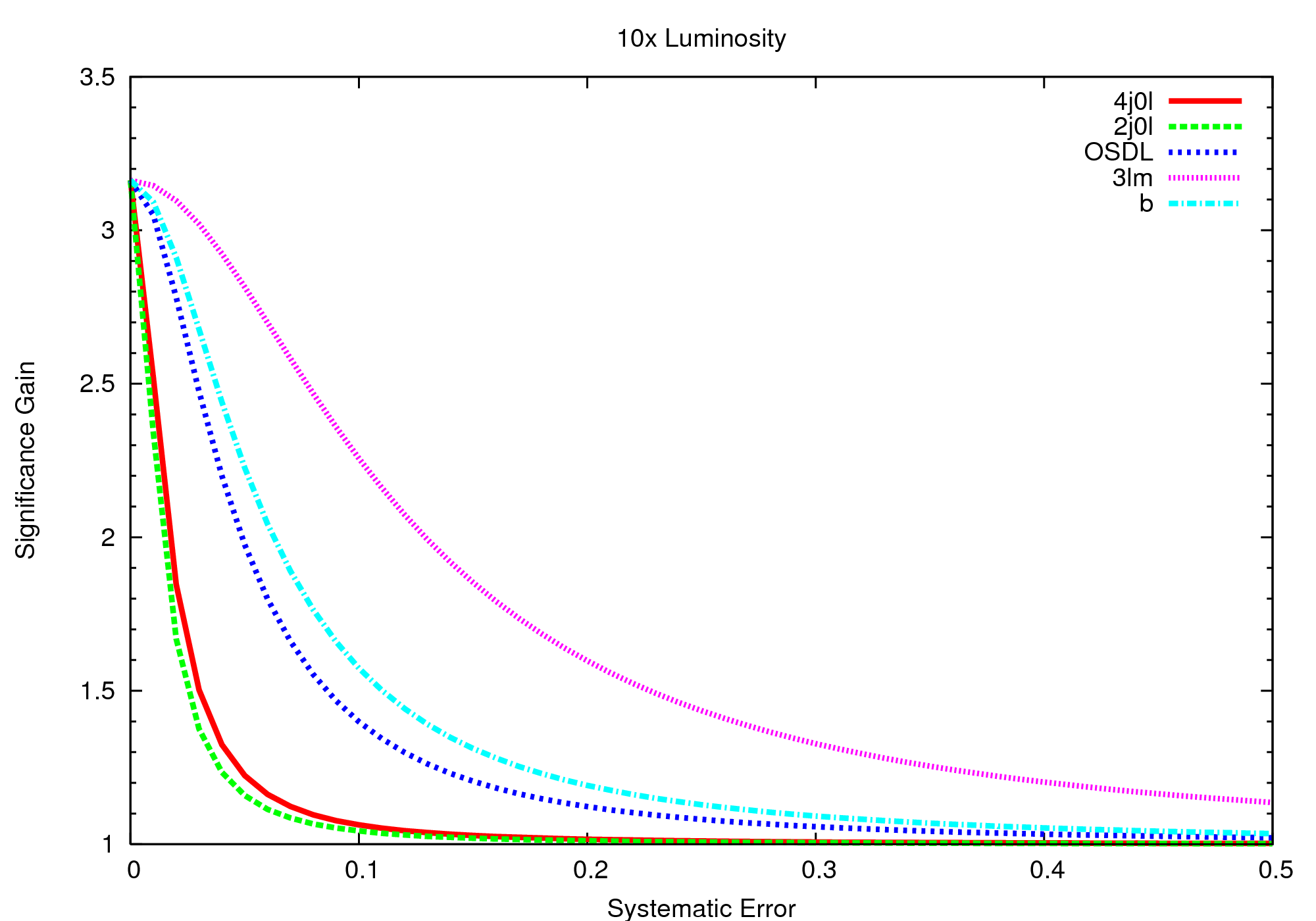}
  \caption{The relative gain in significance due to a tenfold increase in
    integrated luminosity, as a function of the systematic error (in percent) on the
    background cross section for several analysis channels.  For example, if the gain is `2' then the significance doubles. Here we see that 
    analyses with large SM backgrounds are essentially systematics dominated at 1 fb$^{-1}$ and that there reaches can only be improved significantly 
    by the reduction of the background systematic errors.}
  \label{gain}
}

We can understand these results more clearly by examining Table~\ref{counts} which shows the number of expected background events for each of 
the MET analyses assuming an integrated luminosity of 1 $\infb$. This table also shows the corresponding number of signal events required 
to reach the $S=5$ level for each channel assuming a systematic error of 50(20)\% in the estimation of the associated background. 
Here we clearly see that channels with a larger number 
of expected background events benefit the most from a reduction of the corresponding background systematic error, while the reverse is true 
for analyses with smaller backgrounds. 
\TABLE{
    \begin{tabular}{lccc}
      \hline
      \hline
      Analysis & Background & $S=5,~\delta B=50\%$ &  $S=5,~\delta B=20\%$  \\
      \hline
      4j0l  &  709   &  1759  & 721   \\
      2j0l  &  1206  &  2778 & 1129  \\
      4j1l  &  41.6  &  121  & 62   \\
      3j1l  &  7.2  & 44  & 28   \\
      2j1l  &  18.2  & 61  & 36   \\
      OSDL  &  84.7  &  230  & 108     \\
      SSDL  &  2.3  & 17  &  13   \\
      3lj   &  12  &  44  &  28   \\
      3lm   &  72.5  &  198  &  94   \\
      $\tau$ &  51  & 144  &  72   \\
      b     &  69   &  178 & 86   \\
      \hline
      \hline
      \caption{Expected number of background events for each of the ATLAS analyses and the corresponding number of events required to observe a signal 
        with $S=5$ assuming a background uncertainty of either $\delta B=50\%$ or $20\%$. The integrated luminosity is taken to be 1 $\infb$.}
      \label{counts}
    \end{tabular}
}

Correspondingly, the number of signal events required to reach the $S=5$ level for each of the
analyses is shown as a function of the corresponding background systematic error in Fig~\ref{systematics}. These results show that a significant 
gain in the overall model space coverage can likely be obtained through even modest reductions in the background systematic errors.

\FIGURE{
  \includegraphics[width=0.7\columnwidth]{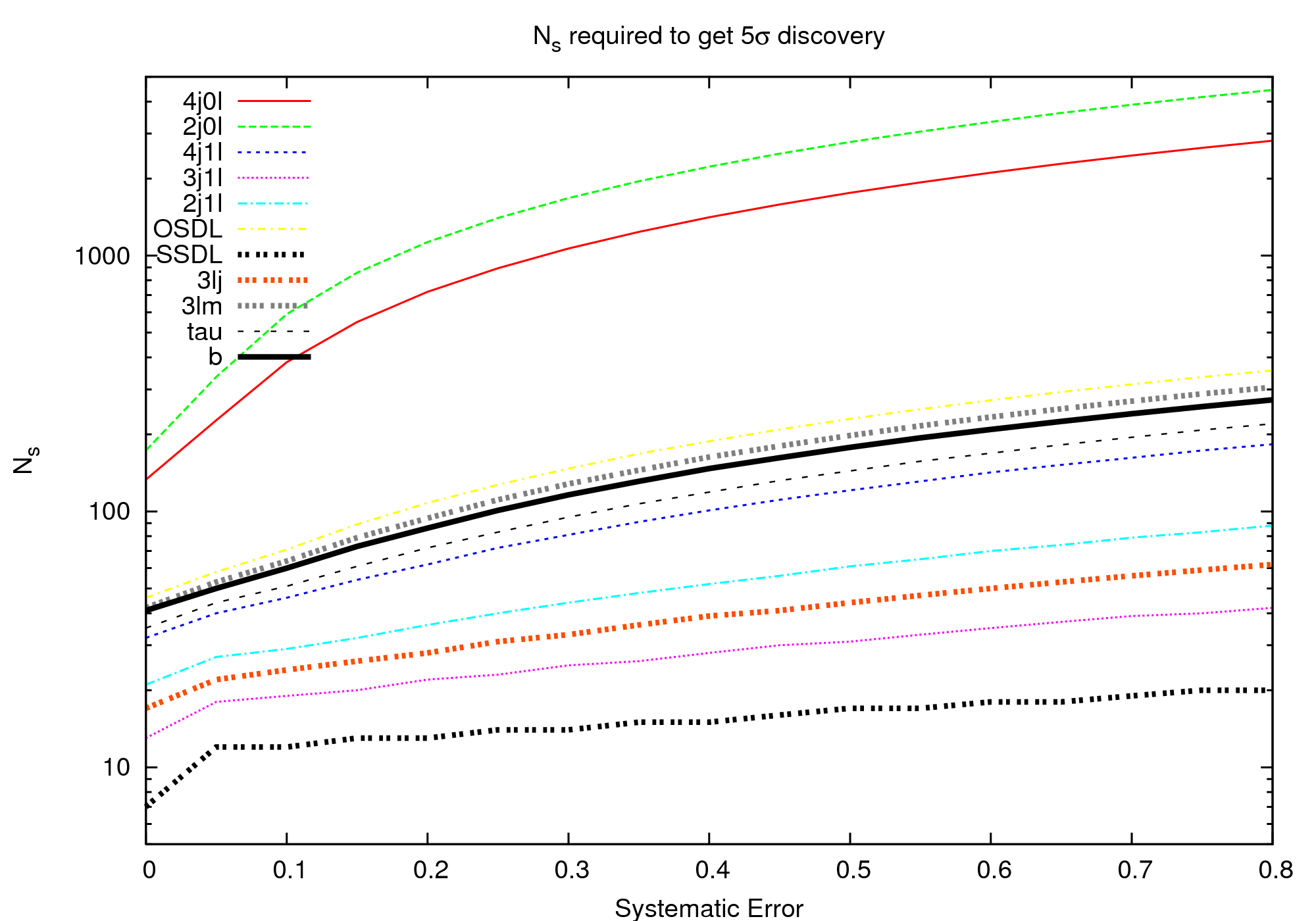}
  \caption{This figure shows how the number of signal events required to reach $S=5$ changes within each analysis as a function of 
    the assumed systematic uncertainty in the SM background.}
  \label{systematics}
}

To further quantify the importance of the background systematic errors in each of the analysis channels, 
we examine the change in the fraction of pMSSM models that are observable in a given analysis when the systematic
uncertainty on the SM background is modified.  As discussed in the previous section, in most of the results we present, including 
those in Tables \ref{mtfl} and \ref{mtflh} above, we have assumed a default $50\%$ systematic
uncertainty in both the QCD and electroweak background rates.  We now study the effect of reducing the systematic 
error on these backgrounds to $20\%$, which may be possible using both the data itself as well as by improving theoretical calculations 
of the SM backgrounds.  

The left panel of Figure~\ref{zn-hist-4j0l} shows the distribution of the significance variable $S$ across our flat-prior model set for the 4j0l
analysis.  In this figure, we compare this distribution for different values of the systematic error on the SM background and integrated luminosity.
For this analysis we see that increasing the integrated luminosity from $1~\infb$ to $10~\infb$ has very little effect on this distribution; 
in particular, the number of pMSSM models
for which $S>5$ hardly changes.  On the other hand, reducing the systematic error on the SM background from 50\% to 20\% shifts the peak in the
distribution to much higher values of $S$, such that many more pMSSM models have $S\geq 5$.  
Clearly, then, the 4j0l search channel is already systematics-dominated at 
$1~\infb$, and further theoretical and experimental work on improving the QCD background determination would be extremely fruitful.
\FIGURE{
  \includegraphics[width=0.45\columnwidth]{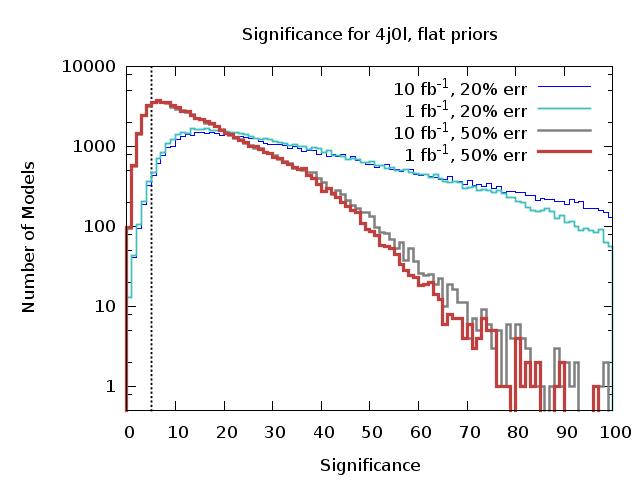}
  \includegraphics[width=0.45\columnwidth]{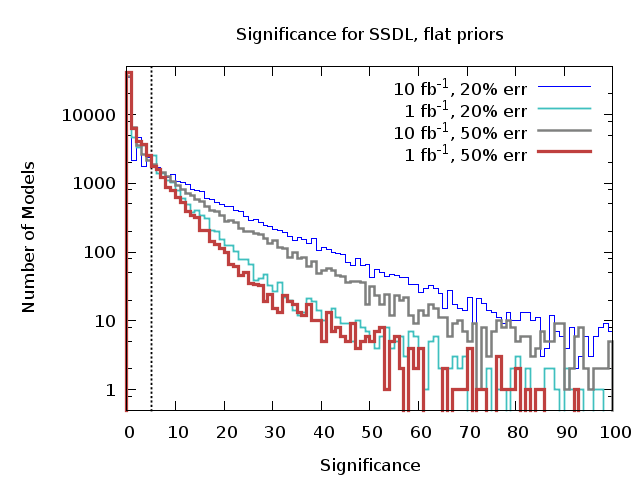}
  \caption{Significance distributions for the 4j0l and SSDL analyses of the flat prior model set for 4 different combinations of 
           integrated luminosity and SM background errors.  The dashed vertical line is located at $S=5$.}
  \label{zn-hist-4j0l}
}
However, for an analysis with a much smaller number of background events, such as SSDL, we find that a reduction of the systematic error
has a smaller impact.  In the right panel of Figure~\ref{zn-hist-4j0l} we show the significance 
distribution for the flat prior model set for this analysis with the same four luminosity-error
combinations.  In this case, one can see that the distribution shifts to higher significance values for $10~\infb$, while a change in the
systematic error has relatively little effect on the distribution.  

We can study the effect of varying the integrated luminosity and systematic error on the remaining analysis channels in the same way.
In Table~\ref{main5020}, we summarize these results by comparing the fraction of pMSSM models that pass each search
analysis for all choices of the luminosity and systematic error. Here we see that, for most analyses, a reduction in the 
background uncertainty goes much further in increasing our model space coverage than does increased luminosity alone.  
Clearly, then, many of the standard SUSY searches at the LHC are systematics limited.

\begin{table}
  \centering
  \begin{tabular}{ | c || c | c | c | c | }
  \hline
  Analysis & $50\%$ error & $50\%$ error & $20\%$ error & $20\%$ error \\ 
           & 1 fb$^{-1}$  & 10 fb$^{-1}$  & 1 fb$^{-1}$   & 10 fb$^{-1}$ \\ \hline \hline
     4j0l  &  88.331  &  88.578  &  98.912  &  99.014  \\ \hline
     2j0l  &  87.616  &  87.774  &  98.75  &  98.802  \\ \hline
     1l4j  &  41.731  &  44.885  &  56.849  &  63.045  \\ \hline
     1l3j  &  64.058  &  70.907  &  69.725  &  81.111  \\ \hline
     1l2j  &  62.942  &  68.419  &  70.646  &  80.641  \\ \hline
     OSDL  &  6.0958  &  6.6796  &  15.262  &  18.659  \\ \hline
     SSDL  &  14.774  &  25.518  &  18.501  &  32.887  \\ \hline
     3lj  &  13.549  &  17.361  &  19.293  &  28.97  \\ \hline
     3lm  &  2.7406  &  2.9135  &  4.8844  &  5.8284  \\ \hline
     tau  &  83.51  &  86.505  &  96.928  &  98.695  \\ \hline
     b  &  73.983  &  76.939  &  91.672  &  94.867  \\ \hline
\end{tabular}

  \caption{The percentage of our pMSSM models that are observable in each analysis for the flat
    prior model set with both 1~$\infb$ and 10~$\infb$ of integrated luminosity and both 50\% and 20\% error assumed for the SM
    background.}
  \label{main5020}
\end{table}

We now ask the very important question of whether or not our pMSSM models are discovered in one, more than one, or even in multiple, searches. 
Furthermore, and perhaps even more importantly, we also want to know if there are any pMSSM models which are {\it missed entirely} by the suite 
of ATLAS inclusive \MET~searches. In such cases, though kinematically accessible sparticles are produced at the LHC, they are not discovered by the 
ATLAS searches. The answers to these questions will give us another good handle on how well the 
ATLAS \MET~searches, designed for mSUGRA, will do at covering the much more general pMSSM parameter space. The answers are to be found  
in Table~\ref{numt} which shows the results for both flat and log prior model samples and for both integrated luminosities of 1 and 10 
$\infb$ assuming $50\%$ background uncertainties. We note that the results for both prior sets are substantially different. 
Specifically, this table shows the fraction of the pMSSM models that have lead to a significance $S\geq 5$ 
in $n$ different ATLAS analysis channels.  For example, we see that for the flat prior model set with a luminosity of 1 $\infb$, $\sim 13.2(15.2)\%$ of 
the models are found by 3(6) different ATLAS analyses. As the integrated luminosity is increased, we 
see that for both the flat and log prior model sets the fraction of models found by a larger number of 
analyses increases as one would expect.  Perhaps even more interesting, we observe that a respectable fraction 
of models are missed by {\it all} of the ATLAS inclusive MET search analyses, even for the larger value of integrated 
luminosity. Of course, as expected, a higher fraction of models are missed in the log prior case due to the reasons
discussed above, \eg, many of the sparticles may be 
significantly more massive. Some significant fraction of these models which are missed may be due to the large background systematic errors discussed 
previously.  We note that if we remove the tau signature from this set of inclusive MET analysis channels, the results in the table are not
appreciably modified. Why are some models missed by the various analyses and not others? We will return to address this question below as the causes 
are not always simple and  obvious.

\TABLE{
  \label{numt}
  \begin{tabular}{ | c || c | c | c | c | }
  \hline
  Number of analyses & Flat, 1 fb$^{-1}$ & Flat, 10 fb$^{-1}$ 
  & Log, 1 fb$^{-1}$ & Log, 10 fb$^{-1}$ \\ \hline \hline
     0  &  0.56754  &  0.36796  &  31.823  &  27.024  \\ \hline
     1  &  1.3458  &  0.98841  &  6.2704  &  6.5374  \\ \hline
     2  &  3.396  &  2.5141  &  8.9525  &  10.072  \\ \hline
     3  &  13.175  &  10.635  &  11.816  &  11.098  \\ \hline
     4  &  22.014  &  18.455  &  16.491  &  16.344  \\ \hline
     5  &  9.5512  &  10.3  &  5.6905  &  6.6135  \\ \hline
     6  &  15.227  &  16.929  &  6.0529  &  7.1456  \\ \hline
     7  &  20.081  &  17.697  &  6.7416  &  6.1954  \\ \hline
     8  &  7.6394  &  11.75  &  3.0083  &  4.371  \\ \hline
     9  &  3.9205  &  6.3569  &  1.5223  &  2.6226  \\ \hline
     10  &  2.0825  &  2.7943  &  1.0511  &  1.1783  \\ \hline
     11  &  1.0013  &  1.2116  &  0.57992  &  0.79818  \\ \hline
\end{tabular}

  \caption{The percentage of models that are observable in $n$ analyses, for each model set, for 1
    and 10 $\infb$ luminosity assuming a 50\% background uncertainty.}
}

It is also of interest to ask `if SUSY signatures are found in {\it only} one of the ATLAS searches, which one is it?'; this can be important for any number of 
reasons including questions about the strict validity of any given analysis. For example, for the flat prior models with 1 fb$^{-1}$ of integrated luminosity 
we find that the 2j0l search is this lone analysis in 75.7\%(84.9\%) of the cases assuming a SM background uncertainty of 50\%(20\%) with the b analysis coming 
in as a distant second at 8.7\%(7.5\%), respectively. These results are thus seen to be somewhat sensitive to the assumed systematic error on the SM background 
and are also found to be sensitive to the choice of the flat or log prior set. Detailed answers to this question can be found in Table ~\ref{oneanalysis}.

\TABLE{
    \begin{tabular}{lcccc}
      \hline
      \hline
      Analysis & Flat$~\delta B=50\%$ & Flat$~\delta B=20\%$ &Log$~\delta B=50\%$ & Log$~\delta B=20\%$ \\
      \hline
      4j0l  &    0.43  &  0   &  0.56  &   0  \\
      2j0l  &   75.7  &  84.6  & 44.1  & 59.9 \\
      4j1l  &     0   &   0   &    0   &   0  \\
      3j1l  &    3.4  &  0  & 18.4  &  11.8   \\
      2j1l  &    3.6  &  5.8 & 10.6 &  11.2   \\
      OSDL  &     0   &   0   &    0   &   0  \\
      SSDL  &    0.56 &   0   &    0   &   0  \\
      3lj   &    0.11  &  0   & 10.1   & 9.9  \\
      3lm   &     0   &   0   &    0   &   0  \\
      $\tau$ &   8.0  & 1.9  &  3.4  &  1.3  \\
      b     &    8.7  &  7.7  &12.3  & 5.9   \\
      \hline
      \hline
      \caption{The identity of the single analysis discovering SUSY signals at the $S=5$ level assuming an integrated luminosity of 1 fb$^{-1}$. The 
               value shown is the fraction, in percent, of models found by a given analysis.}
      \label{oneanalysis}
    \end{tabular}
}

We can further quantify the effect of reducing the systematic errors by reproducing Table~\ref{numt}\ and taking the 
systematic error on the SM background to be $20\%$.  This is shown in Table~\ref{numt20}.  As we can see, this smaller systematic 
error significantly reduces the number of models that are missed in all of the analyses, \eg, only a relatively small number of models from
the flat prior sample would now remain undiscovered by any analysis. Even the log prior sample 
experiences a significant reduction in the fraction of models which are missed entirely.    
To emphasize the power of reduced systematic errors, we compare the number of flat prior models that are missed by all analysis
channels with a luminosity 
of $1(10)~\infb$ and a systematic error on the background of $50\%$, \ie, 369(239),
to the case with a $20\%$ systematic error, \ie, 11(4). We conclude that reducing the systematic error is a very powerful way to increase 
Supersymmetric parameter space coverage. 

\TABLE{
  \label{numt20}
  \begin{tabular}{ | c || c | c | c | c | }
  \hline
  Number of analyses & Flat, 1 fb$^{-1}$  & Flat, 10 fb$^{-1}$  
  & Log, 1 fb$^{-1}$  & Log, 10 fb$^{-1}$  \\ \hline \hline
     0  &  0.016411  &  0.0059733  &  18.688  &  12.629  \\ \hline
     1  &  0.077577  &  0.041813  &  5.3597  &  4.1728  \\ \hline
     2  &  0.57139  &  0.22848  &  7.299  &  8.1241  \\ \hline
     3  &  4.9157  &  2.5939  &  9.4147  &  8.161  \\ \hline
     4  &  22.083  &  13.719  &  21.791  &  17.393  \\ \hline
     5  &  5.9003  &  6.0883  &  6.1707  &  8.7518  \\ \hline
     6  &  11.173  &  14.751  &  7.2285  &  10.377  \\ \hline
     7  &  30.085  &  24.238  &  11.742  &  10.487  \\ \hline
     8  &  9.4376  &  13.201  &  4.5839  &  8.1241  \\ \hline
     9  &  6.051  &  10.57  &  2.9619  &  4.8006  \\ \hline
     10  &  6.5538  &  10.175  &  2.9267  &  4.2836  \\ \hline
     11  &  3.1359  &  4.3874  &  1.8336  &  2.6957  \\ \hline
\end{tabular}

  \caption{As in Table~\ref{numt}, but now assuming a 20\% systematic error on the SM
    background instead of 50\%.}
}

\subsection{Properties of Unobservable Models}

These results now suggest the more specific issue of how and why any of the pMSSM models are not observable in the various ATLAS SUSY
search analyses. Of course with so many models under discussion finding specific reasons in every case is not possible. However, in the detailed 
discussion below we will endeavour to find all of the most important culprits which will cover the vast majority of the missed model cases. Since 
in some cases some subtle issues are involved and the physics is more complex than that encountered in, \eg, mSUGRA models, a thorough discussion 
of all the issues is mandated. 

A useful piece of information in addressing the question of why models are unobservable is what are the various individual SUSY 
contributions to the the relevant signals for any given analysis. For example, in the conventional mSUGRA scenario, apart from events 
which originate from hard ISR, the common wisdom is that gluino pair production is almost exclusively the source of the 4j0l signal since the gluinos 
are usually more massive than the squarks and each gluino essentially decays to the 2j+MET final state. This assumption, \eg, forms the basis of 
the Tevatron squark and gluino searches discussed above. However, in the pMSSM models we consider here, we find that  
the situation is far more complicated since the sparticle spectra do not follow any particular pattern. Figure~\ref{42jprocs} shows
the origin of the 4j0l and 2j0l signals for both prior cases. Here we see, \eg, in the flat prior case, that associated 
squark-gluino production can easily be the major contributor among the various sources for both of these signatures in many of the models. This 
can easily happen when squarks are more massive than gluinos, which they very often are in this model set. In such a case, gluinos can  
commonly decay to 2j+MET while squarks will decay to 3j+MET. Note, however, that in the log prior case the fractions of the initial 
SUSY states contributing to these same signatures is now completely different as, among other reasons, the sparticle spectra 
are somewhat more compressed. Thus the squark-gluino mass ordering, spectrum degeneracy, the number of steps in the decay cascade, as 
well as the amount of ISR can all play a role in generating the (4,2)j0l final states.

\FIGURE{
    \includegraphics[width=0.45\textwidth]{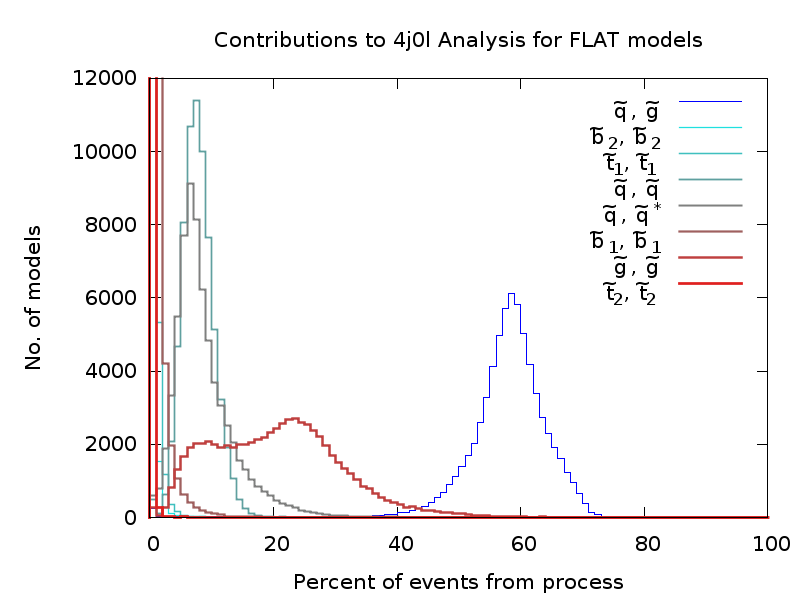}
    \includegraphics[width=0.45\textwidth]{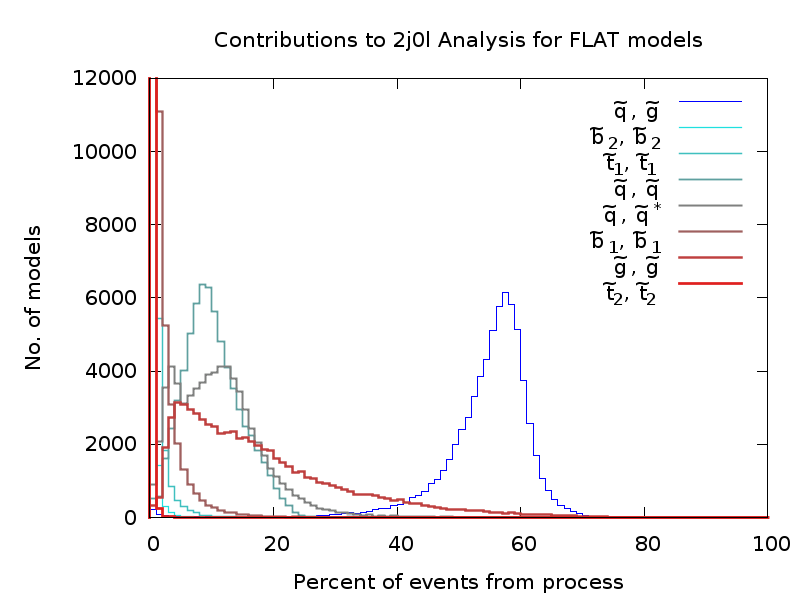}
    \\
    \includegraphics[width=0.45\textwidth]{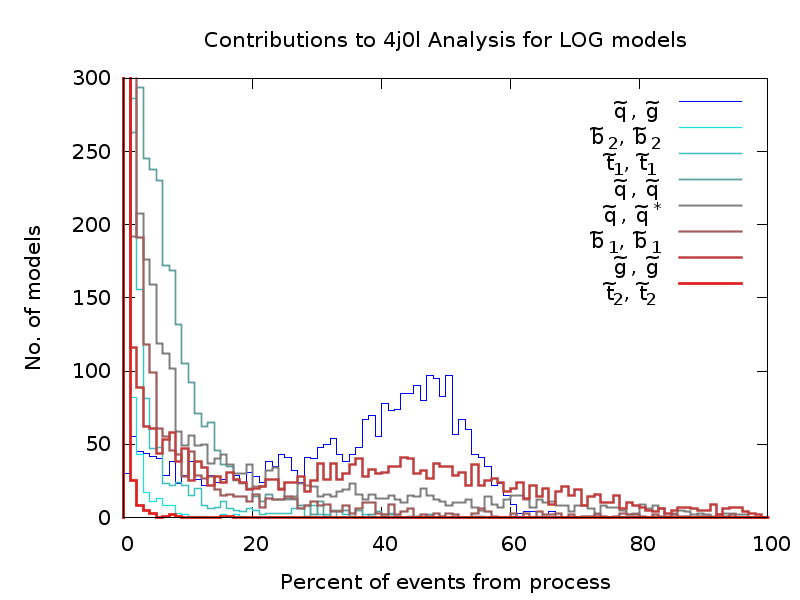}
    \includegraphics[width=0.45\textwidth]{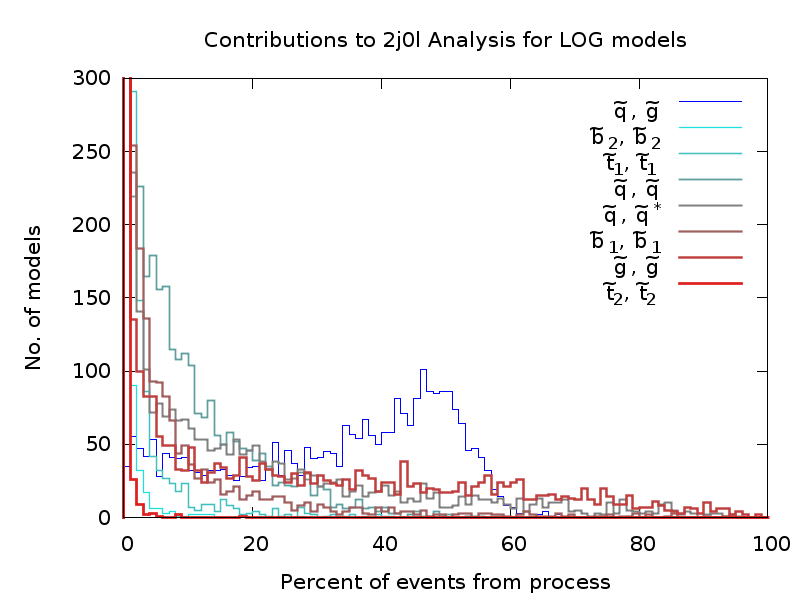}
    \caption{Contributions to the events passing the 4j0l and 2j0l analysis
      cuts from various SUSY production processes as indicated for both flat and log priors.}
  \label{42jprocs}
}

Perhaps the obvious question to ask about the models which are not found by the ATLAS analyses is `how much does the overall SUSY mass scale contribute to preventing 
these pMSSM models from being found?'  Are, \eg, the gluinos and squarks, which are most commonly at the top of SUSY decay chains, just too heavy 
to be produced with sufficient rates to yield a viable signal that is large enough to pass selection cuts?  As we will see, large squark and/or gluino masses, 
while playing a role in the signal significances, are not always the most important determining factor as to whether or not a given pMSSM model is discovered. 
Figures~\ref{znvsg4j},~\ref{znvsg2j} and 
\ref{znvsgjh} address this issue for the specific case of the gluino mass, \ie, perhaps if the gluinos are too massive models will be completely missed. 
In Figures~\ref{znvsg4j} and ~\ref{znvsg2j} we see the significance of the 4j0l and 2j0l analyses, respectively, as a function 
of the gluino mass for both the flat and log prior model sets assuming an integrated luminosity of $1~\infb$ and a 50\% background systematic. 
Both analyses show a qualitatively similar 
behavior. Overall, we see that $S$ tends to decrease as the gluino mass increases.  This is not a surprise and
is especially noticeable in the log prior case as the gluino mass range extends out to $\sim 3$ TeV. We note, however, that for any 
given value of the gluino mass, the range of values of $S$ can extend over two orders of magnitude, so, clearly,  
the gluino mass itself is not the sole determining factor for the overall signal significance. Does this situation change as more luminosity is  
accumulated? Figure~\ref{znvsgjh} shows how the values of S respond to an increase in the integrated 
luminosity for the case of flat priors for both the 4j0l and 2j0l
channels. We see that there is only a marginal increase in the typical value of 
$S$, indicating that increasing the integrated luminosity will not necessarily lead to the analyses capturing all of the missed models; this is as 
expected from the discussion of the background systematic errors above.  

One thing to note about these figures is that there are a number of models whose significance value, $S$, lies rather close to either side of the 
$S=5$ boundary. Clearly for such models variations in the signal generation process, or even statistical fluctuations, may push their significance either 
below or above this boundary. Thus, these models near the observation boundary may or may not be observable; in this paper we will strictly assume that the 
resulting values for $S$ as will be seen by ATLAS is exactly as generated here. Another thing to note is the gluino mass reach implied by the log prior results. 
Here we see that the 2j0l analysis appears to be sensitive to gluino masses even as large as 3 TeV for some pMSSM model cases these models, however, may have
lighter squarks which are being observed rather than the heavy gluinos.

\FIGURE{
  \includegraphics[width=0.45\columnwidth]{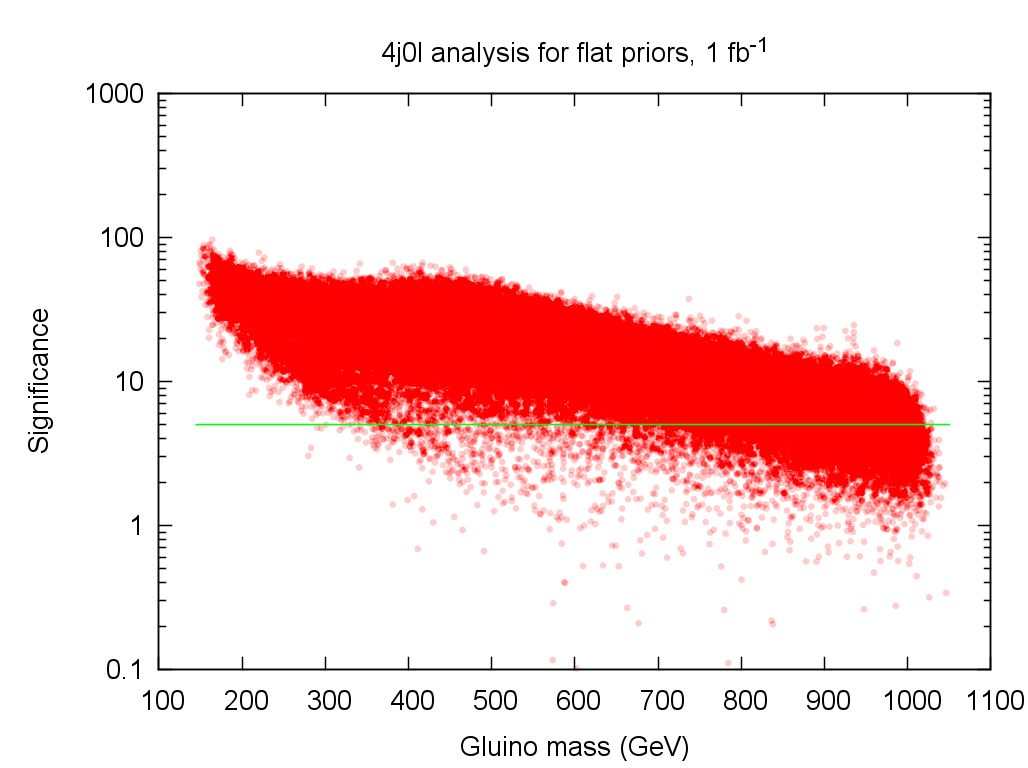}
  \includegraphics[width=0.45\columnwidth]{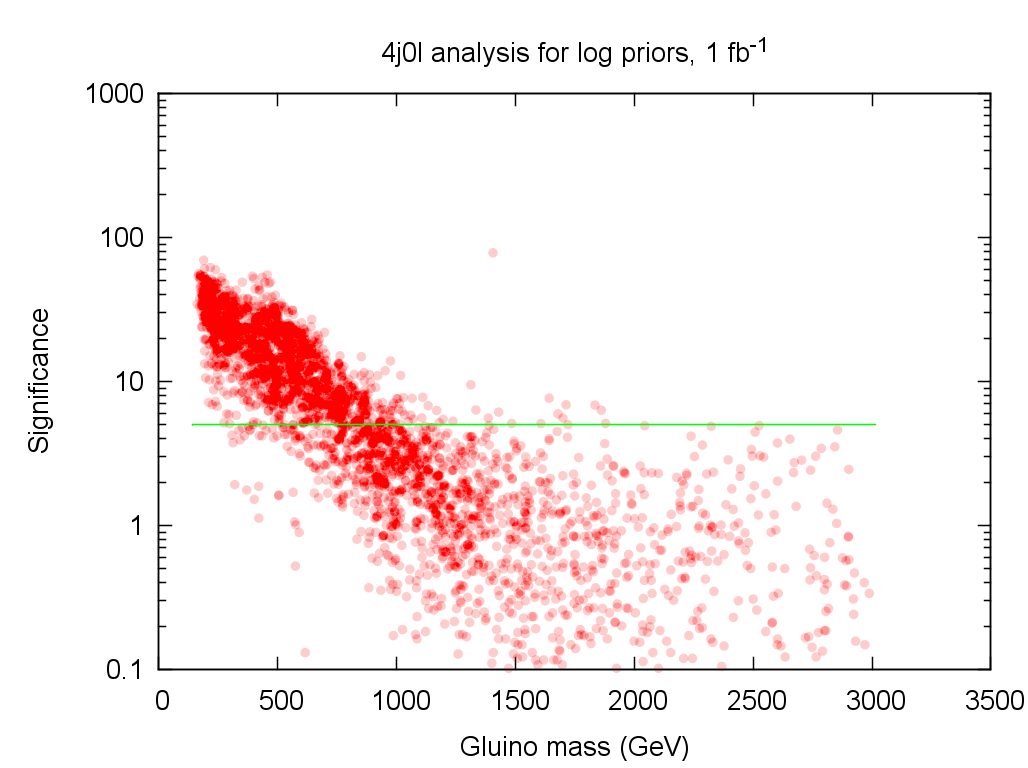}
  \caption{The significance of the 4j0l analysis as a function of gluino
    mass for the flat(log) prior set in the left(right) panel. The horizontal line denotes $S=5$.}
  \label{znvsg4j}
}

\FIGURE{
  \includegraphics[width=0.45\columnwidth]{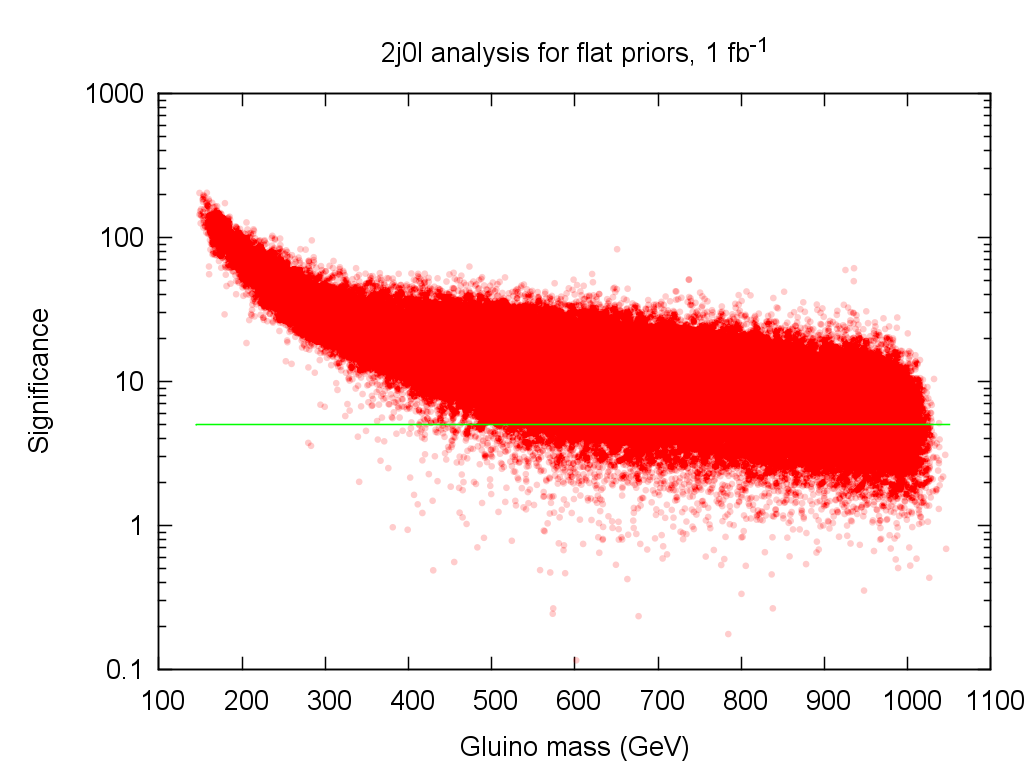}
  \includegraphics[width=0.45\columnwidth]{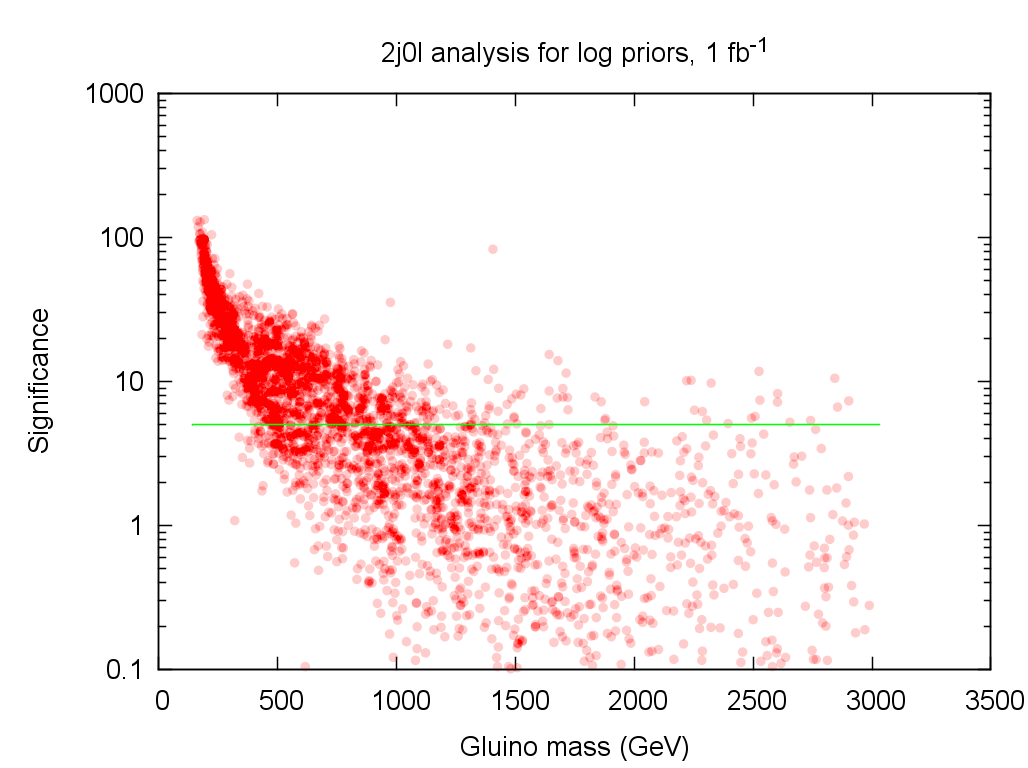}
  \caption{The significance of the 2j0l analysis as a function of gluino
    mass for the flat(log) prior set in the left(right) panel.}
  \label{znvsg2j}
}

\FIGURE{
  \includegraphics[width=0.45\columnwidth]{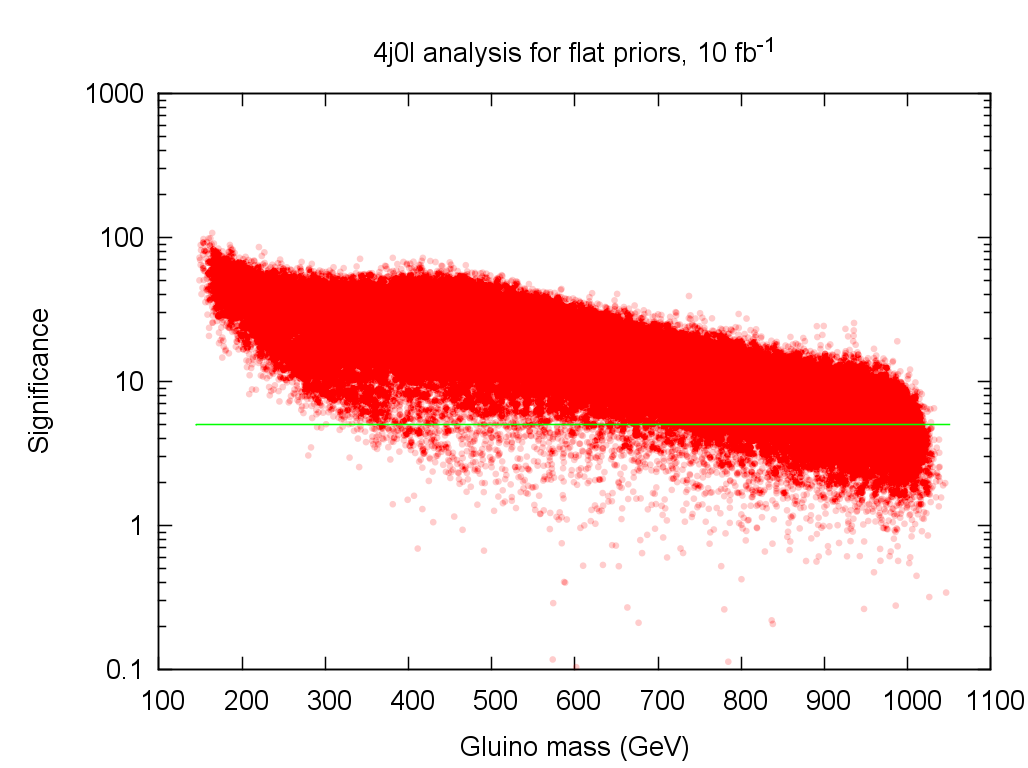}
  \includegraphics[width=0.45\columnwidth]{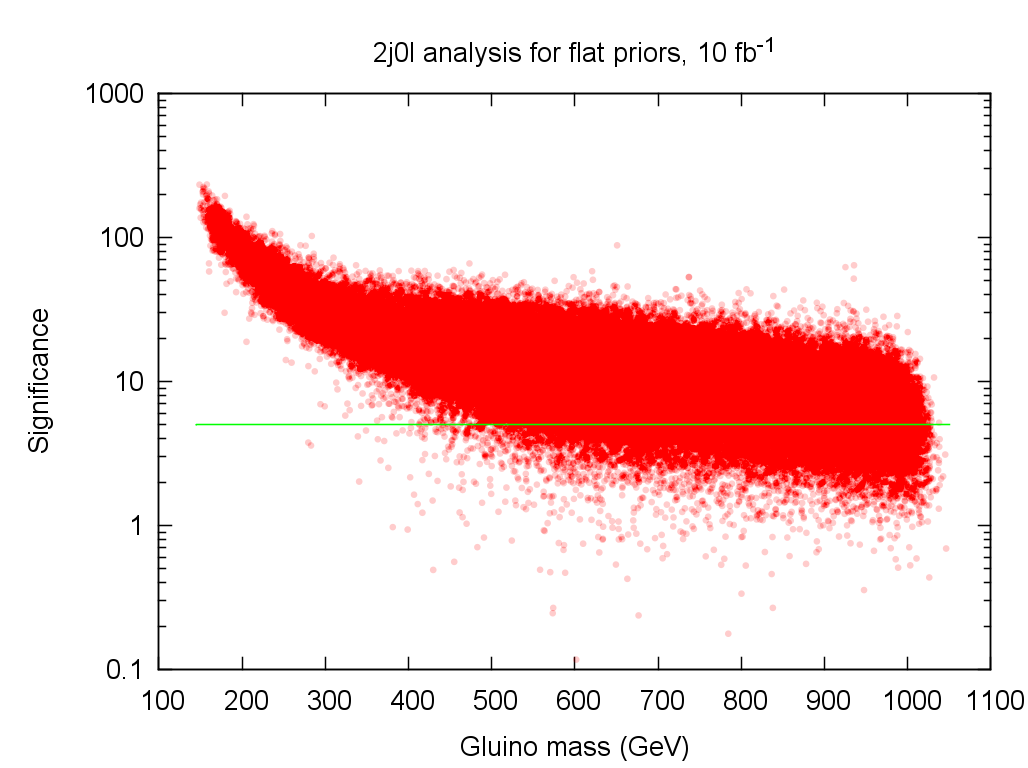}
  \caption{The significance as a function of gluino
    mass for 10 $\infb$ luminosity for the 4(2)j0l analysis in the left(right) panel.}
  \label{znvsgjh}
}

Does a similar result hold for the squarks? Is the squark mass scale an important factor in model observability?  
Figure~\ref{zvslsm}(\ref{zvsavglsm}) shows a comparison of the values of significance, $S$, for both the 4j0l and 2j0l analyses 
with flat priors assuming an integrated luminosity 
of $1~\infb$ as a function of the lightest(average) squark mass within the first two squark generations. As anticipated, $S$ in this case 
shows only a weak decrease as the squarks become more massive. We also see, as was the case for gluinos, that for any 
particular value of the squark mass, the range of values of $S$ spans more than an order of magnitude. This supports our suspicion that 
effects other than just the overall squark mass scale play a major role in determining the signal significance and in preventing models from 
being discovered by these analyses.

\FIGURE{
  \includegraphics[width=0.45\columnwidth]{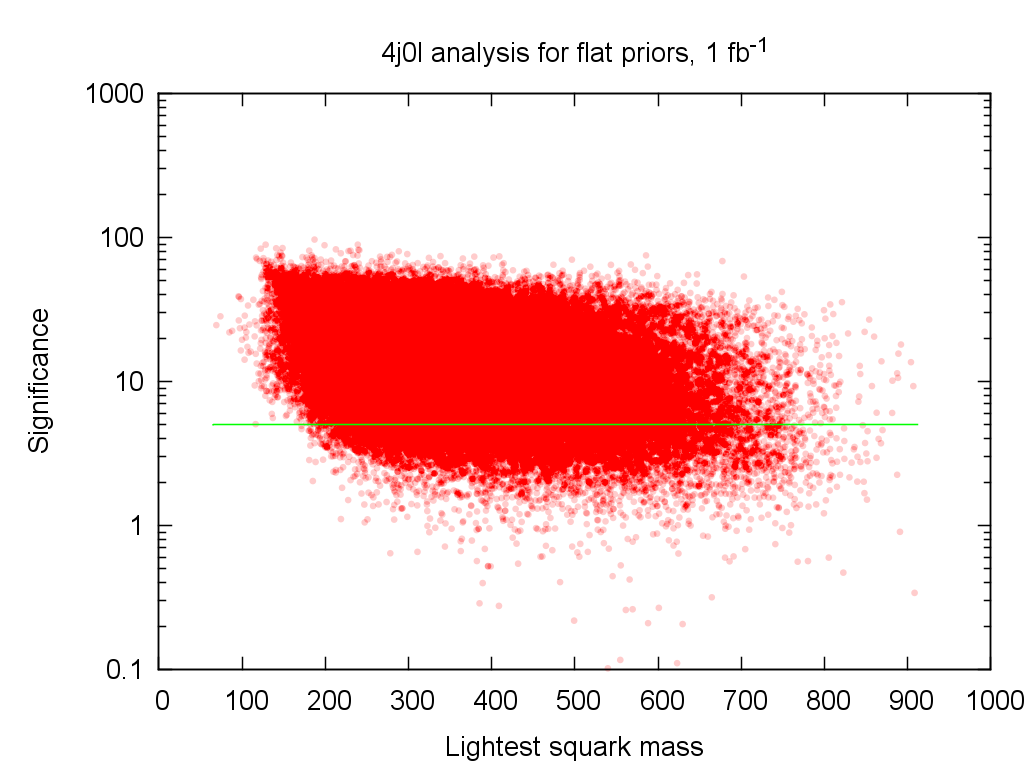}
  \includegraphics[width=0.45\columnwidth]{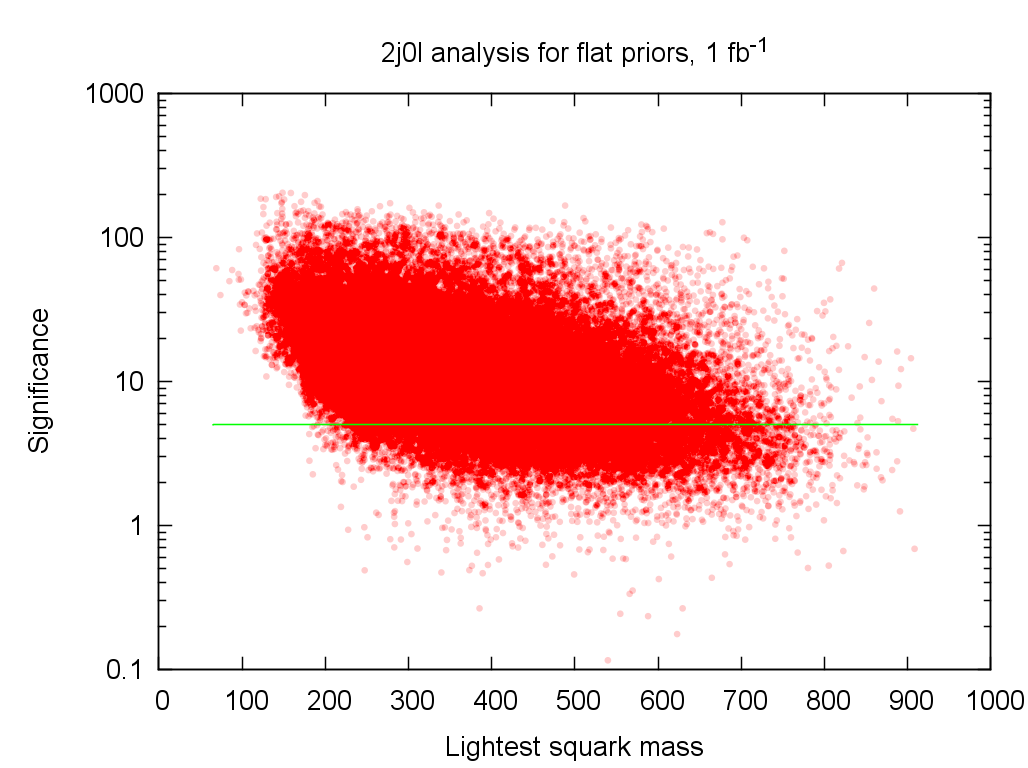}
  \caption{The significance versus the lightest 1st or 2nd generation squark
    mass for the 4(2)j0l analysis in the left(right) panel.}
  \label{zvslsm}
}

\FIGURE{
  \includegraphics[width=0.45\columnwidth]{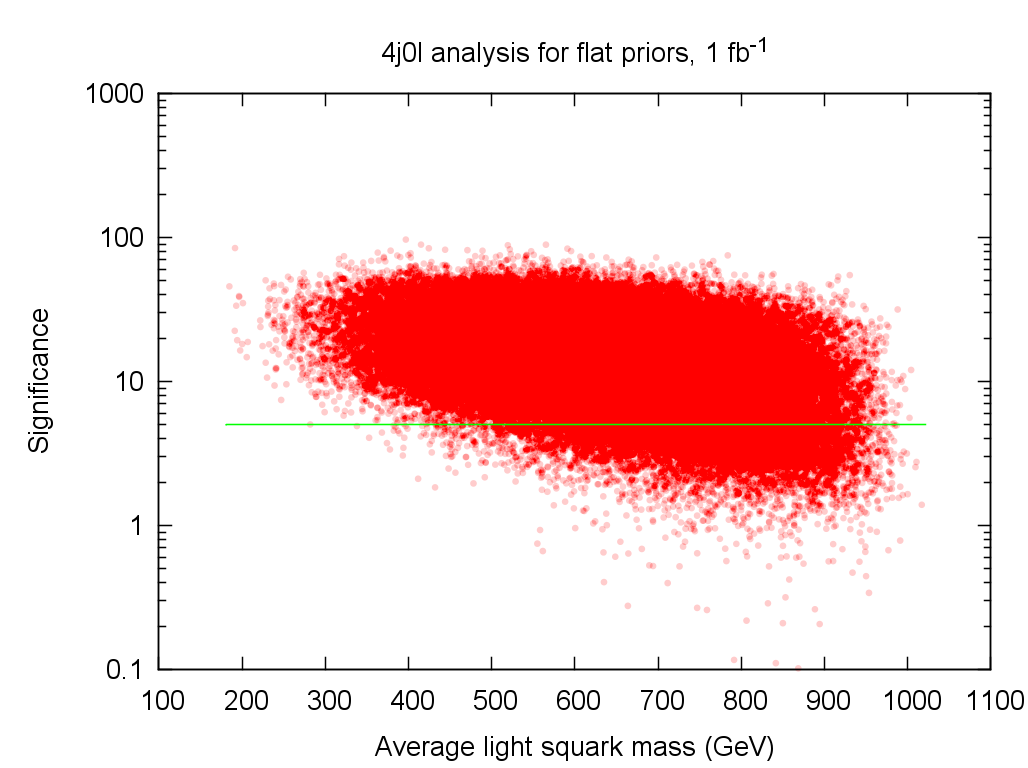}
  \includegraphics[width=0.45\columnwidth]{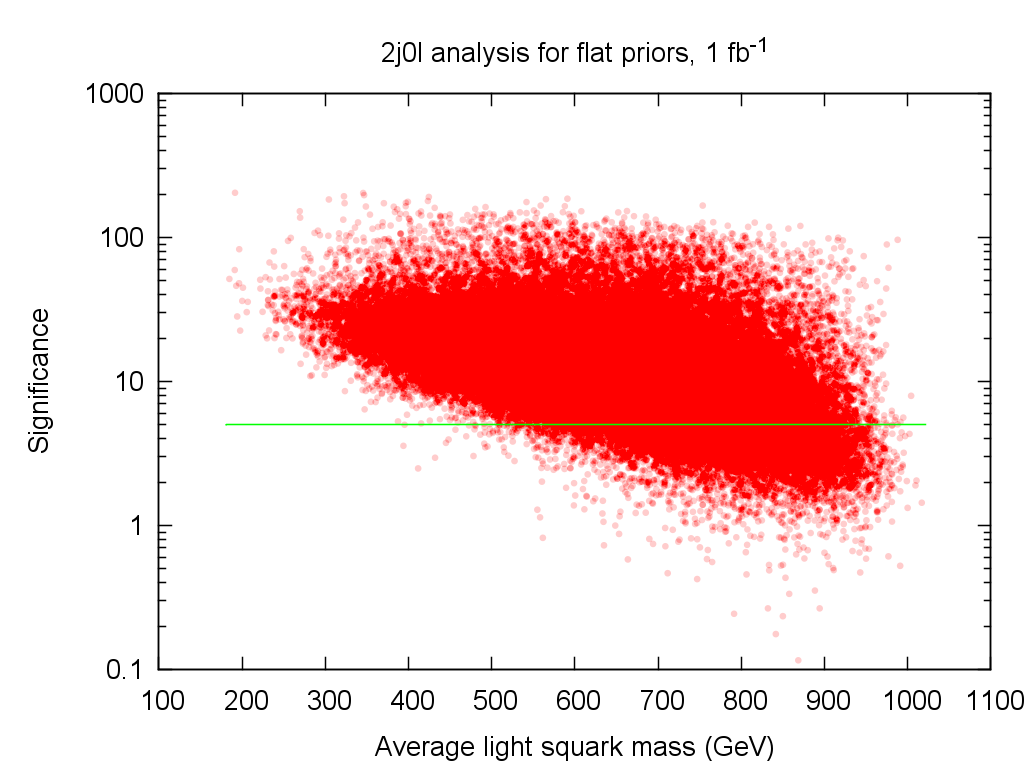}
  \caption{The significance versus the average 1st and 2nd generation squark
    mass for the 4(2)j0l analysis in the left(right) panel.}
  \label{zvsavglsm}
}

At this point it is instructive to consider the relative distributions of pMSSM models which are observed (or not) by the 4j0l and 
2j0l ATLAS analyses in the gluino mass versus average light squark mass plane. This is relevant as squark and gluino production will generate all 
of these MET signals. These results are shown for the flat prior model sample with both low and high integrated luminosities in 
Figures \ref{avglsmvgm4j} and \ref{avglsmvgm2j}. In these figures, the models that are observable in the respective 
analyses are represented as green points whereas those that are missed by the analyses are shown in red. 
Examining these figures we see that most, but not all, of the missed models lie in the upper right-hand corner of this plane where both 
the squark and gluino masses are large. This is just what we would naively expect since in this case both squark and gluino production 
would be kinematically suppressed and a smaller number of events would result. It is important to note, however, that there are also a
significant  number of obviously 
interesting models that have relatively light squark and gluino masses but which are not detected in either of these analyses. Here we again observe that 
increasing the integrated luminosity does not particularly help in most of these cases, even those with rather light squark and/or gluino masses due to the 
large SM background systematic errors.

%

\FIGURE{
  \includegraphics[width=0.45\columnwidth]{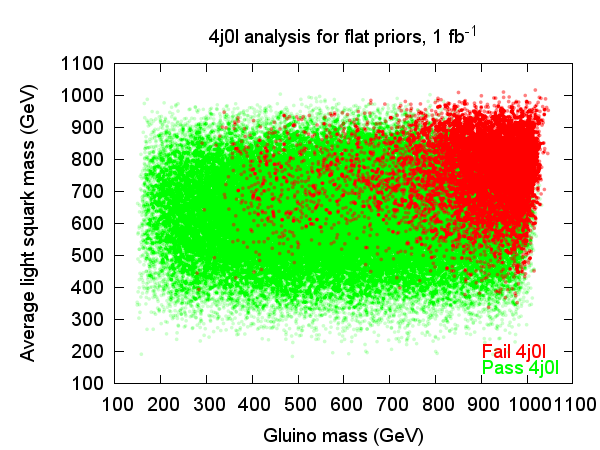}
  \includegraphics[width=0.45\columnwidth]{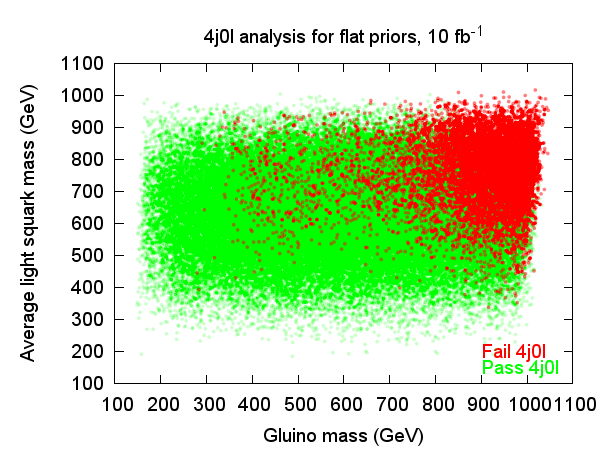}
  \caption{The pMSSM models from the flat prior set in the
    gluino mass - average 1st and 2nd generation squark
    mass plane.  The models that pass (fail) the 4j0l analysis are shown in green(red). 
    The left(right) panel corresponds to an integrated luminosity of 1(10) $\infb$.
  }
  \label{avglsmvgm4j}
}

%
%

\FIGURE{
  \includegraphics[width=0.45\columnwidth]{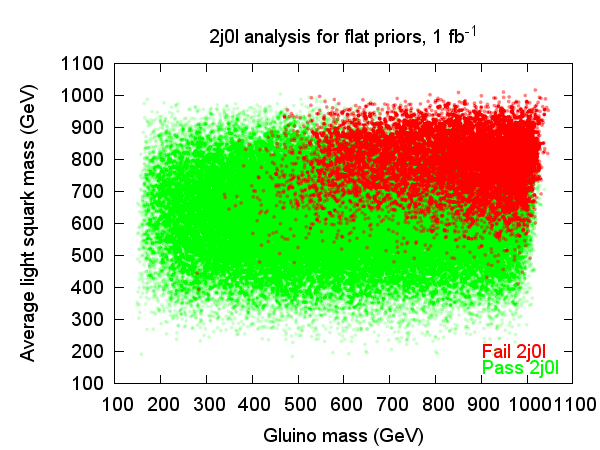}
  \includegraphics[width=0.45\columnwidth]{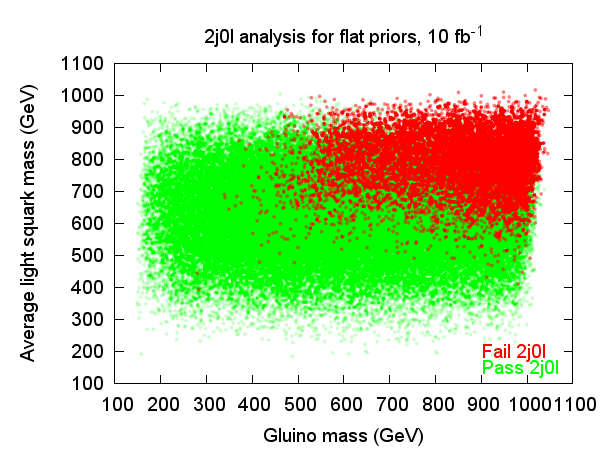}
  \caption{The same as in Figure~\ref{avglsmvgm4j}, but for the 2j0l analysis.
  }
  \label{avglsmvgm2j}
}

%
%

Interestingly, these figures show, \eg,  that a particular pair of models (model numbers 19933 and 53105) have gluino masses below 300 GeV and light 
squarks and yet they are missed by both the 4j0l and 2j0l analyses. The reason for this is that these models have unusual spectra where the gluinos 
mostly decay through the ${\tilde d_R}$ which then, in turn, universally decays via ${\tilde {\chi_2}^0}$ (which is mostly bino in these cases) 
finally yielding the $2jl^+l^-+$MET final state. Since leptons essentially must always appear in the cascade decays of these two models, the 4j0l and 2j0l
analysis requirement of there being no isolated leptons can never be met. These two models are, however, found to be observable in the lepton plus jets 
analyses. To see this more clearly, Fig.~\ref{both1} shows the set of models which fail the 4(2)j0l analyses and {\it simultaneously} indicates whether or not 
they pass the corresponding 4(2)j1l analysis. Here we see that the two specific models under discussion, as well as others, which are missed by the 4(2)j0l 
analyses are indeed subsequently captured by the corresponding leptonic analyses.

\FIGURE{
  \includegraphics[width=0.45\columnwidth]{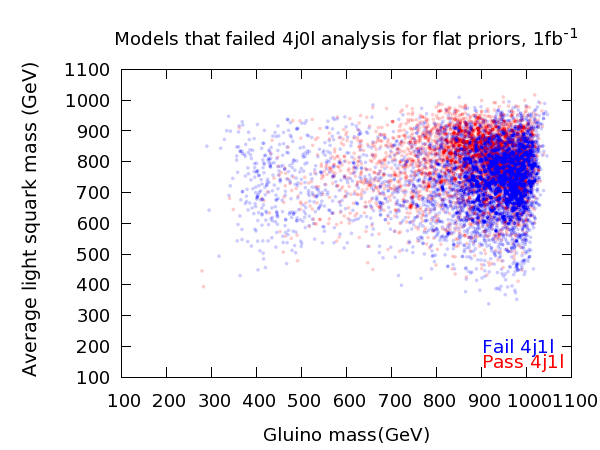}
  \includegraphics[width=0.45\columnwidth]{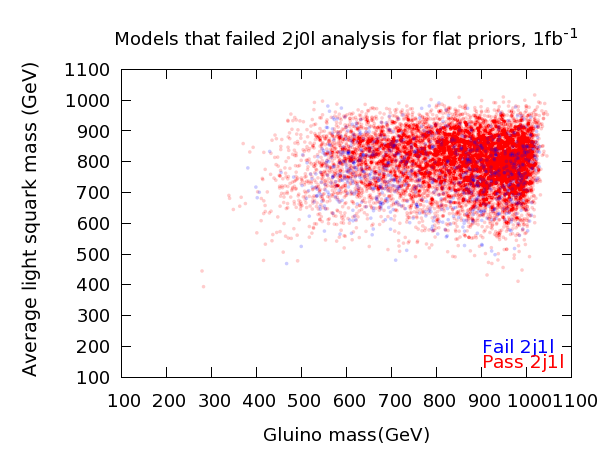}
  \caption{The set of flat prior models that fail the 4(2)j0l analyses and whether they are detected or not in
           the corresponding 4(2)j1l analyses.
  }
  \label{both1}
}

As alluded to in the previous section on model generation, many of our models satisfy the Tevatron search constraints 
even though the squarks and gluinos are fairly light; this occurs when the mass splittings between the squarks and/or gluinos and the 
LSP are relatively small. This configuration easily leads to rather soft jets in the final state and clearly some models will be unobservable in the 
(4,2)j0l analysis channels at the LHC for the same reasons. To see this, it is worth examining which models pass and  
fail the (4,2)j0l analyses as the gluino/squark-LSP mass splittings are varied. This is shown for the case of gluinos with flat priors 
in Figure \ref{gllspsplvgm} (always assuming an 
integrated luminosity of 1 fb$^{-1}$ and $50\%$ systematic background errors). 
Here we see, particularly in the case of the 4j0l channel, that many 
models with light gluinos which are unobservable have small mass splittings with the LSP, 
hence producing rather soft jets. This occurs mainly
for gluino masses $m_{\tilde g}\gtrsim 350$; gluinos lighter than this have large production cross sections associated with hard radiated jets
which can compensate for the soft jets in the decay and pass the kinematic cuts for this channel (we note that squark production could also
be contributing to this channel).
Of course some of these models will again be missed by the 4j0l and 2j0l analyses due to the presence of high $E_T$ leptons as mentioned above. 
To see how this impacts us more clearly, Fig.~\ref{compare1} shows the set of flat prior models that are unobservable in {\it both} of the 4(2)j0l 
analyses as well as the corresponding 4(2)j1l analyses in red while the green points label models passing the $S=5$ significance 
requirements of {\it either} analysis. Still, 
it is clear that many models are unobservable in the (4,2)j0l channel due to the small mass splittings and not due to the presence of leptons.

\FIGURE{
  \includegraphics[width=0.45\columnwidth]{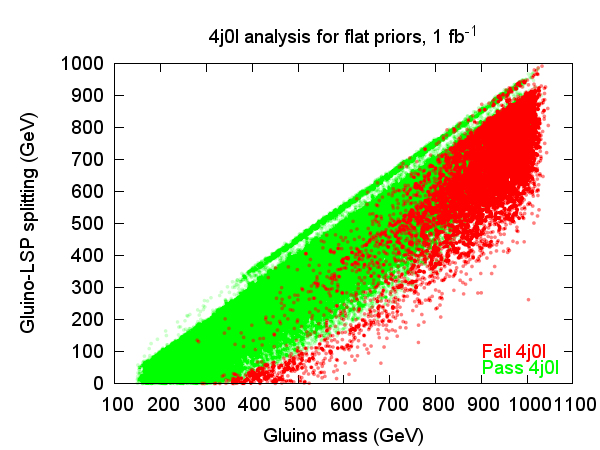}
  \includegraphics[width=0.45\columnwidth]{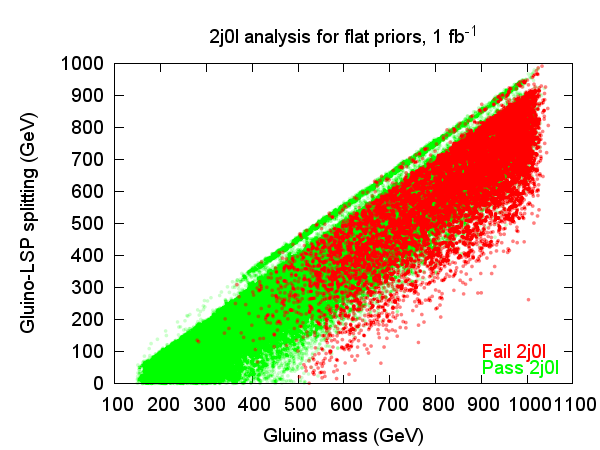}
  \caption{The mass splitting between the gluino and LSP as a function of the gluino mass for the flat prior model sample.
   The models that pass the 4(2)j0l analysis for 1~$\infb$ are shown in
    green, while the ones that fail are displayed in red.  The 4(2)j0l channel is shown in the left(right) panel.
  }
  \label{gllspsplvgm}
}

%
%

\FIGURE{
  \includegraphics[width=0.45\columnwidth]{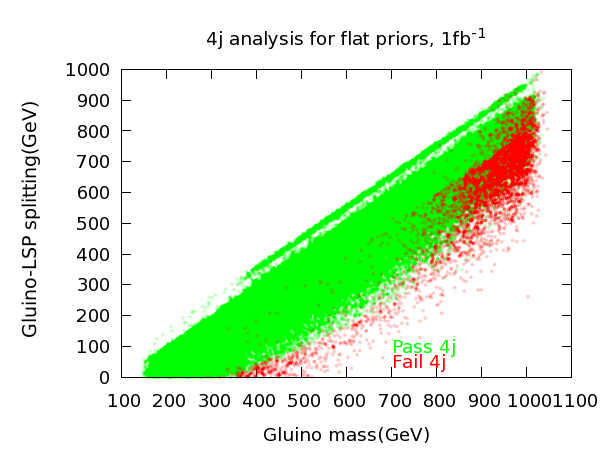}
  \includegraphics[width=0.45\columnwidth]{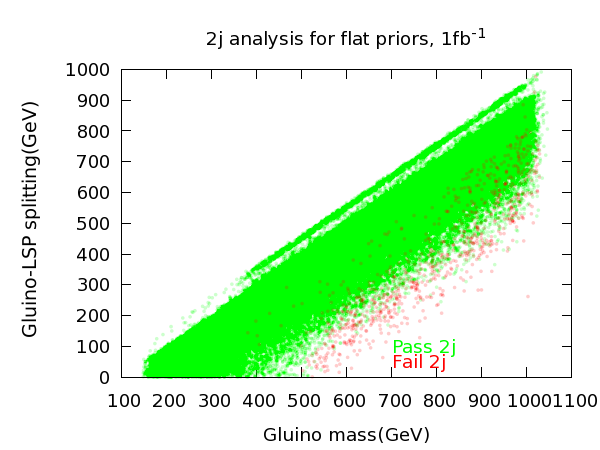}
  \caption{The mass splitting between the gluino and LSP as a function of the gluino mass for the flat prior model sample.
   The models that fail  {\it both} of the 4(2)j0l and 4(2)j1l analyses in shown in red, while the 
   green points label those models passing {\it either} analysis.
   The 4(2)j0l channel is shown in the left(right) panel.
  }
  \label{compare1}
}

\FIGURE{
  \includegraphics[width=0.45\columnwidth]{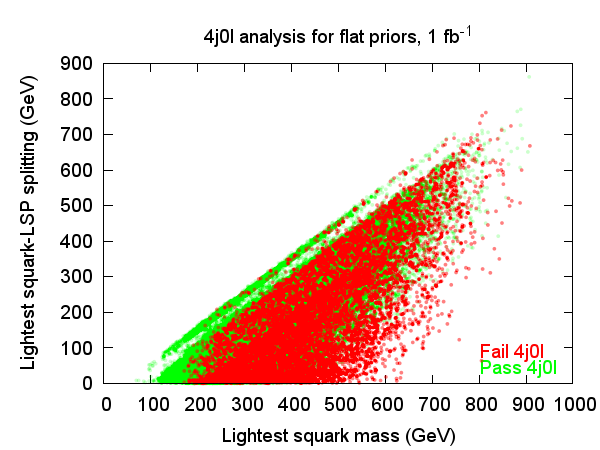}
  \includegraphics[width=0.45\columnwidth]{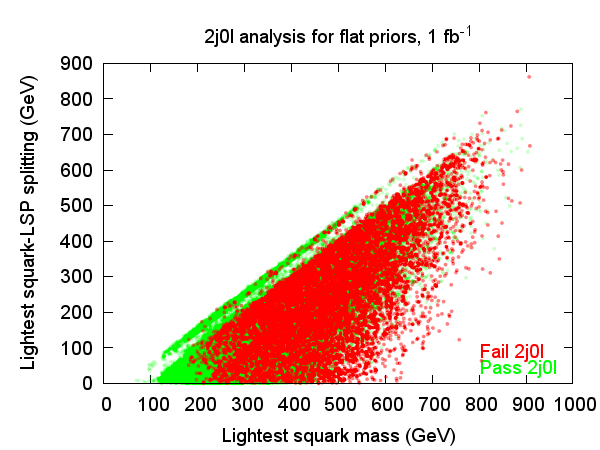}
  \caption{The mass splitting between the lightest first/second generation squark and the LSP as a function of the 
    lightest squark mass for the flat prior model sample.
    The models that pass the 4(2)j0l analysis for 1~$\infb$ are shown in
    green, while the ones that fail are displayed in red.  The 4(2)j0l channel is shown in the left(right) panel.
  }
  \label{avlslspsplvavlsm}
}
%
%

Figure~\ref{avlslspsplvavlsm} shows an analogous behavior to that discussed above for gluinos in the case of the lightest first/second generation 
squark mass splitting with the LSP for both 
the 4j0l and 2j0l channels in the flat prior case. As was found for the gluinos, a respectable number of models which fail these analyses are observed 
to have light squarks with small mass splittings with the LSP leading to soft jets in their decay products. 
Certainly, a sizable fraction of such models will not be observed in the (4,2)j0l analyses for this reason but others again may be missed due to the presence of 
leptons in their cascade decays as is shown in Fig.~\ref{compare2}. 

\FIGURE{
  \includegraphics[width=0.45\columnwidth]{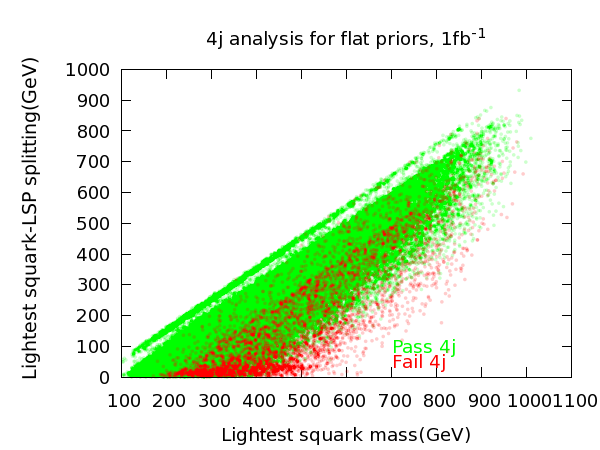}
  \includegraphics[width=0.45\columnwidth]{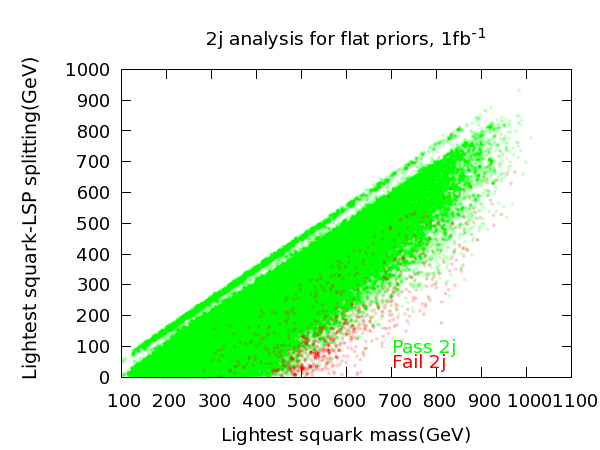}
  \caption{The set of flat prior models failing {\it both} of the 4(2)j0l and 4(2)j1l analyses are shown in red, while green points label models 
    passing {\it either} analysis. The results are shown in the plane of the mass splitting between the lightest first/second generation squark and the LSP
    and the lightest squark mass.
  }
  \label{compare2}
}

So far we have found three `obvious' reasons why some of our pMSSM model points fail to be observed by the 4j0l/4j1l and 2j0l/2j1l analysis
channels: ($i$) 
low signal cross sections for particular channels which can be correlated with ($ii$) heavy colored states at the top of decay 
chains causing kinematic suppression or unexpected decay patterns. The fact that these characteristics render the models unobservable
can also in large part be attributed to the rather large
systematic errors associated with the sizable SM backgrounds in both the 4j0l and 2j0l analyses. As we saw above, a larger  
systematic uncertainty associated with the SM background requires a greater number of signal events to reach the $S=5$ discovery 
level. The size of this SM background uncertainty was found to play a {\it major} role in models being missed by the 4j0l and 2j0l 
analyses. ($iii$) Furthermore, small mass splittings between the various colored states in the spectrum and the LSP can lead to the 
production of significantly softer final state objects that have a more difficult time passing the various analysis thresholds. 

Let us now turn to other search channels. Figure~\ref{lepvsjet} shows the set of flat prior models that {\it fail} the 4j1l and 2j1l analyses 
assuming an integrated luminosity of 1 $\infb$ and the standard $50\%$ background systematic error. Here we have examined whether a given model 
fails because of the jet cut requirements (as in the corresponding 4j0l and 2j0l analyses) or because of the leptonic cuts for these specific 
analyses. As we would expect, most of the models failing the jet criteria 
correspond to cases with large squark and/or gluino masses, particularly so for the 4j1l case where the jet requirements are somewhat 
stronger. For either analysis, however, we see that most of the models are missed due to their failure to pass the leptonic cuts and are
observed in the  zero lepton channels; in many cases this is 
simply due to the absence of the required lepton with either sufficient $E_T$ or lack of isolation from the final state jets.

\FIGURE{
  \includegraphics[width=0.45\columnwidth]{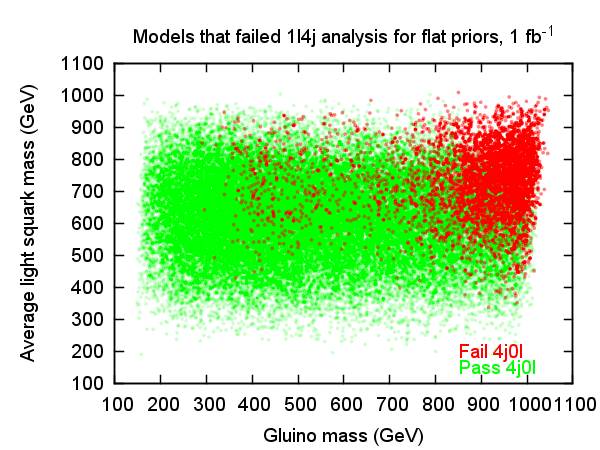}
  \includegraphics[width=0.45\columnwidth]{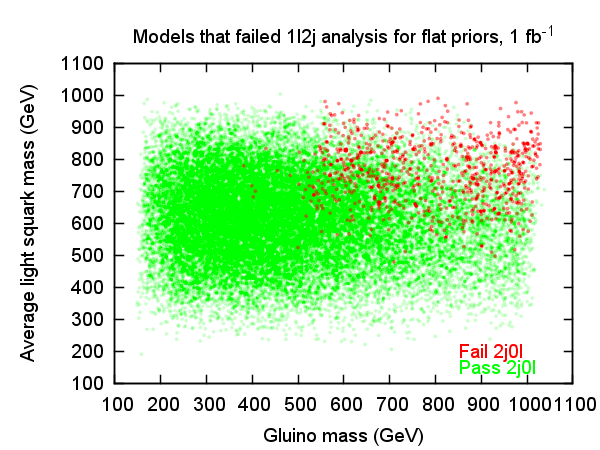}
  \caption{The set of pMSSM models which are unobservable in the 4(2)1l analysis channel in the left(right) panel shown in the plane
    of the average 1st/2nd generation squark mass and the gluino mass.  The models which are observed in the
    corresponding 4(2)j0l channel are shown in green, while those that fail
    these analyses are shown in red.
  }
  \label{lepvsjet}
}
%
%

\subsection{The Effect of Cuts}

It is instructive to consider how our pMSSM model samples `respond' as each of the 
individual experimental cuts are applied for a given analysis.  This provides
another direct indicator of why models are observable or not.  For each
analysis and each model, we keep track of the number of signal events after
each cut is applied in sequence.  With these event numbers, and the number of
background events after all the cuts have been applied, we compute a significance at each step and check if
it is greater than 5.  Since we compare the number of signal events
after each kinematic cut to the number of background events after all cuts (as this is
the only result for the background we were provided), this
significance is somewhat artificial.  Nonetheless it is still illustrative in showing
the relative impact of the cuts.  Note that for the analyses that have a
$\MEFF$ cut, we apply that cut to the signal at each step, and only
consider the effect of the remaining cuts here.

The accompanying Tables 
\ref{4j0lcuts}--\ref{bcuts} 
show the results of these considerations for both the flat and log prior model samples assuming an integrated 
luminosity of either 1 or 10 $\infb$ and a $50\%$ systematic error on the SM background as usual. 
Note that some care must be used in reading these Tables as in many cases the effectiveness of a given cut may strongly depend upon 
the order in which it has been implemented. Here the cuts are applied in the order as
given by the ATLAS SUSY study \cite{Aad:2009wy}.  The cut numbers listed in the Tables correspond to those in Section 3.2. 
Further note that the flat and log prior model sets can respond somewhat differently to any particular cut or set of cuts so it 
is important to study both of these cases seperately in what follows.

We see in Tables~\ref{4j0lcuts} and \ref{2j0lcuts} that the models easily pass the jet cuts for both the familiar 4j0l and 2j0l searches 
but requiring the absence of isolated leptons (cuts number 5 and 4, respectively) 
takes a respectable toll on the fraction of models found. In the 2j0l case, the stronger 
cut on \MET\ (cut 2) is also seen to lead to a significant weakening in the model space reach. The cut study for the 4(3,2)j1l analysis 
channels is shown in Table~\ref{4j1lcuts}(\ref{3j1lcuts},\ref{2j1lcuts}). 
In all three cases the combined requirements of (only) a single isolated lepton, multiple high $E_T$ 
jets as well as \MET~are all seen to lead to significant reductions in the signal events for these separate analyses.  

In the case of the OSDL search, as shown in Table~\ref{OSDLcuts}, the simultaneous requirement of  opposite sign dileptons and 
four hard jets (cuts 1 and 2) eliminates 
more than $\sim 80\%$ of the model set. In addition, the \MET~and transverse sphericity cuts (cuts 3 and 4) are seen to reduce the signal 
further by another factor of $\simeq 2$. In the SSDL analysis shown in Table~\ref{SSDLcuts},  we again see that the lepton 
and jet requirements remove 
almost $\sim 80\%$ of the model set, but here the \MET\ requirements (cut 3) are more easily met in the surviving model subset. For the trilepton 
analyses, shown in Tables~\ref{3ljcuts} and \ref{3lmcuts} 
the 3 lepton requirement alone (cut 1) is seen to eliminate most of flat prior model sample. Requiring an extra high-$E_T$ jet (cut 2) in the 3lj 
analysis and removing dilepton pair masses near the $Z$ (cut 5) in the 3lm analysis both reduce the number of remaining models to rather small 
numbers in these channels.  For the $\tau$ analysis presented in Table \ref{taucuts}, the transverse mass cut (cut 7) is seen to be the
most restrictive.
In the case of the b analysis shown in Table~\ref{bcuts}, the 
double b-tagging requirements (cut 6) has by far the most impact.

\TABLE{
  \begin{tabular}{ | c || c | c | c | c | }
  \hline
  Cut & Flat, 1 fb$^{-1}$ & Flat, 10 fb$^{-1}$ & Log, 1 fb$^{-1}$ & Log, 10 fb$^{-1}$ \\ \hline \hline
     4j0l 1  &  99.724  &  99.74  &  70.063  &  69.756  \\ \hline
     4j0l 2  &  98.575  &  98.623  &  62.341  &  62.149  \\ \hline
     4j0l 3  &  94.722  &  94.893  &  53.032  &  52.917  \\ \hline
     4j0l 4  &  93.361  &  93.516  &  51.481  &  51.256  \\ \hline
     4j0l 5  &  88.331  &  88.578  &  48.166  &  48.08  \\ \hline
\end{tabular}

  \caption{The percent of models that pass the 4 jet 0 lepton analysis after each
    subsequent cut is applied.  Note that the background after all cuts is used
    to determine significance.}
  \label{4j0lcuts}
}
\TABLE{
  \begin{tabular}{ | c || c | c | c | c | }
  \hline
  Cut & Flat, 1 fb$^{-1}$ & Flat, 10 fb$^{-1}$ & Log, 1 fb$^{-1}$ & Log, 10 fb$^{-1}$ \\ \hline \hline
     2j0l 1  &  99.485  &  99.507  &  68.054  &  67.651  \\ \hline
     2j0l 2  &  95.029  &  95.109  &  53.843  &  53.73  \\ \hline
     2j0l 3  &  94.174  &  94.284  &  52.151  &  51.92  \\ \hline
     2j0l 4  &  87.616  &  87.774  &  47.391  &  47.378  \\ \hline
\end{tabular}

  \caption{Same as in Table~\ref{4j0lcuts} but for the 2 jet 0 lepton analysis channel.}
  \label{2j0lcuts}
}
\TABLE{
  \begin{tabular}{ | c || c | c | c | c | }
  \hline
  Cut & Flat, 1 fb${-1}$ & Flat, 10 fb$^{-1}$ & Log, 1 fb$^{-1}$ & Log, 10 fb$^{-1}$ \\ \hline \hline
     4j1l 1  &  84.57  &  86.203  &  65.303  &  67.651  \\ \hline
     4j1l 2  &  84.119  &  85.805  &  64.21  &  66.617  \\ \hline
     4j1l 3  &  69.575  &  71.657  &  45.839  &  48.043  \\ \hline
     4j1l 4  &  61.35  &  63.941  &  35.226  &  37.703  \\ \hline
     4j1l 5  &  53.349  &  56.064  &  27.68  &  29.874  \\ \hline
     4j1l 6  &  41.731  &  44.885  &  18.371  &  20.421  \\ \hline
\end{tabular}

  \caption{Same as Table~\ref{4j0lcuts} but for the 4 jet, 1 lepton, analysis channel.}
  \label{4j1lcuts}
}
\TABLE{
  \begin{tabular}{ | c || c | c | c | c | }
  \hline
  Cut & Flat, 1 fb$^{-1}$ & Flat, 10 fb$^{-1}$ & Log, 1 fb$^{-1}$ & Log, 10 fb$^{-1}$ \\ \hline \hline
     3j1l 1  &  97.018  &  98.813  &  87.2  &  92.393  \\ \hline
     3j1l 2  &  96.888  &  98.74  &  86.848  &  91.95  \\ \hline
     3j1l 3  &  86.937  &  91.157  &  73.484  &  80.096  \\ \hline
     3j1l 4  &  79.014  &  84.358  &  56.735  &  66.581  \\ \hline
     3j1l 5  &  73.5  &  79.283  &  48.131  &  57.386  \\ \hline
     3j1l 6  &  64.058  &  70.907  &  36.601  &  45.975  \\ \hline
\end{tabular}

  \caption{Same as Table~\ref{4j0lcuts} but for the 3 jet, 1 lepton, analysis channel.}
  \label{3j1lcuts}
}
\TABLE{
  \begin{tabular}{ | c || c | c | c | c | }
  \hline
  Cut & Flat, 1 fb$^{-1}$ & Flat, 10 fb$^{-1}$ & Log, 1 fb$^{-1}$ & Log, 10 fb$^{-1}$ \\ \hline \hline
     2j1l 1  &  92.79  &  95.365  &  80.465  &  84.897  \\ \hline
     2j1l 2  &  92.512  &  95.135  &  79.69  &  84.158  \\ \hline
     2j1l 3  &  88.778  &  91.835  &  75.458  &  80.133  \\ \hline
     2j1l 4  &  81.617  &  85.552  &  59.591  &  66.433  \\ \hline
     2j1l 5  &  72.18  &  76.952  &  45.98  &  52.474  \\ \hline
     2j1l 6  &  62.942  &  68.419  &  33.498  &  40.473  \\ \hline
\end{tabular}

  \caption{Same as Table~\ref{4j0lcuts} but for the 2 jet, 1 lepton, analysis channel.}
  \label{2j1lcuts}
}
\TABLE{
  \begin{tabular}{ | c || c | c | c | c | }
  \hline
  Cut & Flat, 1 fb$^{-1}$ & Flat, 10 fb$^{-1}$ & Log, 1 fb$^{-1}$ & Log, 10 fb$^{-1}$ \\ \hline \hline
     OSDL 1  &  49.557  &  51.343  &  39.492  &  40.916  \\ \hline
     OSDL 2  &  17.558  &  18.641  &  11.354  &  12.223  \\ \hline
     OSDL 3  &  8.1799  &  8.942  &  5.0423  &  5.8346  \\ \hline
     OSDL 4  &  6.0958  &  6.6796  &  3.8434  &  4.2467  \\ \hline
\end{tabular}

  \caption{Same as Table~\ref{4j0lcuts} but for the OSDL analysis channel.}
  \label{OSDLcuts}
}
\TABLE{
  \begin{tabular}{ | c || c | c | c | c | }
  \hline
  Cut & Flat, 1 fb$^{-1}$ & Flat, 10 fb$^{-1}$ & Log, 1 fb$^{-1}$ & Log, 10 fb$^{-1}$ \\ \hline \hline
     SSDL 1  &  40.204  &  50.761  &  26.869  &  38.368  \\ \hline
     SSDL 2  &  21.331  &  31.97  &  13.681  &  21.233  \\ \hline
     SSDL 3  &  14.774  &  25.518  &  8.8505  &  15.879  \\ \hline
\end{tabular}

  \caption{Same as Table~\ref{4j0lcuts} but for the SSDL analysis channel.}
  \label{SSDLcuts}
}
\TABLE{
  \begin{tabular}{ | c || c | c | c | c | }
  \hline
  Cut & Flat, 1 fb$^{-1}$ & Flat, 10 fb$^{-1}$ & Log, 1 fb$^{-1}$ & Log, 10 fb$^{-1}$ \\ \hline \hline
     3lj 1  &  24.228  &  29.156  &  19.394  &  23.708  \\ \hline
     3lj 2  &  13.549  &  17.361  &  8.6389  &  11.078  \\ \hline
\end{tabular}

  \caption{Same as Table~\ref{4j0lcuts} but for the trilepton + jet analysis channel.}
  \label{3ljcuts}
}
\TABLE{
  \begin{tabular}{ | c || c | c | c | c | }
  \hline
  Cut & Flat, 1 fb$^{-1}$ & Flat, 10 fb$^{-1}$ & Log, 1 fb$^{-1}$ & Log, 10 fb$^{-1}$ \\ \hline \hline
     3lm 1  &  7.5996  &  8.1386  &  6.488  &  7.127  \\ \hline
     3lm 2  &  6.6299  &  7.0141  &  5.5712  &  6.2777  \\ \hline
     3lm 3  &  6.6299  &  7.0141  &  5.5712  &  6.2777  \\ \hline
     3lm 4  &  6.4106  &  6.8424  &  5.2186  &  5.6499  \\ \hline
     3lm 5  &  2.7406  &  2.9135  &  2.8561  &  2.9542  \\ \hline
\end{tabular}

  \caption{Same as Table~\ref{4j0lcuts} but for the trilepton + missing energy analysis channel.}
  \label{3lmcuts}
}
\TABLE{
  \begin{tabular}{ | c || c | c | c | c | }
  \hline
  Cut & Flat, 1 fb$^{-1}$ & Flat, 10 fb$^{-1}$ & Log, 1 fb$^{-1}$ & Log, 10 fb$^{-1}$ \\ \hline \hline
     tau 1  &  99.976  &  99.985  &  81.382  &  82.386  \\ \hline
     tau 2  &  99.969  &  99.981  &  79.795  &  80.945  \\ \hline
     tau 3  &  99.949  &  99.97  &  78.597  &  79.394  \\ \hline
     tau 4  &  99.739  &  99.827  &  75.141  &  76.551  \\ \hline
     tau 5  &  95.99  &  96.913  &  59.379  &  61.263  \\ \hline
     tau 6  &  90.925  &  92.913  &  51.516  &  53.102  \\ \hline
     tau 7  &  83.51  &  86.505  &  44.006  &  45.606  \\ \hline
\end{tabular}

  \caption{Same as Table~\ref{4j0lcuts} but for the $\tau$ analysis channel.}
  \label{taucuts}
}
\TABLE{
  \begin{tabular}{ | c || c | c | c | c | }
  \hline
  Cut & Flat, 1 fb$^{-1}$ & Flat, 10 fb$^{-1}$ & Log, 1 fb$^{-1}$ & Log, 10 fb$^{-1}$ \\ \hline \hline
     b 1  &  100  &  100  &  94.958  &  95.458  \\ \hline
     b 2  &  100  &  100  &  94.958  &  95.458  \\ \hline
     b 3  &  100  &  100  &  94.781  &  95.052  \\ \hline
     b 4  &  100  &  100  &  92.63  &  92.947  \\ \hline
     b 5  &  100  &  100  &  89.598  &  90.251  \\ \hline
     b 6  &  73.983  &  76.939  &  42.948  &  44.572  \\ \hline
\end{tabular}

  \caption{Same as Table~\ref{4j0lcuts} but for the $b$ jet analysis channel.}
  \label{bcuts}
}

\subsection{Discussion of `Difficult' Models}

It is interesting to understand why some specific models are unobservable in {\it all} of the ATLAS MET search channels. A good sample of 
such cases to study is provided 
by the set of 11 models from the flat prior scan that are missed by all of the analysis channels, assuming an integrated 
luminosity of 1 $\infb$ with a $20\%$ SM 
background systematic error.{\footnote  {Note that only 4 of these specific models remain undiscovered when the integrated luminosity is increased by a factor 
of 10.}} To this end, we display and discuss some of the details of the mass spectra for these
11 specific models. Four of these 
models (labeled as model number 14602, 43704, 62912, and 63694) are undetected due to the presence of long-lived
charginos, resulting in a correspondingly small \MET signature (the spectra for the
latter three models are shown in Figure~\ref{bubbles1}).  Three  
more models (7888, 17158, and 47787) are unobservable due to their compressed sparticle
spectra (the spectrum of one of these is also shown in Figure~\ref{bubbles1}). The remaining four models (5700, 7105, 25692, and 35678) are missed for 
more subtle reasons described below.

\FIGURE{
  \includegraphics[width=0.45\columnwidth]{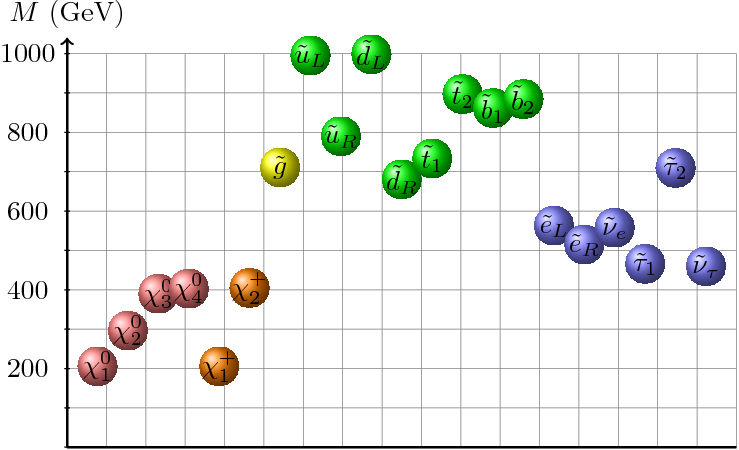}
  \includegraphics[width=0.45\columnwidth]{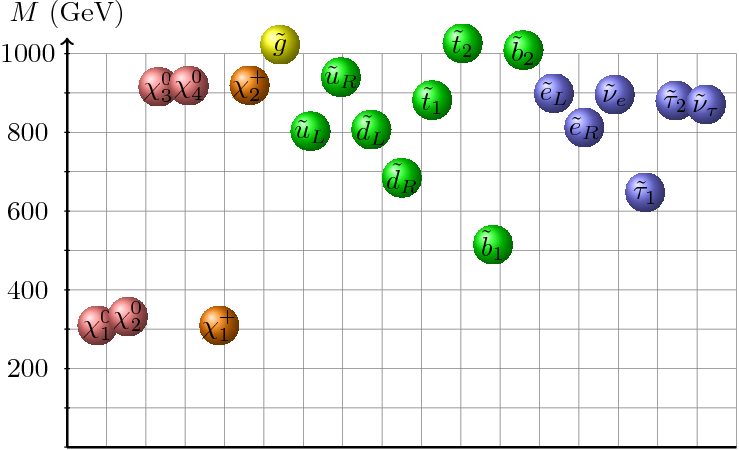}
  \includegraphics[width=0.45\columnwidth]{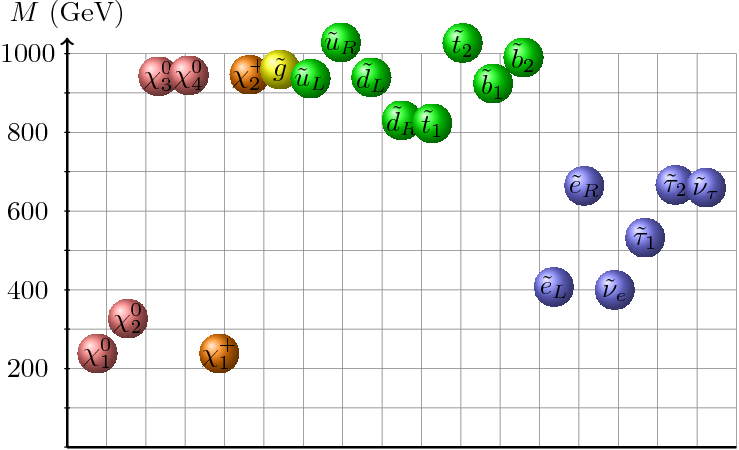}
 \includegraphics[width=0.45\columnwidth]{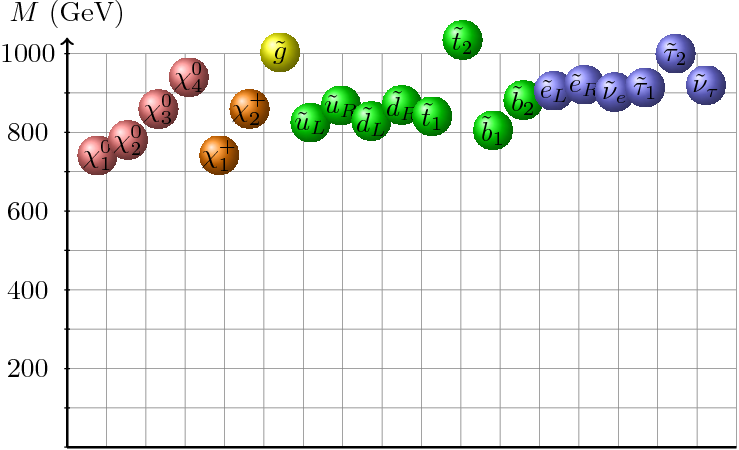}
  \caption{The spectra for four of the eleven models that are unobservable in all analysis channels.
    The first three (from left to right, top to bottom: 43704, 62912, and
    63694) are missed due to the presence of long-lived charginos, while the
    last (bottom right: 17158) is missed due to a compressed spectrum.  The colored balls represent masses for (left to right)
$\tilde\chi_1^0\,, \tilde\chi_2^0\,, \tilde\chi_3^0\,, \tilde\chi_4^0\,, \tilde\chi_1^+\,, \tilde\chi_2^+\,, \tilde g\,,
\tilde u_L\,, \tilde u_R,, \tilde d_L\,, \tilde d_R\,, \tilde t_1\,, \tilde t_2\,, \tilde b_1\,, \tilde b_2\,, \tilde e_L\,,
\tilde e_R\,, \tilde \nu_e\,, \tilde\tau_1\,, \tilde\tau_2\,, \tilde\nu_\tau$.
} 
   \label{bubbles1}
}

One way to better understand why a specific model is unobservable is to try to find a `sister' model (or models) within our 
pMSSM set that has as similar spectra as possible 
to the missed model and yet is observable in at least one of the ATLAS MET analyses. Comparisons between the failed and 
passed models may then reveal the underlying 
cause that renders the model to be undiscoverable. Model 14602 provides a good example of this
approach and Figure \ref{comp1} compares the spectrum of this model and its sister, 43001. 
Both of these models have qualitatively similar cross sections for the production of squarks and gluinos which initiate the long decay cascades. 
However, a side-by-side comparison of these two models shows that 14602 has consistently lower values of $S$ for each of the analyses and yet both models  
have similar preselection jet and lepton spectra as well as having long-lived charginos (which are Higgsino-like and Wino-like, respectively).

\FIGURE{
  \includegraphics[width=0.45\columnwidth]{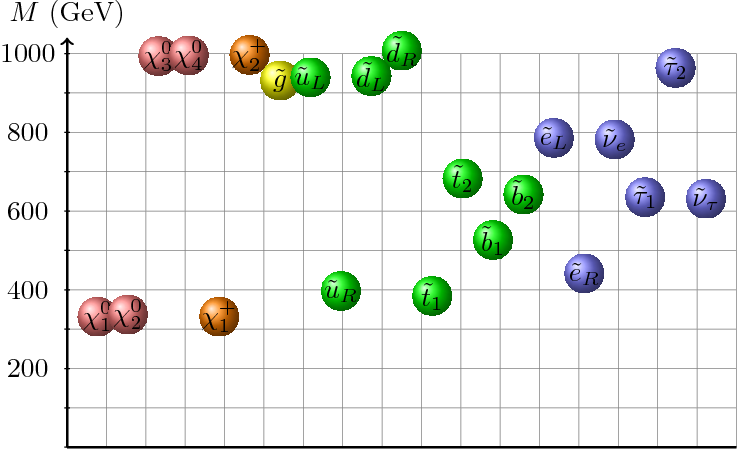}
  \includegraphics[width=0.45\columnwidth]{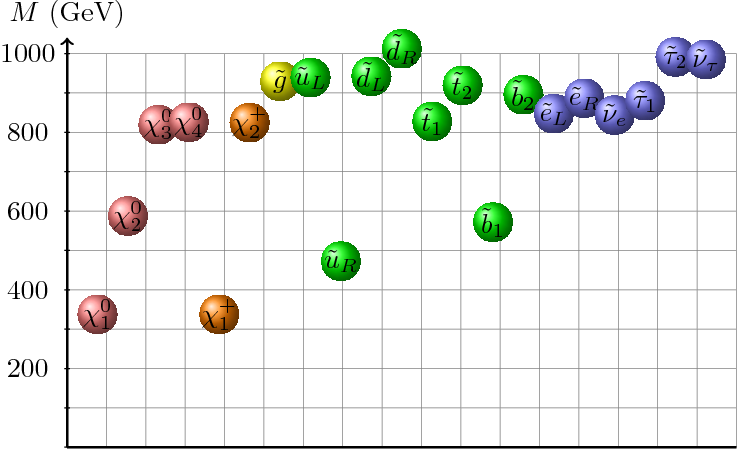}
  \caption{A comparison of the spectra of sister models 14602 (left) and 43001 (right).}
   \label{comp1}
}

The only significant difference between the two models is in their preselection \MET\ distributions as can be seen in Figure \ref{met14062}. Here we see that this 
distribution peaks at much lower values for model 14602 and has a correspondingly diminished high energy tail in comparison to model 43001. Due 
to the presence of large branching fractions in the gluino cascades that lead to a stable chargino in model 14602, there is insufficient \MET\ to pass 
the ATLAS analysis cuts. This is related to the suppressed couplings of the first and second generation squarks to light Higgsinos. However, 
model 14602, with 20\% background systematics, has reasonable values of $S$ in some of the search analyses and 
a factor of 10 or so increase in the integrated luminosity allows this model to 
be discovered in the 3j1l, $\tau$ and $b$ channels. The corresponding examination of the other 3 models with long-lived charginos yields somewhat 
similar results. 

\FIGURE{
  \includegraphics[width=0.7\columnwidth]{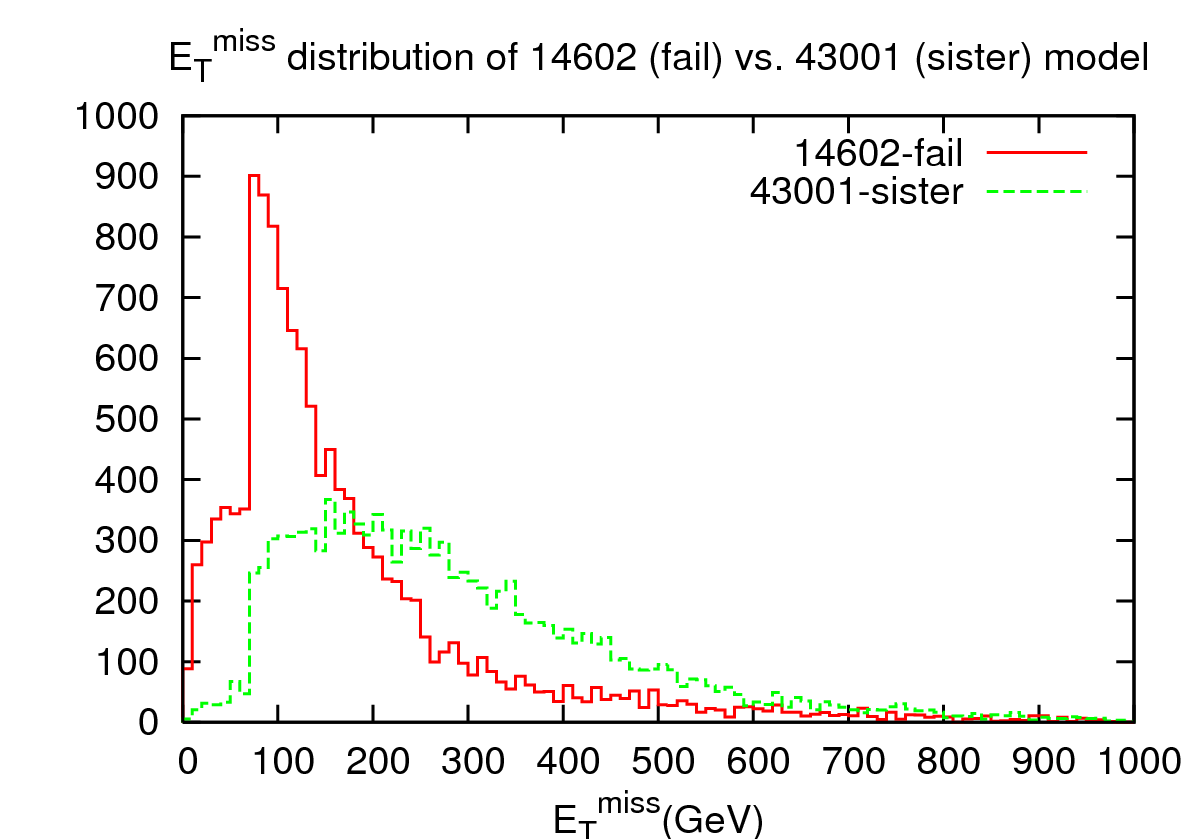}
  \caption{A comparison of the $\MET$ distributions of model 14062 and its sister 43001.}
  \label{met14062}
}

Undiscovered models 7888 and 47787, as well as their discovered co-sister 42790 (Figure \ref{comp2}), all show a relatively heavy and compressed sparticle spectrum 
except that the gluino is slightly more massive and well-separated from the squarks in model 42790 and the squarks are more degenerate with the LSP for 
the two undetected models. Interestingly, the missed models both have larger cross sections for squark and gluino production (due to their lighter gluinos) 
than does their sister model by over a factor of 2. Thus the initial, pre-cut event rates for the missed model 
are not an issue here. However, the larger gluino-squark mass splitting for 
model 42790 allows for a higher $p_T$ jet from the decay $\tilde g \to \tilde q+j$ than do the two missed models and so it is found by the 2j0l analysis. 
The degeneracy of the squarks with the LSP makes it difficult for any of these models to generate additional high $p_T$ jets. Nonetheless, model 7888 would pass 
the 2j0l analysis at 10 times higher integrated luminosity. In addition, model 17158 is seen to have a very massive and highly compressed spectrum and fails the MET 
searches for qualitatively similar reasons.  

\FIGURE{
  \includegraphics[width=0.45\columnwidth]{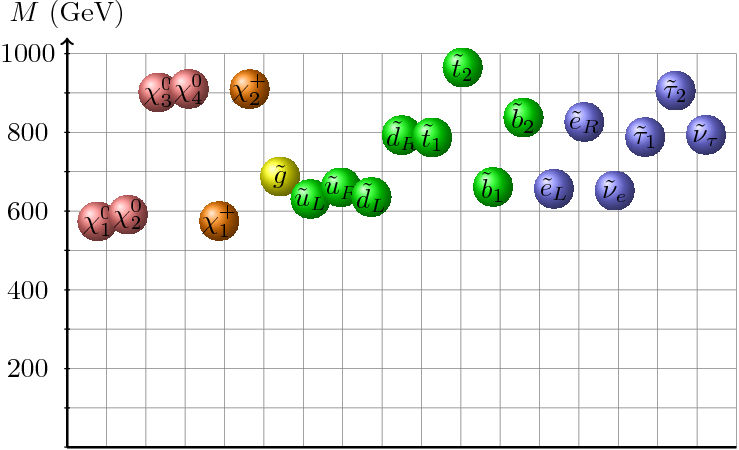}
  \includegraphics[width=0.45\columnwidth]{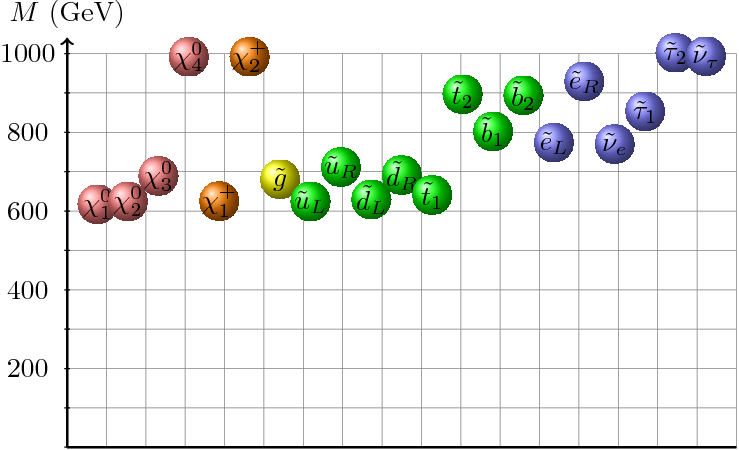}
  \includegraphics[width=0.45\columnwidth]{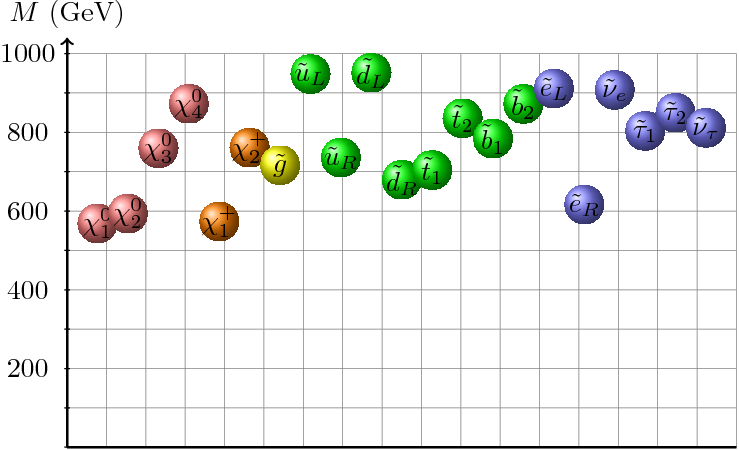}
  \caption{A comparison of the spectra of models 7888 (top left) and 47787 (top right) with their sister 42790 (bottom).}
   \label{comp2}
}

The remaining undetected models are somewhat more difficult to analyze. Model 5700 (with its sister model 28575 shown in Figure \ref{comp3}) is the most 
straightforward case to study and the 
gluino is sandwiched between the squarks in the mass spectrum. The resulting mass 
splitting between the heavier $\tilde u_{L(R)}$ and the gluino is only about half of that of 
the sister model. The essential difference between these two models is the placement of the lightest squark in the spectrum and the relative splittings between 
this squark, the gluino, and the LSP. The model cannot produce 3rd or 4th jets with sufficient $E_T$ to pass the 4j0l selection. 
Note that the splittings are somewhat larger for the sister model. 
In addition, the sister sparticle spectrum makes the decay products arising from stop and sbottom production easier to observe. 
This is another case where a luminosity increase to 10 $\infb$ leads to a discovery for a missed model.  

\FIGURE{
  \includegraphics[width=0.45\columnwidth]{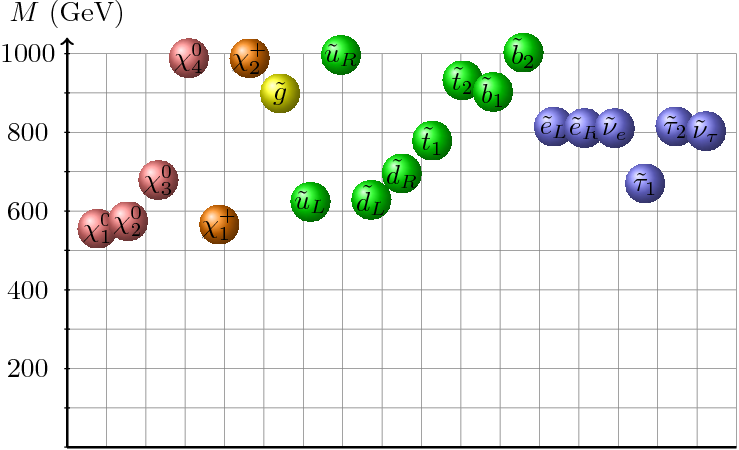}
  \includegraphics[width=0.45\columnwidth]{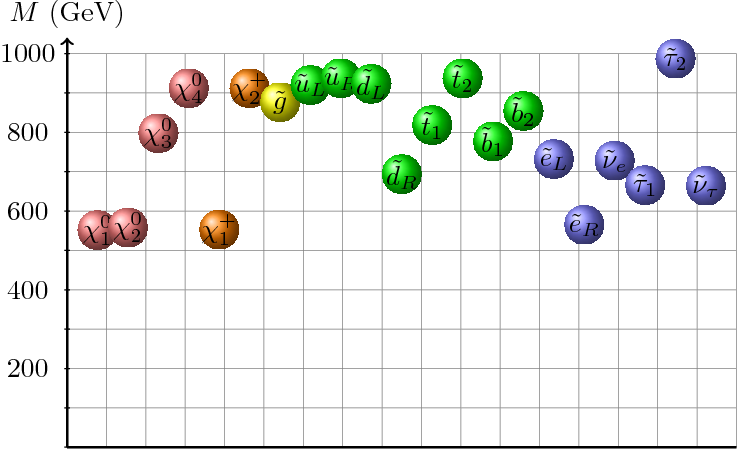}
  \caption{A comparison of the spectra of sister models 5700 (left) and 28575 (right).}
   \label{comp3}
}

Comparing the undetected model 25692 with its sister 1446 (see Figure \ref{comp4}), 
we see that the squarks are somewhat lighter in the sister case allowing for
both larger production cross sections as well as more gluino decay 
modes with larger branching fractions into final states that can populate the 2j0l
channel. We find that increased luminosity would be useful in this case as
well. 

\FIGURE{
  \includegraphics[width=0.45\columnwidth]{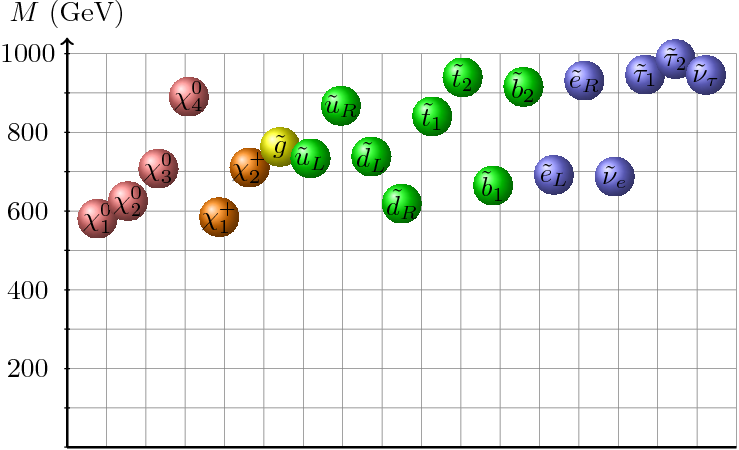}
  \includegraphics[width=0.45\columnwidth]{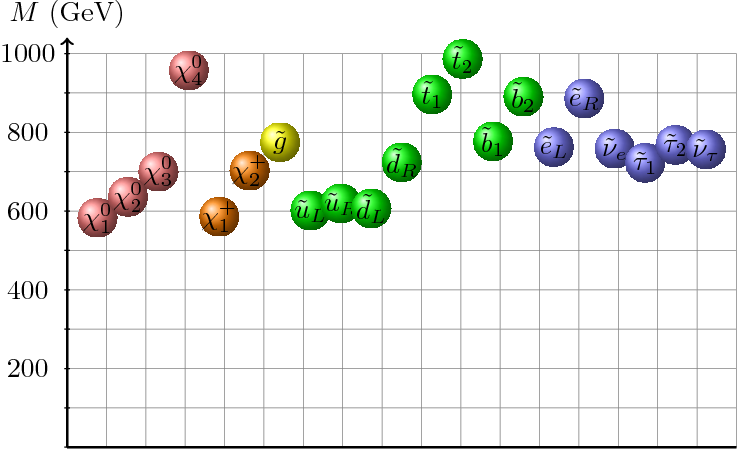}
  \caption{A comparison of the spectra of sister models 25962 (left) and 1446 (right).}
   \label{comp4}
}

For model 35678 and its sister model 9396, shown in Figure \ref{comp5}, the electroweak gaugino sectors are almost identical. 
However, the gluino is heavier than all the first- and second-generation 
squarks in the sister case while the ($\tilde u_L,\tilde d_L$) are heavier than the gluino for model 35678. The lighter slepton spectrum in the sister model 
allows for an enhancement in the number of high $p_T$ leptons produced so that this model can be found in the lepton plus jets channels (but does not do as well in 
the 4(2)j0l analyses as does 35678). Both the 4(2)j0l analyses would allow
model 35678 to be discovered with an integrated luminosity of 10 $\infb$. 

\FIGURE{
  \includegraphics[width=0.45\columnwidth]{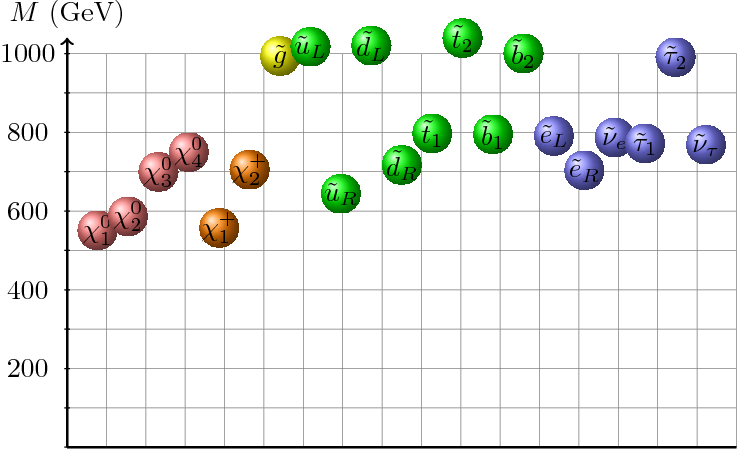}
  \includegraphics[width=0.45\columnwidth]{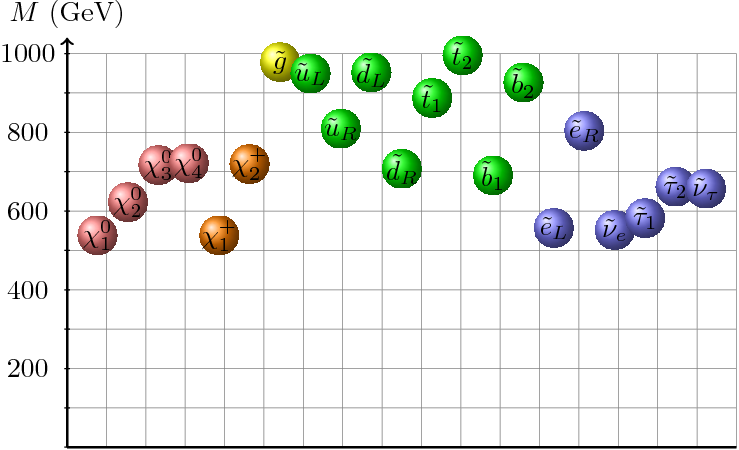}
  \caption{A comparison of the spectra of sister models 35678 (left) and 9396 (right).}
   \label{comp5}
}

Model 7105 has a sister model 53923 whose spectrum is shown in Figure \ref{comp6}; the sister has somewhat larger gluino and squark production rates. This 
sister model also has lighter sleptons which produce a larger fraction of final states with higher $E_T$ leptons. Both models are found to fail the 4j0l 
analysis yet the sister model passes the 2j0l channel. 
It has a higher amount of \MET~since all the squarks are lighter than the gluino and have substantial branching fractions into the LSP. Increased luminosity 
would be useful in this case as well. 

\FIGURE{
  \includegraphics[width=0.45\columnwidth]{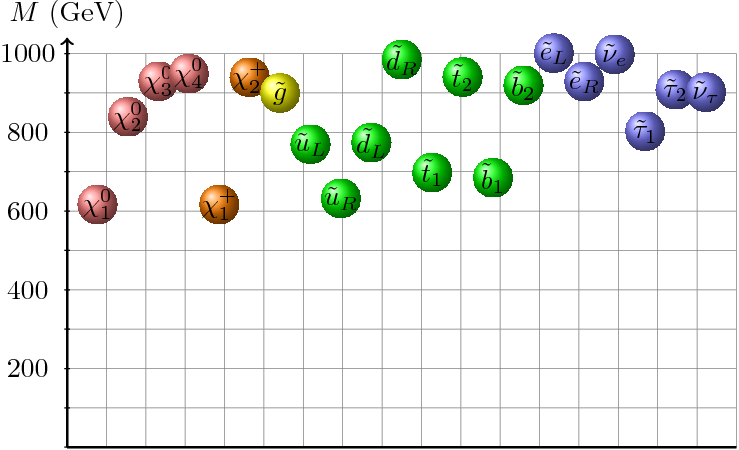}
  \includegraphics[width=0.45\columnwidth]{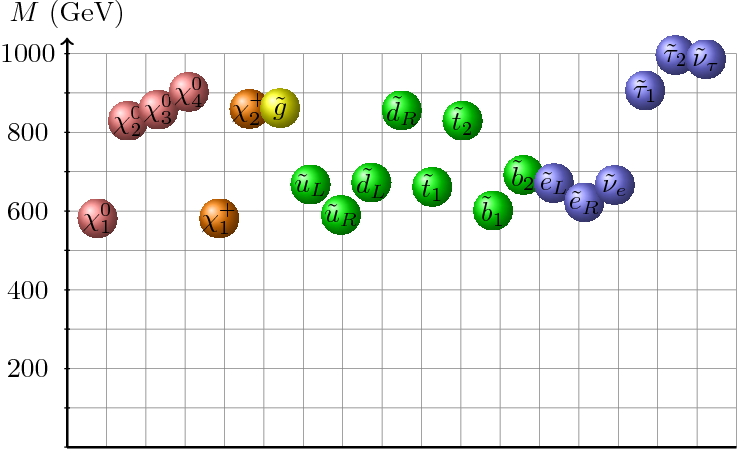}
  \caption{A comparison of the spectra of sister models 7105 (left) and 53923 (right).}
   \label{comp6}
}

\subsection{Classic Decay Modes and the SUSY Mass Scale}

There are a number of other interesting questions that we can address with this large data set. For example, 
a final state that has received much attention for its usefulness in determining sparticle masses \cite{Barr:2010zj} is $jl^+l^-+$MET which originates 
from the cascade decay of an initial colored sparticle, usually with the second neutralino and a slepton
appearing in the chain. One can ask how often 
this specific final state occurs in the decays of the various initial squarks and gluinos in our pMSSM 
model sets; the results are shown in Figure~\ref{chi2} for both the flat and log prior model sets combined.  This figure shows the fraction of the
model sample that leads to this particular final state as a function of the minimum value for the branching fraction for this decay.  For example, the
fraction of the models in our set that have a 
$\tilde u_R$ initiated decay to an $Xjl^+l^-+$MET final state  
with a branching fraction of at least $20(5)\%$ is only $\sim 5(9)\%$! For $\tilde u_L$ cascades, which are commonly studied in this regard, we see that 
the branching fraction for this final state is significantly smaller, only $\sim 1.5(5)\%$.  
Clearly, unlike the case of mSUGRA, this final state does not appear to occur very frequently with a large branching fraction 
in the decays of squarks or gluinos in our 
pMSSM model sample. From this we can conclude that other final states would need to be employed in most cases for measuring sparticle masses.

\FIGURE{
  \includegraphics[width=0.7\columnwidth]{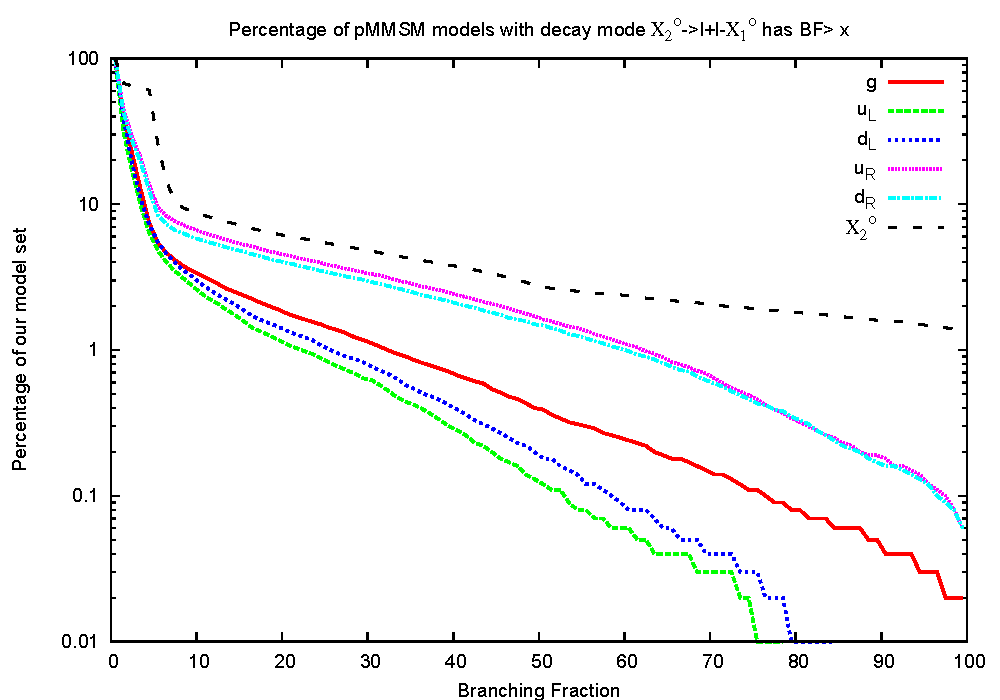}
  \caption{The fraction of pMSSM models that lead to the $X+(j)l^+l^-+$MET signature, passing through the second neutralino, as a function of 
   the minimum branching fraction for this final state.  The various sparticle initial states are color coded as indicated. 
  }
  \label{chi2}
}

%

Another question we address is what are the number of steps in the decay topology necessary to reach a specific final state such 
as, \eg, $Xl^+l^-+$MET, from a given initial colored sparticle at the top of the decay chain. For the case of the gluino, this 
result is shown in Figure~\ref{gluino}. In this figure, branching fraction of the gluino into this inclusive final state
is shown as a function of 
the number of decay chain steps (weighted by the branching fraction so not necessarily an integer) necessary 
to reach this specific final state. The colors reflect model points which do(green) or do not(red) pass the OSDL 
analysis requirements; note that most of the models which pass the OSDL analysis have large branching fractions. While this final state 
may be reached in as few as 2 steps (via gluino loop decay to $g\tilde \chi_2^0$ followed by the 3-body decay $\tilde \chi_2^0 
\to l^+l^-\tilde \chi_1^0$), it is interesting to see that there are some model points where 6 or 7 steps are required. This demonstrates 
that the decay topologies in the pMSSM framework can be much more complex than
those found in mSUGRA, with implications for SUSY searches and mass
measurements at the LHC.

\FIGURE{
  \includegraphics[width=0.7\columnwidth]{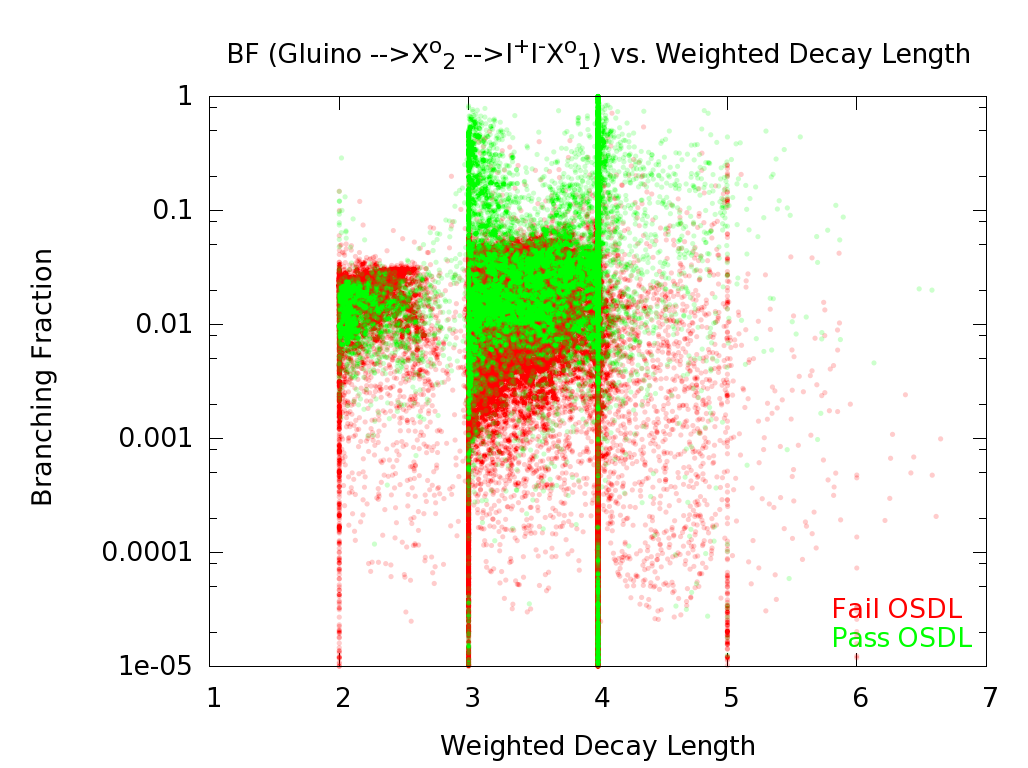}
  \caption{The average branching fraction as a function of the weighted decay length
    to reach the $l^+l^-\tilde\chi_1^0$ final state via the
    $\tilde\chi_2^0$ in decays of the gluino.  The models that pass the OSDL analysis cuts are shown
    in green, while those that fail are in red.
  }
  \label{gluino}
}

%
%

Finally, we investigate whether certain global observables can be used to determine the effective SUSY mass scale. Long ago{\cite {Hinchliffe:1996iu,trobens}}, 
it was observed within mSUGRA that the $\MEFF$ variable can be used to determine the overall scale of the colored sparticles in the SUSY 
mass spectrum. In particular, it was observed that within these models $\MEFF\simeq 1.5 \MMIN \pm (10-15)\%$ where $\MMIN$ is the mass of 
the lightest sparticle in the set $\tilde g, \tilde u_{L,R}, \tilde d_{L,R}$, which are the dominant sources of jets and MET. We can now see whether 
this sort of relationship holds in the much more general context of the pMSSM. The answer to this question is shown in Fig.~\ref{meffpic} which displays  
$\MEFF$ as a function of both $\MMIN$ and $M_{\tilde g}$. The data generated when performing the 4j0l analysis (before imposing the $\MEFF$ cut itself) 
was used to obtain the results shown in these figures for the $\sim68$k pMSSM flat prior model set. The points are also color-coded to show whether they passed (green) 
or failed (red) the 4j0l analysis for an integrated luminosity of 1 fb$^{-1}$ and an assumed $50\%$ SM background uncertainty.  Indeed, we see that 
there is a reasonably strong correlation between $\MEFF$ and $\MMIN$ though somewhat less so in the case of $\MEFF$ and $M_{\tilde g}$. There are, 
however, several differences with the mSUGRA results: ($i$) our range of sparticle masses extends to significantly lower values than one finds 
in mSUGRA due to the strong Tevatron constraints on $m_{\tilde g,\tilde q}$ in the mSUGRA framework. 
For small values of $\MMIN$ we see that $\MEFF/\MMIN\simeq 3$ which is quite far from the expected value of $\simeq 1.5$. 
However, for significantly larger values of $\MMIN\gtrsim 600$ GeV,  we do find that the relation $\MEFF/\MMIN\simeq 1.5$ holds. $(ii)$ The 
relationship between $\MEFF$ and $\MMIN$ is thus not quite linear over the entire mass range of our model set. However, since $\MEFF\geq 350$ 
GeV is {\it required} to pass the 4j0l selection criteria before the $\MEFF$ cut is actually applied (and the points at low values of $\MMIN$ are seen 
to mostly pass this analysis) we can obtain the approximate linear relationship $\MEFF \simeq 1.2\MMIN+350$ GeV.
($iii$) The spread in values of $\MEFF$ at any given value for $\MMIN$ is significantly wider than would be expected in mSUGRA with many pMSSM models 
falling quite far from the middle of the range. Note also that the unobservable models tend to have $\MEFF$ values somewhat further away from the mid-range. 
($iv$) At small values of $\MMIN$ we see that there is a sort of a gap or bifurcation in the distribution. 
This is connected to the identity of the lightest colored sparticle with the lower(upper) lobe corresponding to light gluinos(squarks). A study of 
other kinematic variables{\cite {trobens}} used to determine the SUSY mass scale with this pMSSM model set could prove interesting.

\FIGURE{
  \includegraphics[width=0.45\columnwidth]{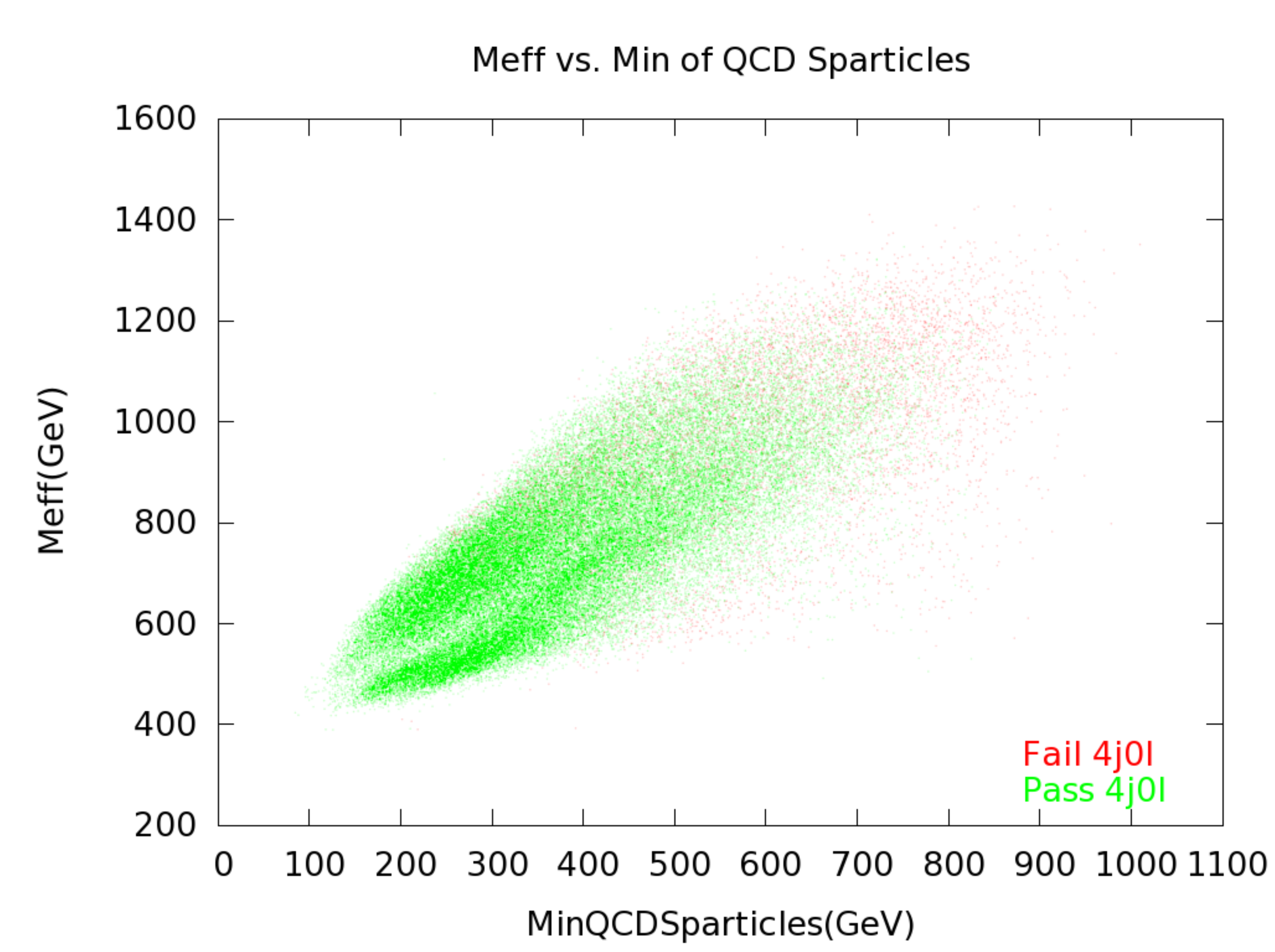}
  \includegraphics[width=0.45\columnwidth]{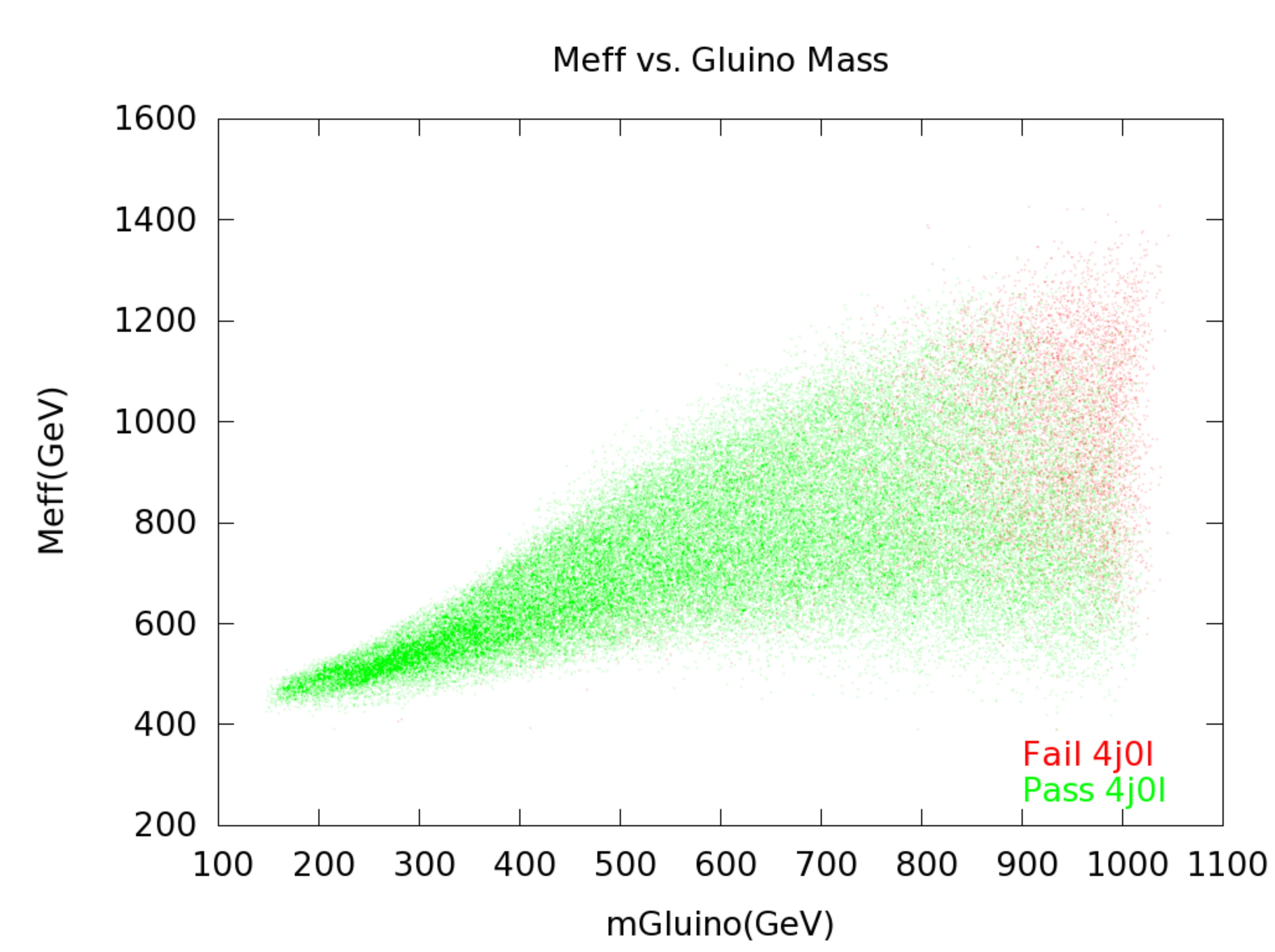}
  \caption{Values of $\MEFF$ as a function of the lightest colored sparticle mass (left) and the mass of the gluino (right) as described in the text. 
  }
  \label{meffpic}
}


\section{Detector-Stable Sparticles}
\label{stabsec}

``Long-lived'', ``metastable'', or ``detector-stable'' particles, {\it i.e.}, those particles which generally decay
outside the detector when produced at colliders, can provide a striking signal of new physics (see~\cite{Fairbairn:2006gg} and references therein).

These detector-stable particles are found to be quite prevalent in both of our pMSSM model sets (to an extent that will be quantified below), 
but are not considered in the inclusive search analyses 
above. Given the nature of the analysis below, we can conveniently combine both the flat and log prior samples together to make a single common study.

Therefore we now discuss the phenomenological consequences of these detector-stable sparticles; the subsequent discussion, however,
will be quite heuristic in comparison with the investigation of the
inclusive SUSY search analyses discussed above.  We will first explain our
criteria for ``detector-stability''.  We will then discuss the
various species of sparticles, when, if ever, such sparticles can be
detector-stable, and the prospects for discovering these detector-stable sparticles at the LHC. We will not discuss specific analyses, \eg, 
searches for R-hadrons in the analysis presented below. Our main point here is that such long-lived states are relatively common in our model 
sets and that suches for long-lived states are an important supplement to the \MET~searches discussed above.

\subsection{Criteria for Stability}

A necessary first step is to specify what precisely qualifies as a 
``long-lived'' or ``detector-stable'' particle.  We note that for a particle at rest, its
lifetime is given by $\frac{\hbar}{\Gamma}$ where $\Gamma$ is the
total width of the particle.  This translates to the particle traveling a
distance of $c\tau \sim \frac{c \beta \gamma \hbar}{\Gamma}$ in the detector.

There are several issues in determining a value of
$\Gamma$, below which a particle will be considered (in this
discussion) to be stable.
Perhaps the most obvious is that the energy of the particle, 
and hence $\gamma$, will vary from event to event. 
Figure~\ref{mp-beta-gamma} shows the distribution of $\beta \gamma$ for
detector-stable charginos in our pMSSM model set; this distribution is, of
course, also sensitive to the mass distribution of the detector-stable
charginos.
\FIGURE{
  \label{mp-beta-gamma}
  \includegraphics[width=0.8\columnwidth]{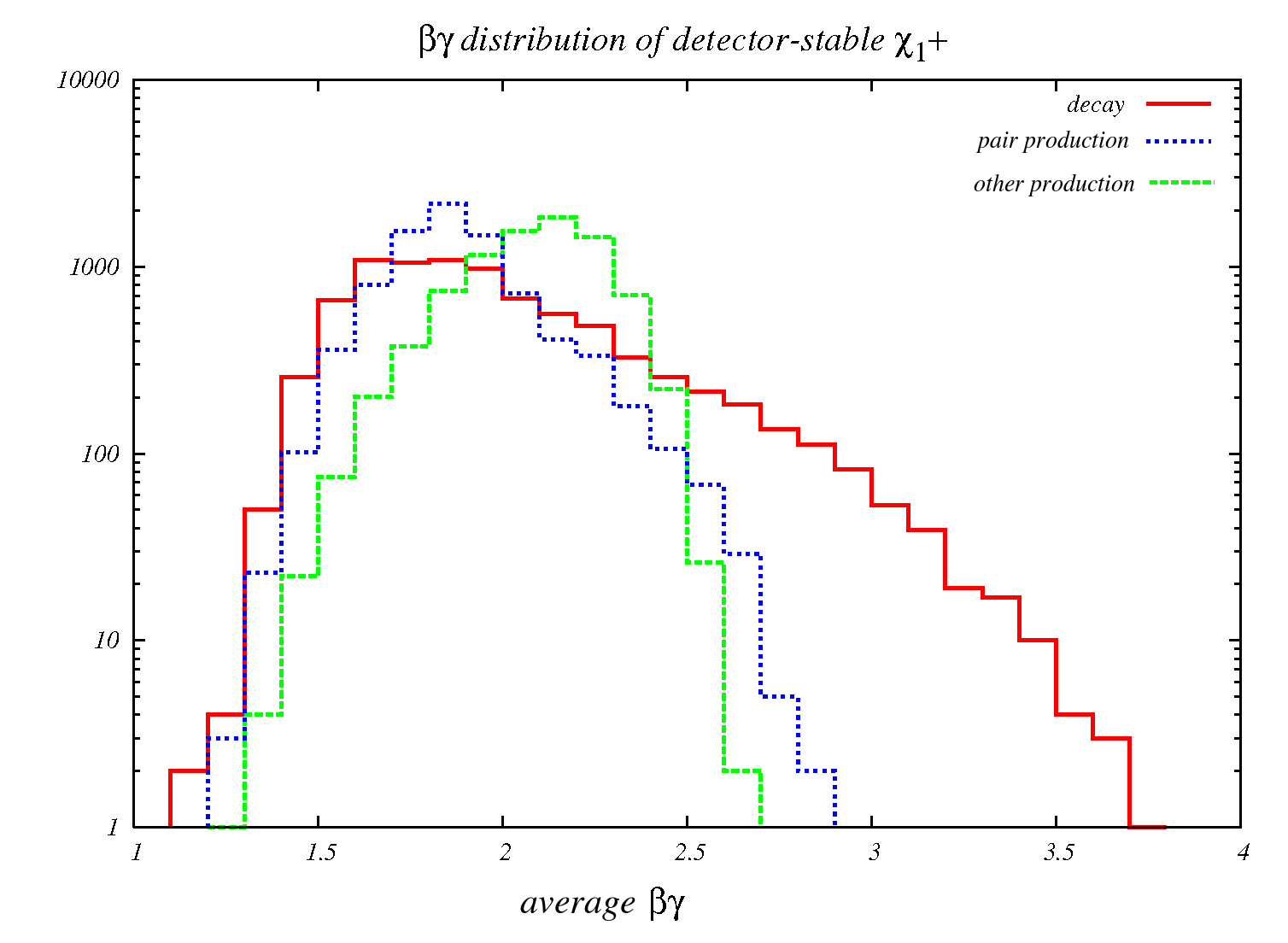}
  \caption{The distribution of $\beta \gamma$
    for detector-stable charginos in our model set.  The curve
    labeled ``decay'' refers to charginos produced in the cascade decay chains of
    other sparticles.  ``Pair production'' refers to charginos directly produced
    in the $pp \to \tilde\chi_1^+ \tilde\chi_1^-$ hard process; ``other production''
    refers to charginos produced in all other hard processes.}
}
We see that $\beta\gamma$ is $\lesssim 4$ for charginos in this model
set; we expect this condition to hold at least roughly for the
other species of stable particles as well.  
Therefore, we define a particle with width $\Gamma$ to be
``long-lived'', or ``detector-stable'' if 
\begin{equation}\Gamma < \Gamma_{\text{stable}},
  \text{~with~} \Gamma_{\text{stable}}=10^{-17}\,\text{GeV}.
\end{equation}
For this value of $\Gamma_{\text{stable}}$, $\frac{c \hbar}{\Gamma}
\sim 20$ m, and $\frac{c \beta \gamma \hbar}{\Gamma_{\text{stable}}}\sim
  20-60$ m.  The effect on our quantitative results from adjusting the
  definition of $\Gamma_{\text{stable}}$ will be discussed below.

We also note, from Figure~\ref{mp-beta-gamma}, that the distribution 
of $\beta \gamma$ is quite different for charginos produced in cascade decay chains, 
compared with charginos produced directly in the hard process.  In particular,
these charginos may be more highly boosted, and, of course, be produced 
in events without an accompanying stable chargino.  Such considerations may be 
very important for stable particle searches at the LHC.

Additional complications in assigning a threshold for stability
arise from the probabilistic nature of
decays; a full analysis taking such effects into account is beyond the
scope of this work.  Rather we discuss the prevalence of various
detector-stable particles in our pMSSM model set in the next section, as
well as the physics responsible for these sparticles' long lifetimes.
Next, we will quantify the prospects for
discovering or ruling out the detector-stable sparticles in our model
set at the LHC.

It may be worthwhile to note that while we have considered models with
absolutely stable charged particles ({\it i.e.}, charginos when the mass
splitting with the LSP is less than the electron mass) to be 
excluded, we did not implement any constraints based on the effect a
long-lived sparticle could have on BBN (see, for
instance~\cite{Fairbairn:2006gg},~\cite{Raklev:2009mg},
or~\cite{Steffen:2006hw}) when we generated our model sample.

\subsection{Detector-Stable Sparticles and R-Hadrons}

\TABLE{
\begin{tabular}{|l|c|c|c|c|c|} \hline\hline
Sparticle & $10^{-15}$ GeV &  $10^{-16}$ GeV &
 $10^{-17}$ GeV &  $10^{-18}$ GeV & $10^{-19}$ GeV \\ \hline
 $\tilde\chi^\pm_1$ & $ 9853$ & $ 9728$ & $ 8642$ & $ 7683$ & $ 6658$ \\
 $\tilde{\tau}_1$    & $ 179$ & $ 179$ & $ 179$ & $ 179$ & $ 179$ \\
 $\tilde{t}_1$    & $ 67$ & $ 66$ & $ 66$ & $ 65$ & $ 65$ \\
 $\tilde{c}_R$ & $ 49$ & $ 49$ & $ 49$ & $ 49$ & $ 49$ \\
 $\tilde\chi^0_2 $& $ 78$ & $ 40$ & $ 19$ & $ 11$ & $ 4$ \\
 $\tilde{\mu}_R$  & $ 17$ & $ 17$ & $ 17$ & $ 17$ & $ 17$ \\
 $\tilde{b}_1$    & $ 12$ & $ 12$ & $ 11$ & $ 9$ & $ 9$ \\
 $\tilde{c}_L$  & $ 8$ & $ 8$ & $ 8$ & $ 8$ & $ 8$ \\
 $\tilde{s}_R$ & $ 8$ & $ 8$ & $ 8$ & $ 8$ & $ 8$ \\
 $\tilde{g}$         & $ 17$ & $ 10$ & $ 5$ & $ 2$ & $ 0$ \\
\hline\hline
\end{tabular}
\caption{The number of models in our pMSSM model set in which 
  the specified sparticle has a
  width less than the value given at the head of each column.  This
  gives some idea of the effect of the specific choice of
  $\Gamma_{\text{stable}}=10^{-17}$ GeV.}
\label{summa}
}

Table~\ref{summa} shows the number of detector-stable sparticles of
each type for different choices of $\Gamma_{\text{stable}}$; elsewhere
we will always take $\Gamma_{\text{stable}} = 10^{-17}$ GeV as noted
above.  In what follows we will discuss the physics that can lead to
detector-stable sparticles or R-hadrons, discussing gauginos first, and
then sfermions.

If colored sparticles are long-lived, they can hadronize to form
R-hadrons\cite{Fairbairn:2006gg, Raklev:2009mg, Farrar:1978xj, 
Chanowitz:1983ci, Hewett:1996ru,
Farrar:1984gk, Buccella:1985cs, Farrar:1994xm, Baer:1998pg, Berger:2000mp, 
Sjostrand:2002ip, Hewett:2004nw, Kraan:2004tz, Kilian:2004uj, 
Mackeprang:2006gx, Mackeprang:2009zs, Buckley:2010fj}, a color singlet state carrying one unit
of R-parity.  
We expect R-hadrons to form when the width of a
colored particle is roughly $\Gamma \lesssim \Lambda_{\text{QCD}}$.  In what follows,
we will give the number of models in which various colored sparticles 
have total widths less than $100$ MeV, taking this to be a rough indication of the number of models which would have significant R-hadron
production.
As the colored sparticles in our pMSSM model set have masses $\gg \Lambda_{QCD}$, 
the lifetime of the produced R-hadron should be roughly that
of its constituent long-lived colored sparticle\cite{Fairbairn:2006gg,
  Sjostrand:2002ip,Kilian:2004uj}, so it is reasonable to use the same
criterion for detector stability for colored and uncolored
sparticles.

\subsection{Detector Stability of Gauginos}

\subsubsection*{Charginos}
The most prevalent detector-stable particles in this pMSSM model set are
charginos.  This is due to the large number of models for which the
lightest neutralino (the LSP) is mostly Higgsino or Wino, as is shown
in Table~\ref{LSP Identity}.  As is well known
(see, for example, \cite{Haber:1984rc, Martin:1997ns, Chung:2003fi, 
Pape:2006ar, Dreiner:2008tw, Drees:2004jm, Baer:2006rs}) 
the Wino-like neutralino (with mass $\approx M_2$)
is nearly degenerate with a Wino-like chargino.  Likewise there are
two nearly degenerate Higgsinos (with mass $\approx |\mu|$) which are
in turn nearly degenerate with a Higgsino-like chargino.  There are
no models in our sample where the heavier of the two chargino species 
is stable.

\TABLE{
\begin{tabular}{|l|c|r|} \hline\hline 
LSP Type & Definition & Fraction \\ 
& & of Models \\ \hline
Bino & $|Z_{11}|^2 > 0.90$ & 0.156 \\
Wino & $|Z_{12}|^2 > 0.90$ & 0.186 \\
Higgsino & $|Z_{13}|^2+|Z_{14}|^2 > 0.90$ & 0.393 \\
All other models & & 0.265 \\
\hline\hline
\end{tabular}
\caption{The majority of models in our pMSSM sample have LSPs which are
  relatively pure gaugino/Higgsino eigenstates.  The fraction which
  are of each type is given here; with the definition of each type
  given in terms of the modulus squared
  of elements of the neutralino mixing matrix in the SLHA convention.
  See~\cite{Skands:2003cj} for details.}
\label{LSP Identity}
}

As discussed above, we use a more detailed treatment than 
is given in SUSY-HIT to describe sparticle decay.
In particular, for the case of close mass charginos that have small mass
splittings with the LSP ($\Delta m$), we utilize expressions from
~\cite{Chen:1996ap,Chen:1999yf} to compute their decays exactly.
We find that charginos generally fit our definition of
detector-stability when $\Delta m = m_{\tilde\chi^\pm_1} -
m_{\text{LSP}} < 112$ MeV.
The distribution of the $\tilde\chi^\pm_1$ width as a function of
the $\tilde\chi^\pm_1$  LSP mass splitting is shown in Figure~\ref{gauginos} for our
pMSSM model set.

One sees from the figure that there is very little scatter in $\tilde\chi^\pm_1$ 
widths at low values of $\Delta m$ and that the widths lie along a curve in this case.  This is to be 
expected, as both the $\tilde\chi^\pm_1$ and the $\tilde\chi^0_1$ 
are nearly pure Higgsino and Wino eigenstates, and the widths are not 
dependent at this level on the rest of the SUSY spectrum.
One can also see from Figure~\ref{gauginos} where the three body chargino decay 
to $\mu^+ \nu_\mu \tilde\chi^0_1$ turns on, and where the width becomes highly 
suppressed due to electron mass effects.  
The longest-lived chargino in our model set has $\Delta m = 512$ keV and 
a $c \tau$ of $\approx 5\times 10^5$ light years, which
is large even compared with the size of the ATLAS detector.

\subsubsection*{Neutralinos}

There are $19$ models in our pMSSM sample in which the second lightest
neutralino is detector-stable. There are no models for which the
third or fourth neutralino is detector-stable.
Most, though not all, of the detector-stable second neutralinos are
nearly Higgsino eigenstates, with the LSP being essentially the other neutral
Higgsino eigenstate.  In all cases where the second neutralino is
detector-stable, the mass splitting of this neutralino with
the LSP is less than $\sim 650$ MeV as can be seen from Figure~\ref{gauginos}.
From this figure it is clear that for low $\Delta m$, the neutralino
width is basically a function of $\Delta m$.  We also observe that
unlike the case of charginos, the neutralino width as a function of
$\Delta m$ is a power law; {\it i.e.}, there are no obvious effects from the masses
of decay products other than the LSP. 

\subsubsection*{Gluinos}

There are only $5$ models for which the gluino width is less that
$\Gamma_{\text{stable}}$; we take this to be a rough estimate
of the number of cases for which the R-hadrons produced by such
gluinos would be detector-stable.  All of these models have a mass
splitting between the gluino and LSP of $\lesssim 300$ MeV.
There are, however, $12598$ models where the gluino width 
is less than $100$ MeV as shown in Table~\ref{R hadron}; 
these would be expected to form R-hadrons.  The mass splitting between the LSP
and the gluino is not necessarily small for these models as can be
seen in Figure~\ref{gauginos}.  Note also that there
is a large spread in gluino widths, which is probably due to different patterns
of squark masses, for the gluinos at any given $\Delta m$.

 \FIGURE{
    \label{gauginos}
    \includegraphics[width=0.7\columnwidth]{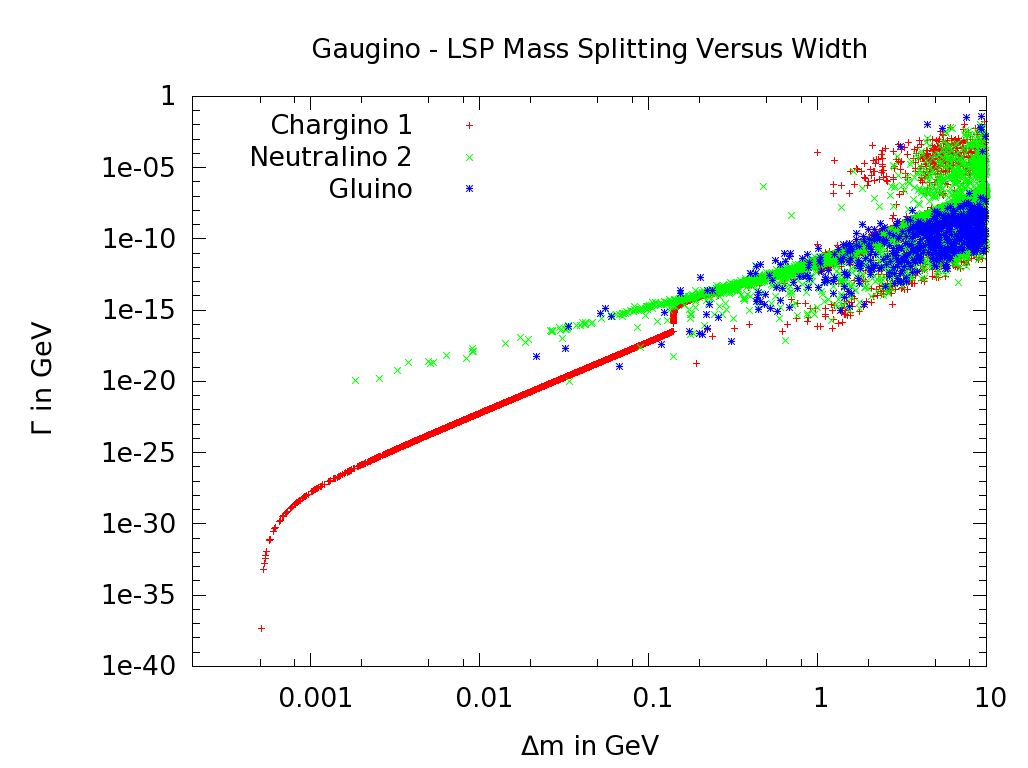}
    \caption{The width as a function of its mass
  splitting with the LSP ($\Delta m$) for the light chargino mass eigenstate (red),
  eigenstate, the  second lightest neutralino mass eigenstate (green), and the
  gluino (blue).  Heavier mass eigenstates for charginos and neutralinos tend
  to have larger mass splittings with the LSP and correspondingly
  large widths.}  
 }   

\subsection{Detector Stability of Sfermions}

The widths of the various species of sfermions as a function of
the sparticle-LSP mass splitting are presented in 
Figures~\ref{selectrons-and-smuons},~\ref{up-squarks},~\ref{down-squarks},~\ref{staus},~\ref{stop-sbottom}, and~\ref{sneutrino}.  
Here, we will discuss salient aspects of the physics that leads to detector-stable
sfermions in our pMSSM model set.

\subsection*{Close Mass Sparticles, Degeneracies, and Important Decay 
Modes}

We remind the reader that in our pMSSM model set, the masses 
of the first and the corresponding second
generation sparticle (such as left selectron and left smuon) are
degenerate.  In addition, these sparticles have the same couplings, 
except for Yukawa couplings that are generally negligible.  
Thus the first and corresponding second generation sfermion also have degenerate widths.
Hence, for example, a point in
Figure~\ref{selectrons-and-smuons} for the left smuon
lies on top of the point for the left selectron for all
models.  

The exception to this picture occurs when, for some sfermion, 
$\Delta m \lesssim m_f$, where $f$ is the corresponding fermion.
This is because for $\Delta m$ slightly greater than $m_f$, the decay
$\tilde{f} \to \tilde\chi^0_1 f$ is kinematically suppressed, and when 
$\Delta m < m_f$, this decay is forbidden.  As this decay is generally
very important for close mass sfermions, this leads to large suppressions 
of the sfermion decay widths.  
We note that the suppression due to small $\Delta m$
may occur for the second generation sfermion but not the first generation
sfermion when $m_{f_1} < \Delta m < m_{f_2}$. We do not find many models
where charged current decays such as 
$\tilde{c} \to \tilde\chi_1^+ s$ have a large effect on the widths of close
mass sfermions.  The exception is in models where the stau has $\Delta m < m_\tau$,
and is not detector-stable due to the decay $\tilde{\tau} \to
\nu_\tau + \tilde\chi^+$.

Generally, in fact, we set the decay width to be zero for first and
second generation sfermions when $\Delta m < m_f$ since we do not 
include four-body decays for these sfermions, 
or CKM suppressed decays in the case of first and second generation squarks.
This has an important consequence for 
Figures~\ref{selectrons-and-smuons},~\ref{up-squarks},~\ref{down-squarks},~\ref{staus},~\ref{stop-sbottom}, and~\ref{sneutrino} 
as a sparticle not appear in the figures if its width is zero.  
A stable second generation sparticle can be identified by noting where one sees a point
for the first generation sparticle with out a nearly degenerate point for the 
second generation sparticle.

For third generation squark decays, some of these additional decay modes
were included in calculating the widths as discussed above. CKM suppressed and four-body stop decays
are included in SUSY-HIT\cite{Djouadi:2006bz}.  
SUSY-HIT does not, however, incorporate these decays for sbottoms and 
we added the CKM suppressed decays 
(though not the four-body decays) for sbottoms. 

We should note that our analysis may somewhat overstate the prevalence of
detector-stable second generation sfermions in the model set. 
If the mixing between
right and left sfermions were not precisely zero, the mass of the lighter 
eigenstate would be raised, possibly so that the
decay $\tilde{f} \to \tilde\chi^0_1 f$ would not be significantly suppressed.
A full understanding of this effect would have required the inclusion
of trilinear coupling terms for the first two generations.

Like gluinos, squarks are colored sparticles and hence can form
R-hadrons as discussed above.  The number of squarks of
each flavor which have widths $< 100$ MeV $\approx \Lambda_{QCD}$
is shown in Table~\ref{R hadron}.

\subsubsection*{Sum Rules and the Effect on which Sparticles are Stable}

The $17$ right-handed smuons
with $\Delta m < m_\mu$ are the only charged sleptons
that we find to be detector stable.
The fact that it is right-handed rather than left-handed smuons which are detector stable
in this model set is in part a consequence of the tree-level slepton mass sum 
rule (see for example~\cite{Haber:1984rc, Martin:1997ns, Drees:2004jm, Baer:2006rs})
\begin{equation}
m^2_{\tilde{l}_L}-m^2_{\tilde{\nu}_l}=-\cos{(2\beta)}m_W^2.
\end{equation}
When $\tan{\beta} > 1$, as in our pMSSM model
set, the electron (muon) sneutrino is always
lighter than the left-handed selectron (smuon).  This means the left-handed selectron (smuon) is at best
the third lightest sparticle and thus generically has a sufficient value of
$\Delta m$ to decay promptly.  In fact, the minimum width for such
particles in the model set is $\sim 2$ keV, far greater than
$\Gamma_{\text{stable}}$.  As the right-handed sleptons are $SU(2)$
singlets, there is no similar effect, and they can have arbitrarily small
$\Delta m$; models with cosmologically stable right-handed sleptons were excluded when our pMSSM
model set was generated.
A similar situation holds for squarks, where there are
fewer models where $\tilde{d}_L$ or $\tilde{s}_L$ have close mass splittings.

  \FIGURE{
    \label{selectrons-and-smuons}
    \includegraphics[width=0.7\columnwidth]{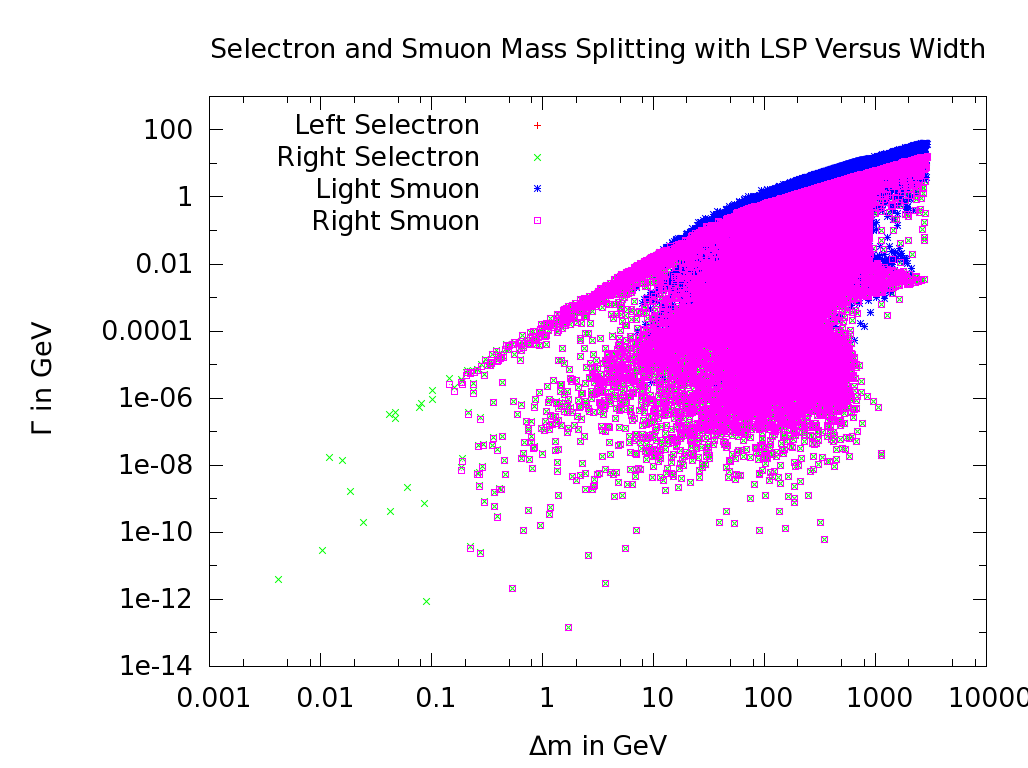}
    \caption{The width as a function of the mass splitting with the
      LSP ($\Delta m$) for the left- and right-handed selectrons and smuons.
      Note that as we do not include four-body smuon decays,
      when $\Delta m < m_\mu$ the smuon width is set to 
      zero; thus for small $\Delta m$ we see points corresponding to
      right-handed selectrons without the corresponding smuon point.}
  }


\begin{table}
\begin{center}
\begin{tabular}{|l|c|} \hline\hline 

  $\tilde{g}$   &     $12598$ \\
  $\tilde{u}_L$ &     $9628$ \\
  $\tilde{c}_L$ &     $9629$ \\
  $\tilde{u}_R$ &     $22667$ \\
  $\tilde{c}_R$ &     $22668$ \\ 
  $\tilde{d}_L$ &     $13595$ \\
  $\tilde{s}_L$ &     $13595$ \\
  $\tilde{d}_R$ &     $27996$ \\
  $\tilde{s}_R$ &     $27998$ \\
  $\tilde{b}_1$ &     $13355$ \\
  $\tilde{b}_2$ &     $431$ \\
  $\tilde{t}_1$ &     $5695$ \\
  $\tilde{t}_2$ &     $1$ \\
\hline\hline

\end{tabular}
\end{center}
\caption{The number of squarks or gluinos of the indicated
  species with widths $< 100$ MeV.  This gives a rough idea of
  the number of models where R-hadrons would be formed; however in
  most cases the R-hadrons decay promptly in the
  detector.}
\label{R hadron}
\end{table}

\FIGURE{
  \label{up-squarks}
  \includegraphics[width=0.7\columnwidth]{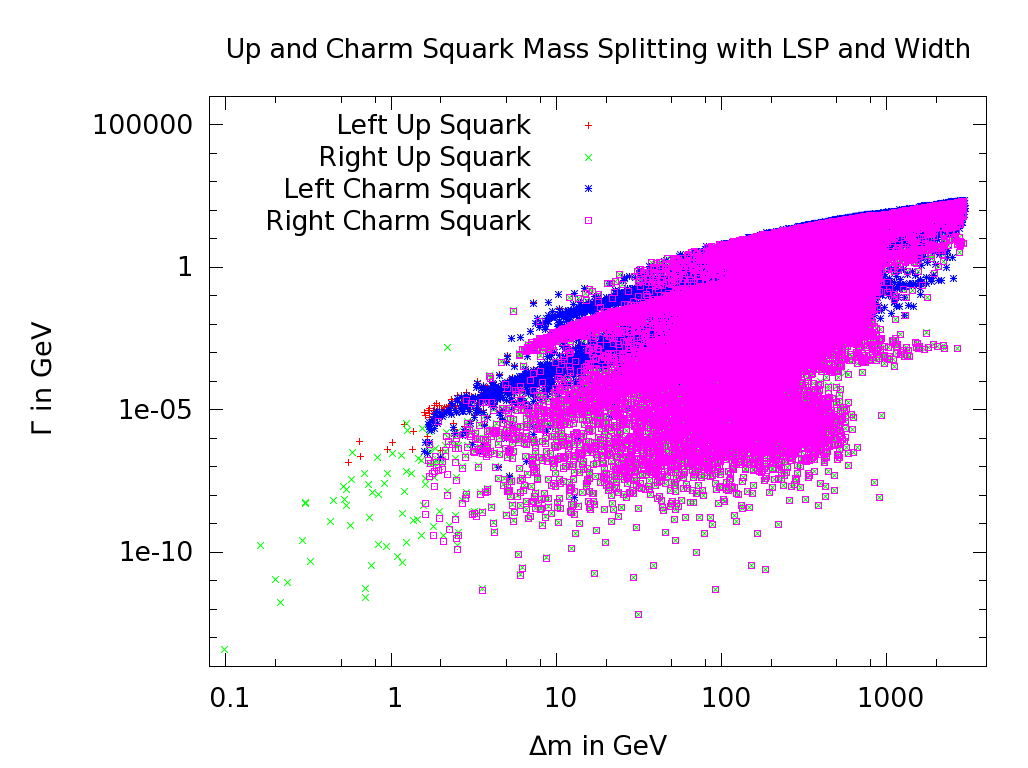}
  \caption{The distribution of widths for up 
    and charm squarks as a function of $\Delta m$, the mass splitting
    between the sparticle and the LSP. }
}
\FIGURE{
  \label{down-squarks}
  \includegraphics[width=0.7\columnwidth]{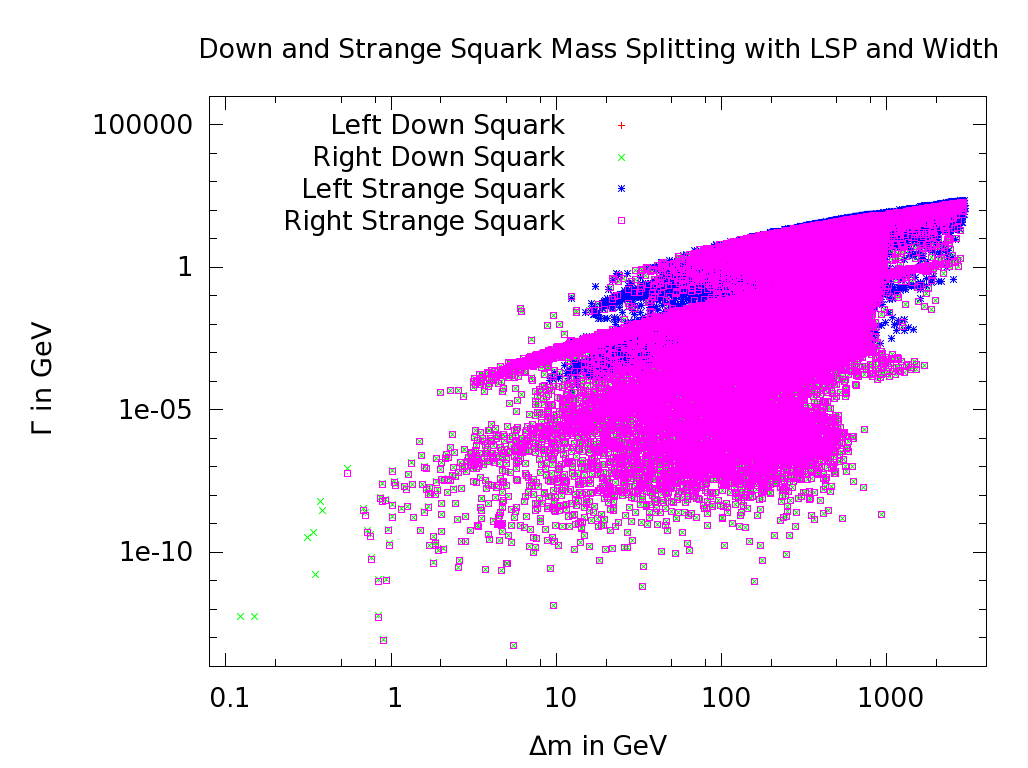}
  \caption{The distribution of widths for down 
    and strange squarks as a function of $\Delta m$, the mass splitting
    between the sparticle and the LSP. }
}

\FIGURE{
  \label{staus}
  \includegraphics[width=0.7\columnwidth]{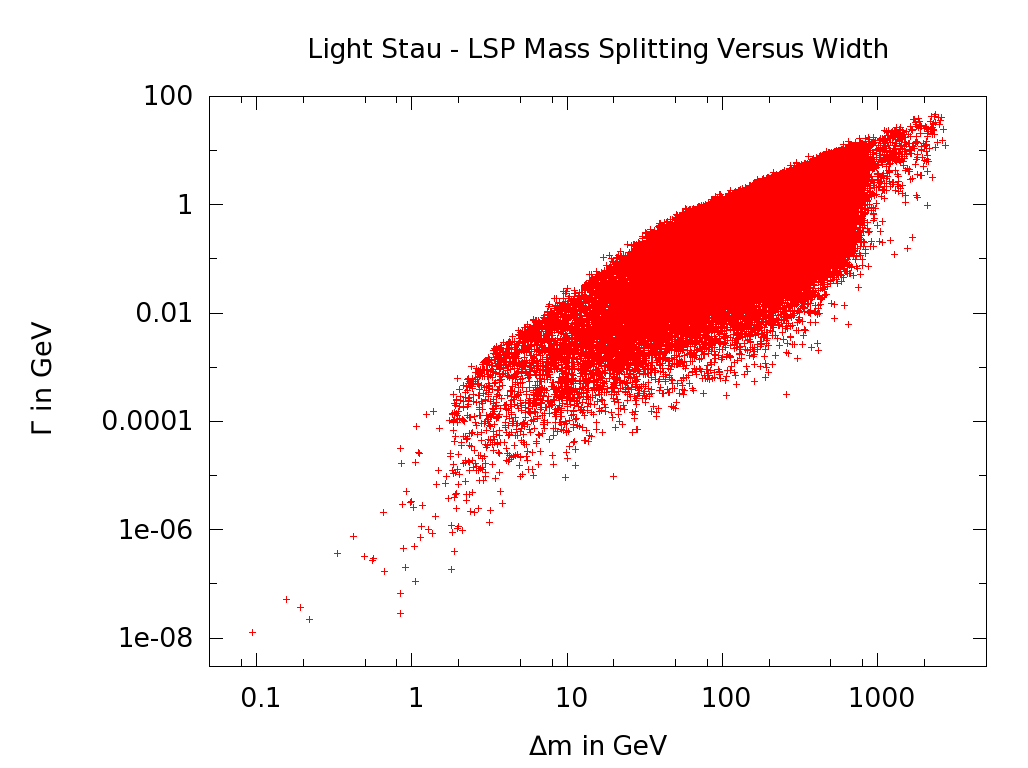}
  \caption{The distribution of widths for the
    lighter stau mass eigenstate as a function of $\Delta m$, the mass
    splitting between the sparticle and the LSP.}
}

\FIGURE{
  \label{stop-sbottom}
  \includegraphics[width=0.7\columnwidth]{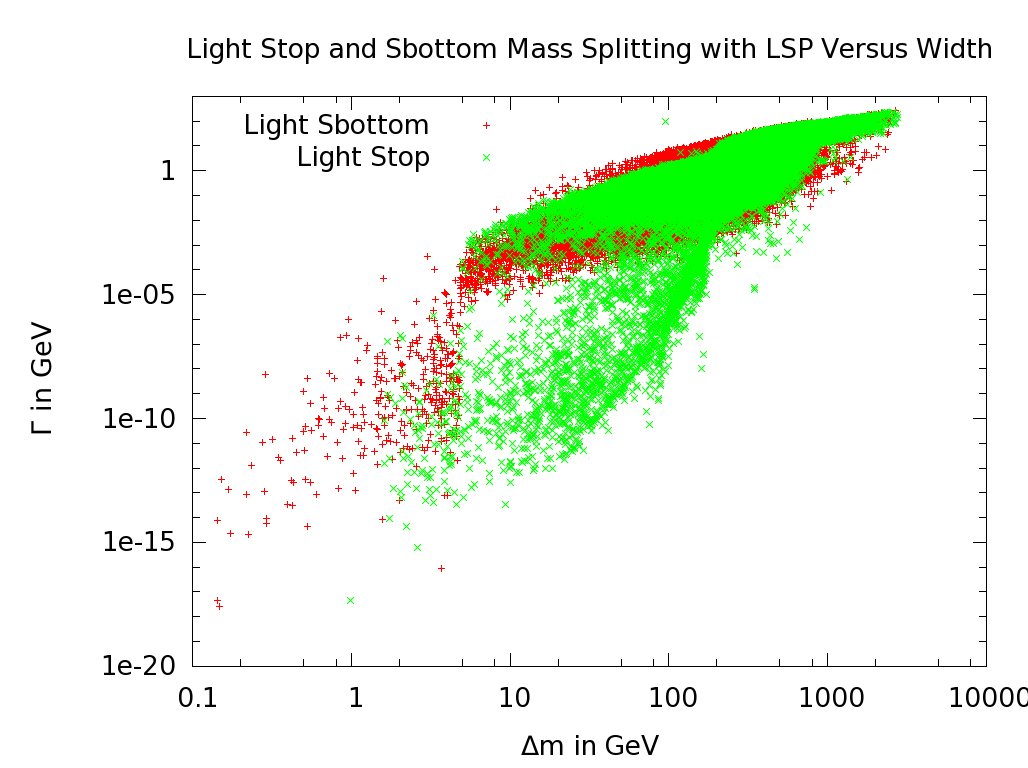}
   \caption{The distribution of widths for the lightest stop and sbottom
    as a function of $\Delta m$, the mass
    splitting between the sparticle and the LSP.  Note: in $65$ models the calculated light stop width is zero,
    representing nearly all of the detector-stable stops in the model
    set.   Likewise there are
    $9$ models in which the calculated light sbottom width is zero.}
}

\FIGURE{
  \label{sneutrino}
  \includegraphics[width=0.70\columnwidth]{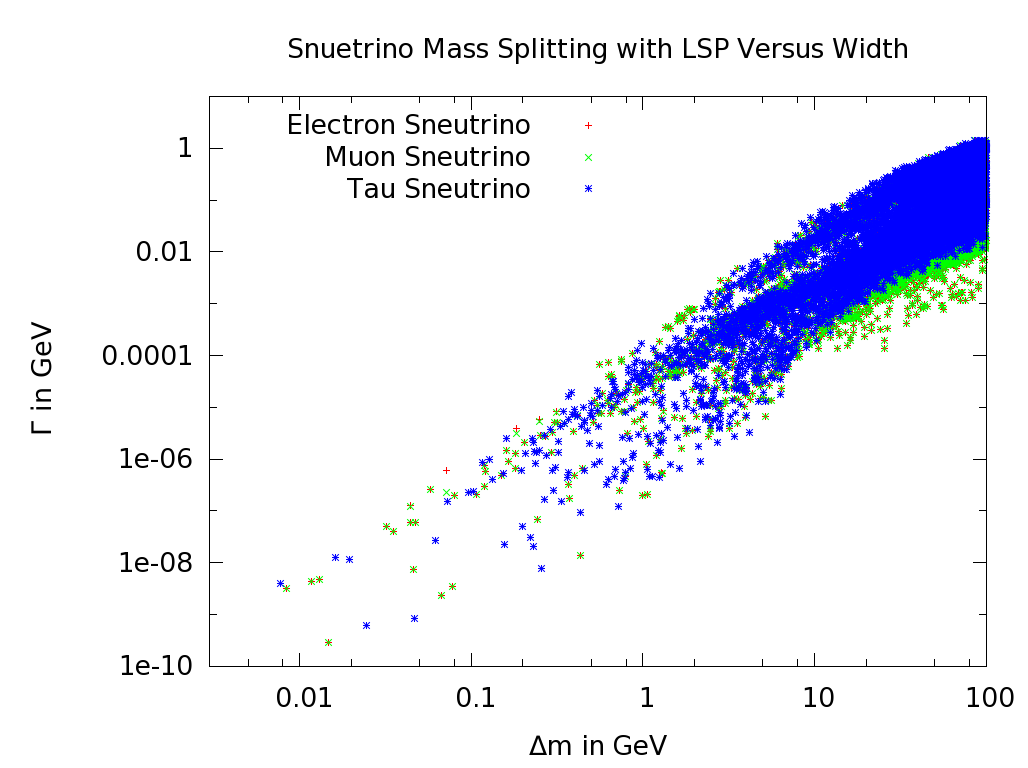}
  \caption{The distribution of widths for the
    three species of sneutrinos as a function of $\Delta m$, the mass
    splitting between the sparticle and the LSP. The minimum width for the
    electron or smuon sneutrinos in our model set is $\approx 3 \times 
    10^{-10}$ GeV.}
}

\subsection{Detector-Stable Sparticle Prospects}

\TABLE{
\begin{tabular}{|l|c|c|} \hline\hline 
Sparticle & LHC Reach $1$ fb$^{-1}$ GeV & LHC Reach $10$ fb$^{-1}$ \\ 
\hline \hline
 $\tilde\chi^+$ (Wino-like)& $ 365$ GeV & $ 467$ GeV \\
 $\tilde\chi^+$ (Higgsino-like)& $ 280$ GeV & $376 $ GeV \\ 
 $\tilde\tau$ (or $\tilde\mu$)     & $ 145$ GeV & $ 198$ GeV \\
 $\tilde t$ (or $\tilde c$)    & $ 562$ GeV & $ 681$ GeV \\
 $\tilde b$ (or $\tilde s$) & $562$ GeV & $681$ GeV \\
 $\tilde g$        & $> 1000$ GeV & $> 1000$ GeV \\
\hline\hline
\end{tabular}
\caption{The approximate $14$ TeV LHC search reach for stable particles of a given 
  type with $1$ and $10$
  fb$^{-1}$\cite{Raklev:2009mg}. These search reaches assume the sparticles are
  produced in the hard subprocess, rather than in cascade decays.  For simplicity, we take the LHC
  reach for sbottoms to be equal to that for stops.}
\label{reach}
}

In Table~\ref{reach} the approximate $14$ TeV LHC reach for each
sparticle, using~\cite{Raklev:2009mg}, is presented for the specified
integrated luminosities.  These bounds are somewhat conservative, 
as they only take pair production into account; it is difficult to determine
a model independent reach including detector stable particles produced in
cascade decays.  
We assume this reach is generation
independent.   (This may not
necessarily be the case for stable up-squarks, for example, as in such a case there
could be significant contributions from t channel production.)
Further, we make the conservative assumption that we can neglect the
$\approx 1$ fb$^{-1}$ of data that will be collected at $7$ TeV.  Nonetheless,
one can obtain a qualitatively accurate picture of the
prospects for LHC discovery for detector-stable sparticles in this
pMSSM model set.

Table~\ref{stable-results} shows the number of detector-stable ($\Gamma <
10^{-17}$ GeV) sparticles of various species in the model set, 
as well as the number of such sparticles which would have evaded
discovery at the LHC with $1$ and $10$ fb$^{-1}$ of integrated luminosity, following
\cite{Raklev:2009mg}.  In addition,
the LHC search reach for charginos, and its effectiveness in discovering or
excluding detector-stable charginos in our model set, is shown in
Figure~\ref{chargino-reach}.

\TABLE{
\begin{tabular}{|l|c|c| c | } \hline\hline 
Sparticle & In Model Set & LHC Reach $1$ fb$^{-1}$ & LHC Reach $10$ fb$^{-1}$ \\ \hline \hline
 $\tilde\chi_1^+$ & $8642$ & $560$ & $72$  \\
 $\tilde\tau_1$     & $179$ & $ 179$ & $179$  \\
 $\tilde t_1$     & $66$ & $4$ & $0$  \\
 $\tilde c_R$ & $49$ & $0$ & $0$  \\
 $\tilde\mu_R$  & $17$ & $16$ & $16$  \\
 $\tilde b_1$    & $11$ & $0$ & $0$  \\
 $\tilde c_L$  & $8$ & $0$ & $0$  \\
 $\tilde s_R$ & $8$ & $0$ & $0$  \\
 $\tilde g$    & $5$ & $0$ & $0$  \\
 \hline\hline
\end{tabular}
\caption{
  The number of stable particles of various types present in our
  pMSSM model set and the number that would not have been
  discovered with $1$ and $10$ fb$^{-1}$ at $14$ TeV, following~\cite{Raklev:2009mg}.  
  Note that the LHC will be
  more efficient at discovering or excluding stable squarks, gluinos, or
  charginos than sleptons.}
\label{stable-results}
}

 \FIGURE{
    \label{chargino-reach}
    \includegraphics[width=0.7\columnwidth]{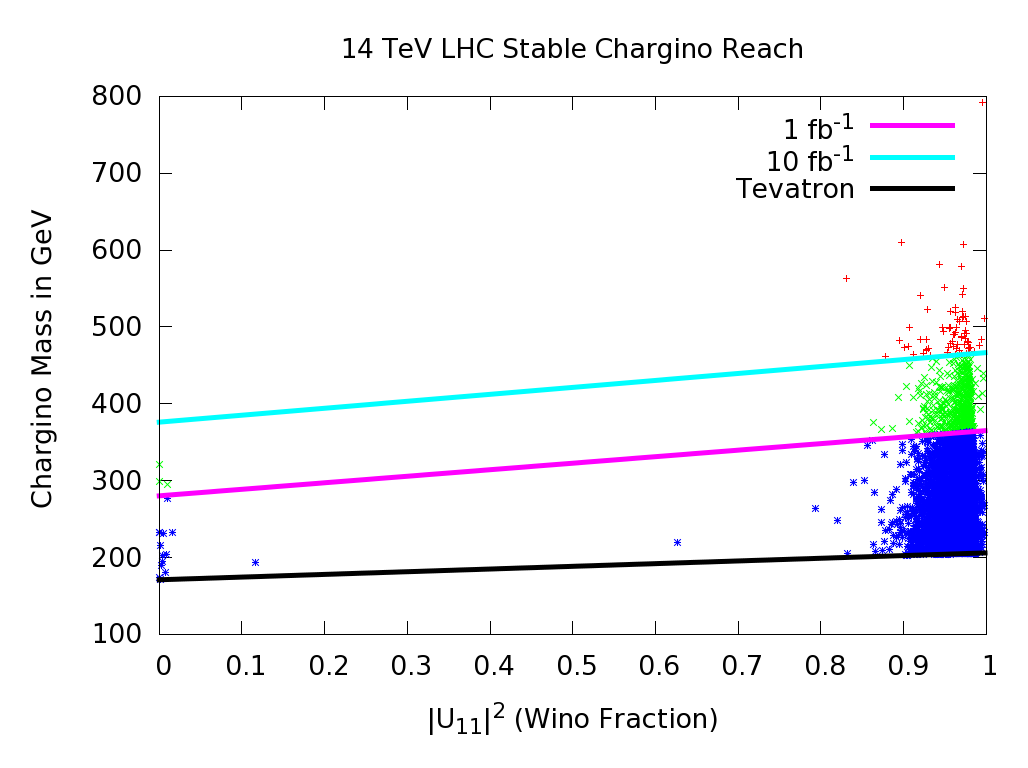}
    \caption{The $14$ TeV LHC search reach for stable wino-like charginos with
      $1$ and $10$ fb$^{-1}$ of integrated luminosity
      following~\cite{Raklev:2009mg}.  As noted above, these reaches 
      assume pair production of the charginos.   
      We also show the Tevatron reach after $1.1$ fb$^{-1}$ 
      \cite{Abazov:2008qu}.}
}

\subsubsection*{Complementarity with Astrophysics}

Since detector-stable particles are nearly degenerate with the LSP,
they provide important channels for co-annihilation in the early
universe.  Thus, we expect models with detector-stable
particles to have lower values for the relic density.  In our pMSSM model set,
there are no models with a detector stable particle with a relic
density greater than $\Omega h^2 = 0.089$.  Thus, the subset of these
models considered in special detail in~\cite{Cotta:2010ej}, all of
which have $\Omega h^2 > 0.1$, do not have detector-stable particles.
This suggests that the discovery of a detector-stable particle at the
LHC would have important consequences for cosmology, as
well as for particle physics.



\section{Conclusions}
\label{conclusions}

In this paper we have investigated the robustness of the ATLAS SUSY searches prepared for the 14 TeV LHC.  We passed an extensive set of
broad-based SUSY models through the full planned SUSY analysis suite, following the analysis procedure designed by ATLAS in detail.  We employed
our previously generated $\sim$ 71k model points in the 19-dimensional pMSSM parameter
space. We simulated the eleven ATLAS SUSY \MET\ analyses, as well as the stable charged particle search, 
which were originally designed for an exploration of mSUGRA-based models. To test our approach, we first applied our analysis to the set of ATLAS 
mSUGRA model benchmark points and successfully reproduced the published results obtained previously by ATLAS. We found that for the case of the $\sim 68$k models 
from the flat  
prior scan, where the squarks and gluinos have masses below $\lesssim 1$ TeV, essentially all of the pMSSM points ($>99\%$) were observable in at least one of the ATLAS 
\MET\ analyses allowing for an uncertainty of $50\%$ in the SM background with 1 fb$^{-1}$ of integrated luminosity. 
Even this excellent level of parameter space coverage was seen to improve when these 
systematic errors 
were reduced to $20\%$.  Furthermore, most of these pMSSM models were
found to give significant signals in several of these \MET\ analysis channels simultaneously. In the 
log prior model sample, totalling
$\sim 3$k models, the success rate for discovery fell to approximately $\sim 68(81)\%$  for an assumed $50(20\%)$ SM background error.  This is also 
quite impressive since the sparticle masses could be as large as 3 TeV in this model set. We emphasize that these statistics apply only to our pMSSM
model set, but believe they are indicative of the performance of the LHC SUSY searches in a broader SUSY parameter space.
In summary, although they were designed for mSUGRA, we found that the ATLAS SUSY 
search analyses are quite powerful in their ability to cover the points in the pMSSM model space.  This is quite reassuring!

Model points that were {\it not} observable by any of the ATLAS search 
analyses were found to be relatively few in number, \ie, below $\sim 0.57(0.02)\%$ in the case of the 
flat prior sample assuming a $50(20)\%$ systematic uncertainty in the background. 
The main reasons why these models were missed can be summarized as follows: ($i$) In the most trivial cases, the colored sparticles, which have the 
largest production cross sections, were found to have kinematically suppressed production rates since these particles were heavy. This was a much more common 
occurrence in the case of the log prior model set where the masses of the squarks and gluinos were allowed to be as large as $\sim 3$ TeV. 

($ii$) Many models contain charginos that are close in mass to the LSP due to the high proportion of the occurrence of Higgsino-like or 
Wino-like LSPs in our model set, unlike in most mSUGRA models. For some models there were substantially large branching fractions for squarks to decay to these 
charginos. In such cases, essentially stable charginos were then found to occur at the end of most decay chains thus leading to a reduction in the average amount 
of \MET\ that was produced in typical events. Since the ATLAS analyses
required a fairly large amount of \MET\ to obtain significant observable signals, these models 
were more easily undetected. 
Reducing the ATLAS \MET\ requirements might allow access to some of these models at the expense of increased SM backgrounds; this  
requires further study. Also for such cases, searches for stable charged particles become of great importance, particularly when these states appear, 
as they more commonly do, at the end of long SUSY cascade decay chains and are
not simply pair produced in isolation. We found that that the $\beta\gamma$ distribution was quite different for stable charginos produced in
cascade decays than for those directly produced in hard processes; the observability of such stable particles requires further study.

($iii$) Some models in our pMSSM set have a rather 
compressed mass spectra.  This results in a significant reduction in phase space which is available in the various decays and, hence, in a corresponding decrease 
in the values of $p_T$ available for the final state jets and leptons. These final state objects
were then too soft to satisfy the necessary analysis cuts. 

($iv$) Processes with large 
backgrounds have an associated correspondingly large systematic uncertainty of $\delta B=(50\%,20\%)$. In order to reach a significance of 5 or more, 
a requirement for 
discovery, a substantially larger number of signal events must be produced. For example, the 4j0l(2j0l) analysis requires 1759(2778) signal events to reach the 
$S=5$ level assuming a luminosity of 1~fb$^{-1}$ provided $\delta B=50\%$. However, a reduction in the background uncertainty to $\delta B=20\%$, 
substantially decreases the required number of 
signal events to only 721(1129) to reach $S=5$. Thus it is clear that a reduction in the systematic uncertainty in the SM 
background is very important in order 
to increase the coverage of the pMSSM model space. In cases where the large background uncertainties were important, we found that increasing the 
luminosity by a factor of 10 was not very useful in increasing the parameter space coverage.  This is expected due to the corresponding dominance of the 
background systematic errors.

The study presented in this paper suggests a number of areas for future work.
In light of the current status of the LHC, repeating this analysis with a 7~\tev\ 
center-of-mass energy and perhaps somewhat lower luminosities is an obvious next step \cite{atlas10}; such
a study is now underway.

The preliminary study of stable particles in the MSSM presented here makes it
clear that more work could be performed in this area.  A more detailed modeling of the
interactions and decays of R-hadrons, for example, is necessary to accurately
predict their signatures at the LHC.  Searches for stable particles produced
in decay chains, rather than pair-produced in the hard process, also deserves
significant study.

As always, it would be interesting to explore ways to optimize the kinematic cuts or
otherwise modify the search analyses to obtain a better performance across the general
MSSM.  This would require generating
the actual background events, so that various distributions could be examined.

In summary, we found that the standard SUSY search analyses, taken together,
provide excellent coverage of the MSSM parameter space at the LHC with relatively small luminosity, at least for
sparticle masses up to $\sim\tev$.  We conclude that the prospects for observing Supersymmetry in the early
running of the LHC are quite good!

\acknowledgments

This project would not have been possible without the assistance and input 
from many people. 

The authors thank the members of the ATLAS SUSY group, in 
particular, S.~Caron, P.~ de Jong, G.~Polesello and G.~Redlinger, for 
discussions and, most importantly, for providing us with additional 
information detailing the ATLAS-generated SM background distributions for 
their SUSY studies and the corresponding results for their mSUGRA benchmark 
analyses.

We also thank the SLAC ATLAS group, as well as the SLAC Computing
Division itself, for assistance with computing 
resources.

We thank T.~Plehn for his help in overcoming the special problems 
we had with the implementation of PROSPINO for our pMSSM model set. We would 
also like to thank A.~Djoudai and J.~Conway for similar help with the 
implementations of SDECAY and PGS, respectively.

We thank L.~Dixon for discussions about the theoretical 
assessment of the size of the systematic errors for the SM backgrounds to 
SUSY signals. 
We thank R.~Cotta for important discussions related to our 
analysis.
We thank J.~Cogan for his assistance with setting up the 
ATLAS analysis code during the earlier phase of this work.

Work supported by the Department of Energy, Division of High
Energy Physics, Contracts DE-AC02-76SF00515 DE-AC02-06CH11357, and
DE-FG02-91ER40684, and by the BMBF ``Verbundprojekt HEP-Theorie'' under
contract 05H09PDE.

\bibliography{susybib}

\begin{thebibliography}{10}
\bibitem{ATLAS_SUSY}
We would like to thank the various members of the ATLAS SUSY group with
  providing us with this information on SM backgrounds as well as for many
  discussions during the course of this analysis.
\end{thebibliography}
\bibliographystyle{JHEP} 

\end{document}